\documentclass[aps,prb,twocolumn,amssymb,amsmath,superscriptaddress]{revtex4-1}

\usepackage{bm}
\usepackage{color}
\usepackage{graphicx}
\usepackage{dcolumn}
\usepackage{graphicx}
\usepackage[colorlinks=true, letterpaper=true, pdfstartview=FitV, linkcolor=blue, citecolor=blue, urlcolor=blue]{hyperref}
\usepackage[normalem]{ulem}
\usepackage{amsmath}
\usepackage{epstopdf}
\usepackage{multirow}
\usepackage{slashed}
\usepackage{float}
\usepackage{pifont}
\usepackage{amsmath}

\setcitestyle{square}

\makeatletter
\renewcommand\@biblabel[1]{#1.}
\makeatother

\begin{document}
	
	\title{Magnetic-Field and Strain Engineering of Modulated Transverse Transport in Altermagnetic Topological Materials}
	
	\author{Xiuxian Yang}
	\affiliation{Jiangsu Key Laboratory of Extreme Multi-Field Materials Physics, School of Physics and Electronic Engineering, Jiangsu Normal University, Xuzhou 221116, China}
	
	\author{Xiaodong Zhou}
	\email{zhouxiaodong@tiangong.edu.cn}
	\affiliation{School of Physical Science and Technology, Tiangong University, Tianjin 300387, China}

	\author{Jingming Shi}
	\affiliation{Jiangsu Key Laboratory of Extreme Multi-Field Materials Physics, School of Physics and Electronic Engineering, Jiangsu Normal University, Xuzhou 221116, China}

	\author{Shifeng Qian}
	\email{qiansf@ahnu.edu.cn}
	\affiliation{Anhui Province Key Laboratory for Control and Applications of Optoelectronic Information Materials, Department of Physics, Anhui Normal University, Anhui, Wuhu 241000, China}	

	\author{Xiaotian Wang}
	\email{xiaotianw@uow.edu.au}	
	\affiliation{Institute for Superconducting and Electronic Materials, Faculty of Engineering and Information Sciences, University of Wollongong, Wollongong 2500, Australia}
	
	\author{Wenhong Wang}
	\affiliation{School of Electronic and Information Engineering, Tiangong University,
	Tianjin 300387, China}

	\author{Yinwei Li}
	\email{yinwei$_$li@jsnu.edu.cn}
	\affiliation{Jiangsu Key Laboratory of Extreme Multi-Field Materials Physics, School of Physics and Electronic Engineering, Jiangsu Normal University, Xuzhou 221116, China}
	
	
	
	
	\date{\today}
	
	\begin{abstract}
	Here, we explore the role of inherent altermagnetic topology in transverse transport phenomena (such as crystal/anomalous Hall, Nernst, and thermal Hall effects) in several famous altermagnets, including tetragonal \textit{X}V$_2$\textit{Y}$_2$O (\textit{X} = K, Rb, Cs; \textit{Y} = S, Se, Te), RuO$_2$, MnF$_2$, as well as hexagonal CrSb and MnTe. Notably, in \textit{X}V$_2$\textit{Y}$_2$O, the first experimentally realized layered altermagnets, transverse transport is governed by altermagnetic pseudonodal surfaces, emphasizing the purely topological contributions to  transverse transport. Interestingly, we demonstrate that strain engineering and magnetic field, two unique methods for selectively controlling crystal and anomalous transport, can substantially enhance the magnitude of these phenomena while preserving the alternating spin characteristics in both real and momentum space. Moreover, due to the spin symmetry breaking via shear strain, a new magnetic phase, fully compensated ferrimagnetism, with isotropic spin splitting, can be induced. Our findings provide effective strategies not only for manipulating transverse transport in altermagnets but also for controlling magnetic phase transitions, offering valuable insights for their potential applications in spintronics and spin caloritronics.

	\end{abstract}
	
	\maketitle
	
	Anomalous transport phenomena, which are a type of transverse transport, including the anomalous Hall effect (AHE), anomalous Nernst effect (ANE), and anomalous thermal Hall effect (ATHE), are fundamental effects in condensed matter physics with important applications in spintronics and spin caloritronics~\cite{Nagaosa2010,XiaoD2006,Qin2011,IgorZutic2004,Gerrit2012}.  The AHE, ANE, ATHE refer, respectively, to the emergence of a transverse electrical current (thermoelectric current, thermal current) in response to a longitudinal electric field (temperature gradient) in the absence of an external magnetic field.  These anomalous transport phenomena are typically observed in ferromagnets and proportional to the spontaneous magnetization~\cite{Jungwirth2002,Fang2003,YaoYG2004,Onoda2006,Onoda2008,Onose_Lor2008,Shiomi_Lor2009,Shiomi_Lor2010}, whereas antiferromagnets (AFMs) have traditionally been thought to lack these effects due to their zero net magnetization.  However, this notion has been challenged in certain noncollinear AFMs with vector or scalar spin chirality\cite{Ikhlas2017,LiXiaoKang2017,GuoGY2017,LC-Xu2020,sugii2019,XD-Zhou2020,Zhou2019,Shiomi2013,Hirschberger2020,ZhangH2021,ZhouJ2016,Hanke2017,FengWX2020}.  The underlying physics is that the AHE, ANE, and ATHE are driven by the breaking of specific symmetries ($\mathcal{TS}$, here $\mathcal{T}$ is time reversal symmetry and $\mathcal{S}$ is spatial symmetries including inversion or translation), rather than the net magnetization itself. Therefore, theoretically, these effects can also occur in more commonly available collinear AFMs if the $\mathcal{TS}$ symmetry is removed~\cite{Libor2020,xdzhou2024}. 


	Recently, a third type of collinear magnetic ground state, termed altermagnetism (AM), has been theoretically proposed based on nonrelativistic spin-symmetry groups~\cite{Libor2022,Libor2022b,Mazin2022}. AM is characterized by zero net magnetization, nonrelativistic  alternating spin splitting, and spin-degenerate nodal topology, which has garnered significant attention in this emerging field~\cite{HayamiS2019,xdzhou2021,MaHY2021,feng2022,H-Bai2023,xdzhou2024,L-Bai2024,Ouassou2023,YXLi2023,DiZhu2023,zhang2024,Fender2025JACS,guo_AM2025}.  Subsequent experiments have confirmed momentum-dependent spin splitting in altermagnets (AMs) $\alpha$-MnTe~\cite{Li2024,krempasky2024}, CrSb~\cite{Reimers2024,DingJY2024,YangGW2024,ZengM2024,zhouzhiyuan2025}, KV$_2$Se$_2$O~\cite{jiang_NP2025}, Rb$_{1-\delta}$V$_2$Te$_2$O~\cite{ZhangFY2024}, and the controversial RuO$_2$~\cite{H-Bai2022,H-Bai2023,Karube2022,feng2022,LinZH2024,Fedchenko2024,chen_AM2025,Jeong2025,Berlijn2017,ZhuZH2019,LiuJY2024,Hiraishi2024,ZhangYC2024,Noh2025}.   Among experimentally confirmed AMs, the room-temperature metallic compounds \textit{X}V$_2$\textit{Y}$_2$O (\textit{X} = K, Rb, Cs; \textit{Y} = S, Se, Te) represent the first known layered AMs~\cite{jiang_NP2025,ZhangFY2024} and have emerged as a potentially candidate for two-dimensional AM~\cite{MaHY2021,QR-Cui2023,Zhuyu2024} due to the possibility of K, Rb or Cs atoms deintercalation via topochemical process. In AMs, the two oppositely-spin sublattices are connected via crystal rotational symmetry instead of inversion or translation symmetry in conventional AFMs.  The resulting combined $\mathcal{TS}$ symmetry breaking activates a ``crystal'' version of transverse transport, which is highly sensitive to the crystal environment and crystal fields~\cite{Libor2020,Samanta2020,xdzhou2021,Mazin2021,ShaoDF2021,feng2022,Libor2022,xdzhou2024}, representing a distinct type of transverse transport known as the crystal Hall effect (CHE). Unlike the conventional AHE, which is $\bm{m}$-odd and scales with the net magnetization $\bm{m}$, the CHE is $\bm{m}$-even, independent of the magnitude of $\bm{m}$~\cite{Libor2020,Libor_NRM2022}.  Interestingly, experimental studies have reported substantial transverse Hall conductivity in AMs~\cite{feng2022,Gonzalez_CrSb_PRL2023,zhouzhiyuan2025,Takagi_FeS2_NM2025,Jeong2025}, further confirming the presence of symmetry-driven crystal transport.

	In this work, we comprehensively investigate how strain engineering and external magnetic fields interact with the altermagnetic nodal topology\textemdash a universal feature in AMs\textemdash to govern transverse transport properties, including crystal and anomalous transport. Utilizing symmetry analysis and state-of-the-art first-principles calculations, we explore the underlying mechanisms of transverse transport in representative AMs, including tetragonal \textit{X}V$_2$\textit{Y}$_2$O, RuO$_2$, MnF$_2$, and hexagonal CrSb and MnTe, as well as their responses to external perturbation, such as strain engineering and magnetic field.  Our results reveal that in \textit{X}V$_2$\textit{Y}$_2$O, the altermagnetic topology plays a crucial role in shaping transverse transport behavior.  Notably, we demonstrate that strain engineering can selectively control of the crystal transport component and even activate from-zero-to-finite responses, while an external magnetic field enhances the anomalous transport contribution, thus allowing independent manipulation of the two transport channels.  Another noteworthy observation is that shear strain can drive a transition from AM to fully compensated ferrimagnetism~\cite{Kawamura2024,LD-Yuan2024,YC-Liu2025,JQ-Feng2025}.

	\begin{figure}
		\includegraphics[width=1\columnwidth]{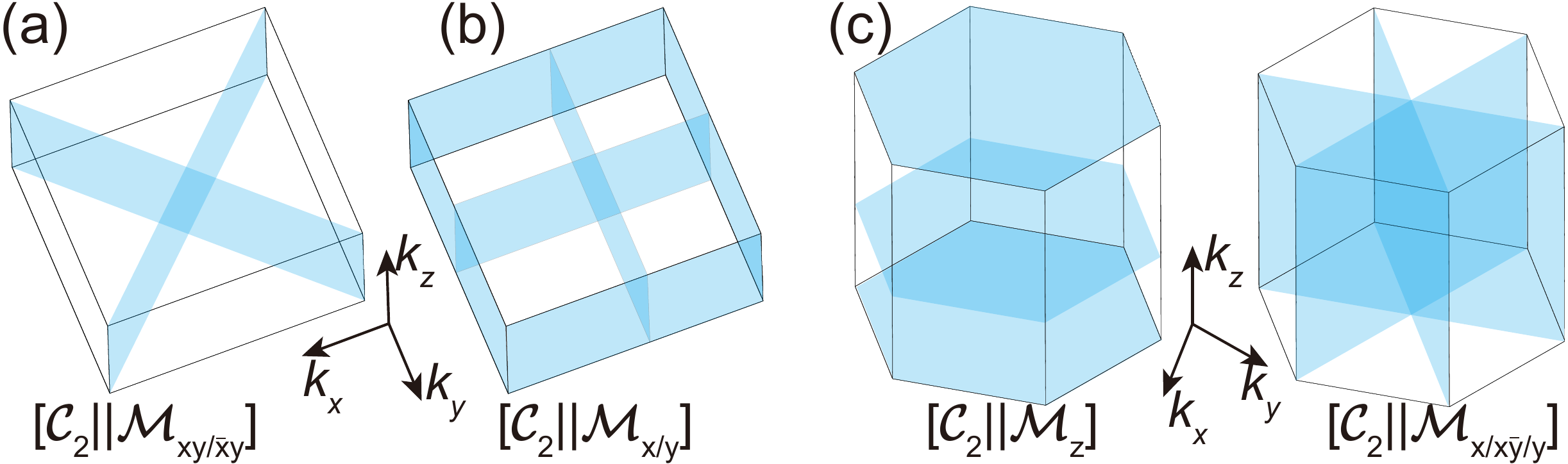}
		\caption{3D first Brillouin zone (BZ) and spin-degenerate nodal surfaces for (a) \textit{X}V$_2$\textit{Y}$_2$O (\textit{X} = K, Rb, Cs; \textit{Y} = S, Se, Te) protected by [$\mathcal{C}_2||\mathcal{M}_{xy/\bar{x}y}$], (b) RuO$_2$ and MnF$_2$ protected by [$\mathcal{C}_2||\mathcal{M}_{x/y}$], (c) CrSb and $\alpha$-MnTe protected by [$\mathcal{C}_2||\mathcal{M}_{z}$] (left panel) and [$\mathcal{C}_2||\mathcal{M}_{x/x\bar{y}/y}$] (right panel), respectively.}
		\label{fig:surface}
	\end{figure}

	We start with the introduction of altermagnetic pseudonodal surfaces guided by the spin-symmetry principles. In AMs, the nonrelativistic spin splitting is not universally forbidden at time-reversal invariant momenta, whereas the spin-degenerate nodal lines, rings, and surfaces are protected by the spin group symmetries~\cite{Mazin2021,Libor2022,Libor2022b,Li2024,xdzhou2024}. For example, in the case of \textit{X}V$_2$\textit{Y}$_2$O, the nonrelativistic symmetry space group contains the following symmetry operations,
	\begin{eqnarray}
		&[\mathcal{E} || \{\mathcal{E}, \mathcal{P}, \mathcal{C}_{2x}, \mathcal{C}_{2y}, \mathcal{C}_{2z}, \mathcal{M}_{x}, \mathcal{M}_{y}, \mathcal{M}_{z}\}] + \notag \\
		\ [\mathcal{C}_{2}||\{&\mathcal{C}_{4z}^+, \mathcal{C}_{4z}^-, \mathcal{C}_{xy}, \mathcal{C}_{\bar{x}y}, \mathcal{PC}_{4z}^+, \mathcal{PC}_{4z}^-, \mathcal{M}_{xy}, \mathcal{M}_{\bar{x}y}\}].
	\end{eqnarray}	
	Here, the $\mathcal{E}$, $\mathcal{P}$, $\mathcal{C}$, $\mathcal{M}$ represent identity, space inversion, rotation, and mirror operations, respectively. The symbols $+$ ($-$) indicate counterclockwise (clockwise) rotations and ``$||$'' is used purely as a separator to distinguish between operations in spin and real spaces. The appearance of nodal surfaces in AMs can be understood by analyzing the action of symmetry operations in momentum space. As an example, consider the action of the operation $[\mathcal{C}_{2}||\mathcal{M}_{xy}]$ on energy band $E(k_x, k_y, k_z, s)$,
	\begin{eqnarray}
		[\mathcal{C}_{2}||\mathcal{M}_{xy}]E(k_x, k_y, k_z, s) = E(-k_y, -k_x, k_z, -s),
	\end{eqnarray}	
	where $s$ denotes spin. Therefore, when $k_x$ = -$k_y$, the energy satisfies $E(k_x, k_y, k_z, s)$ = $E(k_x, k_y, k_z, -s)$, two spins are degenerate.  The action of other operations on the $E(k_x, k_y, k_z, s)$ can be found in the \textcolor{blue}{Supplemental Material (SM), Section I}~\cite{SuppMater}.  As a result, spin-degenerate nodal surfaces at the $\mathcal{M}_{xy}$ and $\mathcal{M}_{\bar{x}y}$ invariant planes are protected by [$\mathcal{C}_2||\mathcal{M}_{xy/\bar{x}y}$] operations, which is highlighted in Fig.~\ref{fig:surface}(a). For RuO$_2$ and MnF$_2$, the [$\mathcal{C}_2||\mathcal{M}_{x/y}$] operations uphold spin degeneracy across the $\mathcal{M}_{x/y}$ invariant planes [Fig.~\ref{fig:surface}(b)]. In the case of CrSb and $\alpha$-MnTe~\cite{Li2024}, the [$\mathcal{C}_2||\mathcal{M}_{z}$] and [$\mathcal{C}_2||\mathcal{M}_{x/x\bar{y}/y}$] operations preserve spin degeneracy at $\mathcal{M}_z$ and $\mathcal{M}_{x/x\bar{y}/y}$ invariant planes, as shown in Fig.~\ref{fig:surface}(c).  These spin group symmetries are shown in the \textcolor{blue}{SM, Section. II} and \textcolor{blue}{III}~\cite{SuppMater}.  When the spin-orbit coupling (SOC) is introduced, the spin group symmetries and the protected nodal surfaces are broken, leading to the emergence of enhanced Berry curvature and crystal transport near the gapped nodal surfaces (pseudonodal surfaces), as revealed below for the representative \textit{X}V$_2$\textit{Y}$_2$O. Additional details for AMs such as RuO$_2$, MnF$_2$, CrSb, and $\alpha$-MnTe are provided in Fig.~\textcolor{blue}{S3}~\cite{SuppMater}. 

	\begin{figure}
	\includegraphics[width=1\columnwidth]{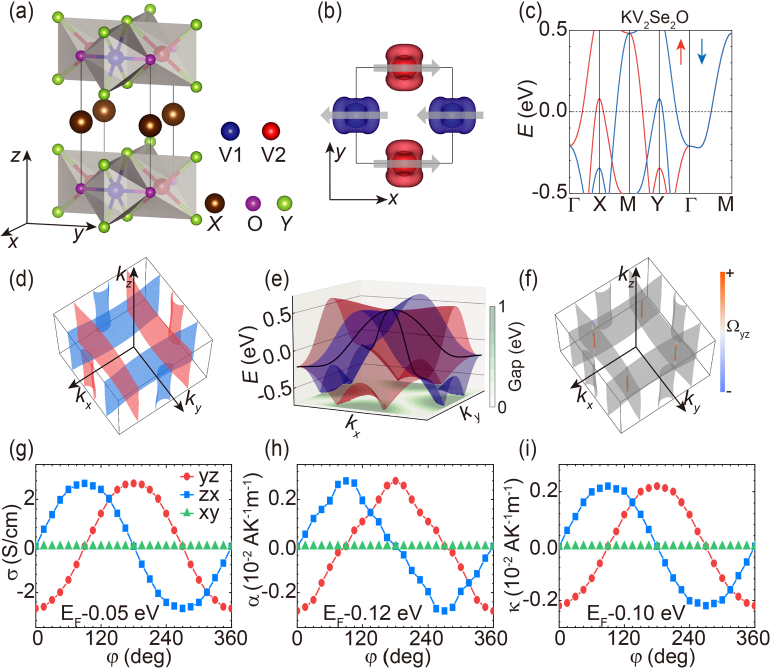}
	\caption{(a) Crystal structure of \textit{X}V$_2$\textit{Y}$_2$O. The blue and red spheres indicate the V atoms with antiparallel magnetic moment.  The brown, violet, and green spheres denote nonmagnetic \textit{X}, O, and \textit{Y} atoms, respectively.  (b) Real-space alternating spin density of \textit{X}V$_2$\textit{Y}$_2$O.  (c-e) Reciprocal-space alternating band structures along high-symmetry points, 3D Fermi surfaces, and 3D band structures in (001) plane, of KV$_2$Se$_2$O without SOC. Red and blue lines are spin-up and spin-down bands. (f) Distribution of Berry curvature ($\Omega_{yz}$) and Fermi surfaces in the 3D BZ at the true Fermi energy. (g-i) CHC, CNC, and CTHC at $T$ = 300 K as a function of azimuthal angle when $\bm{n}$ rotates within the (001) plane.}
	\label{fig:structure}
	\end{figure}

	The \textit{X}V$_2$\textit{Y}$_2$O crystallizes in a tetragonal layered crystal structure with crystallographic space group P4/mmm (No.123), as shown in Fig.~\ref{fig:structure}(a).  The V and O atoms form a rectangular arrangement in a single plane, resembling the anti-CuO$_2$ structure, with the \textit{Y} atoms positioned directly above and below the center of the V$_2$O plane. The \textit{X} atoms are located centrally within the crystal structure, effectively deintercalated via a topochemical process~\cite{ZhangFY2024}, allowing for the potential formation of 2D AMs V$_2$\textit{Y}$_2$O, as predicted theoretically~\cite{MaHY2021,QR-Cui2023,Zhuyu2024}. Among the \textit{X}V$_2$\textit{Y}$_2$O family, KV$_2$Se$_2$O and Rb$_{1-\delta}$V$_2$Te$_2$O have been successfully synthesized, exhibiting a room-temperature metallic altermagnetic state~\cite{jiang_NP2025,ZhangFY2024}, with alternating spin polarization in both real space and reciprocal space as illustrated in Figs.~\ref{fig:structure}(b-e) and Fig.~\textcolor{blue}{S2}~\cite{SuppMater}. For example, the KV$_2$Se$_2$O exhibits spin splitting along the high-symmetry lines $\Gamma$-X-M-Y-$\Gamma$, while along the $\Gamma$-M line, the bands are spin degenerate, forming nodal surfaces in the (110) and ($\bar{1}$10) planes [Fig.~\ref{fig:surface}(a) and Fig.~\ref{fig:structure}(e)]. This alternation spin-polarization plays a crucial role in the emergence of transverse transport phenomena. Here, due to the zero net magnetization, the transverse transport responses are purely crystal transport phenomena.

	Considering the combined spin-polarized electronic properties and SOC effects, anisotropic crystal transport may arise depending on the orientation of $\bm{n}$, which can be prejudged through magnetic group analysis. The crystal transport coefficients are derived using the generalized Landauer-B$\ddot{u}$ttiker formalism~\cite{ashcroft1976solid,Houten1992,behnia2015fundamentals}:
	\begin{equation}\label{eq:LB}
		R^{(n)}_{ij}=\int^\infty_{-\infty}(E-\mu)^n\left(-\frac{\partial f}{\partial E}\right)\sigma_{ij}^{T = 0}(E)dE,
	\end{equation}
	where $\mu$ is the chemical potential, $\sigma_{ij}^{T = 0}(E)$ is zero-temperature crystal Hall conductivity (CHC). The temperature-dependent CHC, crystal Nernst conductivity (CNC), and crystal thermal Hall conductivity (CTHC) are expressed as follows:
	\begin{equation}
	\sigma_{ij}  = R^{(0)}_{xy}, \ \ \alpha_{ij}  = -R^{(1)}_{ij}/eT, \ \ \kappa_{ij}  =  R^{(2)}_{ij}/e^2T. \label{eq:transport}
	\end{equation}
	When $\bm{n}$ $\parallel$ [001], the components of the pseudovector $\bm{R} = [R_{yz}, R_{zx}, R_{xy}]$ are forbidden by the mirror symmetries $\mathcal{M}_{z}$, $\mathcal{M}_{xy}$, and $\mathcal{M}_{\bar{x}y}$ of the $4'/mm'm$ magnetic point group. However, the crystal transport can be induced when the $\bm{n}$ deviates from the [001] direction, driven by experimental conditions such as an external magnetic field or spin-orbit torque, as performed successfully in RuO$_2$~\cite{feng2022}. For example, when $\bm{n}$ $\parallel$ [100] ([010]) direction, the $R_{yz}$ ($R_{zx}$) becomes nonzero due to the role of $\mathcal{M}_{x}$, $\mathcal{TM}_{y}$, and $\mathcal{TM}_{z}$ ($\mathcal{M}_{y}$, $\mathcal{TM}_{x}$, and $\mathcal{TM}_{z}$) from $m'm'm$ magnetic point group.

	We now examine the crystal transport properties using KV$_2$Se$_2$O with $\bm{n}$ $\parallel$ [100] as an example, as shown in Figs.~\ref{fig:structure}(f-i). Additional cases of other AMs disscussed in our work are shown in Figs.~\textcolor{blue}{S2 and S5-S7}~\cite{SuppMater}. Considering the SOC effect, the degenerate nodal surfaces are lifted. This pseudonodal surface, which arises from interband transitions between states with opposite spins~\cite{Mazin2021,xdzhou2024}, leads to substantial Berry curvature with negligible contributions from other trivial momentum regions [Fig.~\ref{fig:structure}(f)]. This highlights a pure altermagnetic topological origin to crystal transport. As the $\bm{n}$ rotates within the (001) plane away from the [100] direction, both $R_{yz}$ and $R_{zx}$ can become nonzero and vary periodically with a period of 2$\pi$ [Figs.~\ref{fig:structure}(g-i)]. Unfortunately, the crystal transport properties of KV$_2$Se$_2$O near Fermi energy do not offer significant advantages over other AMs~\cite{Libor2020,feng2022, xdzhou2024}.  Therefore, for practical application, it is crucial to develop an effective method to modulate the altermagnetic-topology and crystal transport properties.

	\begin{figure}
	\includegraphics[width=1\columnwidth]{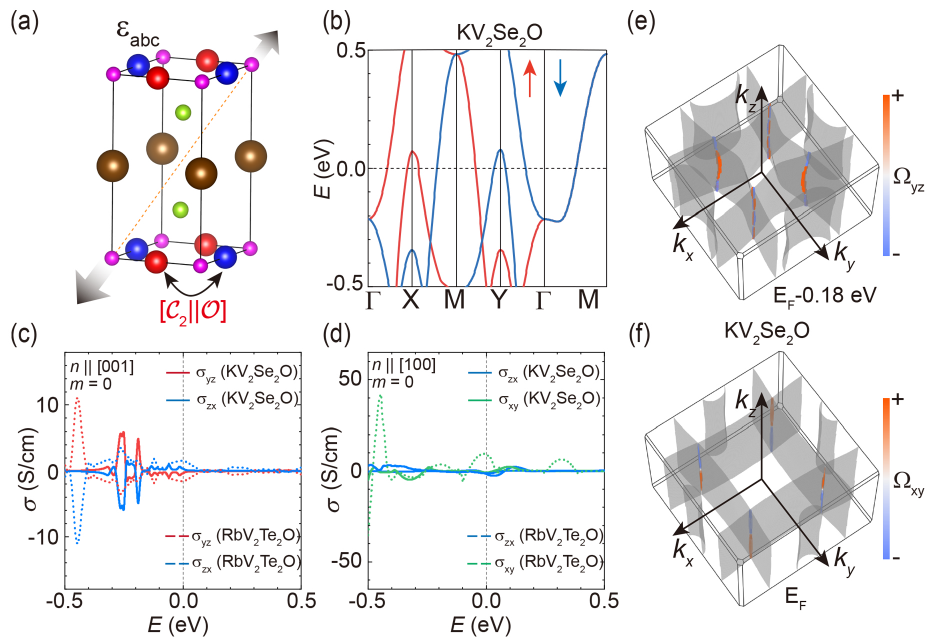}
	\caption{(a) Schematic illustration of the $\varepsilon_{abc}$ shear strain. Nonrelativistic band structures (b), CHC (c-d), and distribution of Berry curvature and Fermi surfaces in the 3D BZ (e-f) of KV$_2$Se$_2$O at 3\% $\varepsilon_{abc}$. Here, the $\bm{n}$ $\parallel$ [001] (c, e) and $\bm{n}$ $\parallel$ [100] (d, f), respectively.  For comparison, the CHC of RbV$_2$Te$_2$O under 3\% $\varepsilon_{abc}$ are also shown in (c-d).}
	\label{fig:AHC-KVSeO}
	\end{figure}

	Here, we propose two distinct yet complementary strategies for controlling transverse transport properties in altermagnetic materials. The first involves applying strain engineering [Fig.~\ref{fig:AHC-KVSeO}(a) and S8~\cite{SuppMater}]. Strain engineering breaks certain crystal symmetries in thin-flake AMs~\cite{zhouzhiyuan2025,strainLeiBC2025,strainGyo2025}, thereby modulating transverse transport. However, it often induces a small net magnetization in experiments, complicating the distinction between crystal and anomalous contributions.  In contrast, our proposed shear strain preserves zero net magnetization while enabling transverse transport, allowing selective control of the crystal transport contribution while suppressing the anomalous contribution. Three shear strain directions are considered: $\varepsilon_{ab}$, $\varepsilon_{ac}$, and $\varepsilon_{abc}$ [Fig.~\textcolor{blue}{S8}~\cite{SuppMater}], corresponding to lattice deformation along the diagonals of the $ab$ and $ac$ planes, and along the body diagonal of the crystal, respectively.  The phonon spectrum calculations [Fig.~\textcolor{blue}{S9}~\cite{SuppMater}] indicate that the structures remain dynamically stable under 3\% shear strain.  Spin group analysis reveals that in \textit{X}V$_2$\textit{Y}$_2$O, $\varepsilon_{ab}$ and $\varepsilon_{abc}$ preserve the AM phase, whereas $\varepsilon_{ac}$ drives a transition from AM to a fully compensated ferrimagnetism, characterized by isotropic band splitting.  Similar behavior is found in other AMs [Section. IV and Fig.~\textcolor{blue}{S10}]~\cite{SuppMater}.

	The band structures and CHC of KV$_2$Se$_2$O for $\varepsilon_{abc}$ with $\bm{n}$ $\parallel$ [001] and [100] are plotted in Fig.~\ref{fig:AHC-KVSeO}.  A modest $3\%$ $\varepsilon_{abc}$ strain causes only slight changes on band structures, maintaining the alternating spin characteristic. However, its profound impact lies in symmetry breaking, which fundamentally alters the allowed transport components and gives rise to CHE that are otherwise forbidden in the unstrained phase. For instance, when $\bm{n}$ $\parallel$ [001], symmetry strictly prohibits all transverse transport components in the pristine system. Upon applying $\varepsilon_{abc}$ strain, this symmetry is lowered, leading to the emergence of finite $\sigma_{yz}$ and $\sigma_{zx}$ responses, which exhibit equal magnitude and opposite sign protected by $\mathcal{M}_{\bar{x}y}$ symmetry [Fig.~\ref{fig:AHC-KVSeO}(c)]. In the case of $\bm{n}$ $\parallel$ [100], where $\sigma_{yz}$ is symmetry-allowed even without strain, the application of $\varepsilon_{abc}$ further induces two new $\sigma_{xy}$ and $\sigma_{zx}$ components [Fig.~\ref{fig:AHC-KVSeO}(d)]. To unveil the microscopic origin of these transport phenomena, we analyze the Berry curvature and Fermi surfaces in the 3D BZ [Figs.~\ref{fig:AHC-KVSeO}(e-f)]. The Berry curvature hotspots are found to concentrate near the pseudonodal surfaces, indicating that the strain primarily couples to the alternating topological structure, thereby tuning the transport response.

	In addition to strain engineering, spin canting---long recognized as a powerful route to exploring unconventional physical phenomena~\cite{cant_suzuki_NP2016,cant_takahashi_Science2018,Borisenko2019,cant_LC_PRB2021,Libor2020,Karube2022,cant_pan_NM2022,cant_das_NC2022,cant_li_NC2023,cant_EK_SA2023,cant_Zhang_NP2024,cant_Zhang_NP2024_2,cant_leenders_N2024}---is proposed as an alternative strategy that can strongly couple to the alternating topological structure and further enhance the transport response. It can be induced by an external magnetic field, with the canting angle adjustable through doping or temperature~\cite{Borisenko2019,YangR2020}.  Spin canting induces a net magnetic moment along the canting direction while maintaining collinear antiferromagnetism in the original direction.  This dual effect enhances relativistic spin splitting, particularly in the nodal surfaces, while preserving alternating spin characteristics [Figs.~\textcolor{blue}{S18-S20}~\cite{SuppMater}], thereby enabling effective tuning of altermagnetic topology and then the transverse transport properties.  When the spin cants slightly from the $x$-axis towards the $z$-axis, the corresponding magnetic point group transitions to $2'/m'$, which contains a combined symmetry $\mathcal{TM}_y$ that preserves $\bm{R}_{yz}$ and $\bm{R}_{xy}$, but forces $\bm{R}_{zx}$ to vanish.

	\begin{figure}
	\includegraphics[width=1\columnwidth]{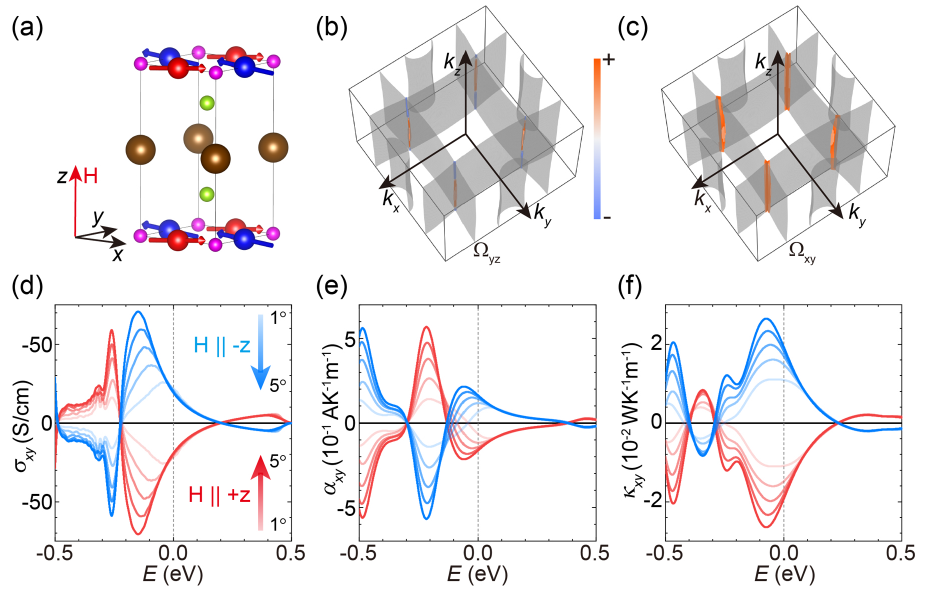}
	\caption{(a) Schematic illustration of spin canting induced by an external magnetic field ($H$). (b-c) Distribution of Berry curvatures ($\Omega_{yz}$ and $\Omega_{xy}$) and Fermi surfaces in the 3D BZ for KV$_2$Se$_2$O calculated at $E_F$-0.04 eV, respectively. (d-f) Magnetic field dependence of the Hall, Nernst, and thermal Hall conductivities for KV$_2$Se$_2$O. For comparison, the black solid lines in (d-f) represent the case with canting angle of $0^\circ$, where the $\sigma_{xy}$ vanishes.}
	\label{fig:ano_berry}
	\end{figure}

	Considering experimental feasibility, we focus on the effects of small canting angles ($1^\circ$-$5^\circ$) on transverse transport, as illustrated in Fig.~\ref{fig:ano_berry}. The effective magnetic fields required to induce these canting angles are detailed in the Fig.~\textcolor{blue}{S17} and Tab.~\textcolor{blue}{S4}~\cite{SuppMater}.  Spin canting primarily affects the $\bm{R}_{xy}$ component, with negligible impact on $\bm{R}_{yz}$ [Fig.~\textcolor{blue}{S22-S23}~\cite{SuppMater}].  In the absence of canting, the $\sigma_{xy}$ component is absent limited by the $\mathcal{TM}_z$ symmetry.  As the spin cants, the $\sigma_{xy}$ will appear and strengthens with increasing canting angle, particularly in the hole doping regime (-0.3$\sim$0.0 eV), exhibiting a linear dependence on the net magnetization $\bm{m}$.  To gain further insight, we decompose the transverse response into crystal and anomalous contributions.  Our results show that the $\bm{R}_{yz}$ component is almost constant at small canting angles, confirming its origin from crystal transport, whereas $\bm{R}_{xy}$ varies linearly with $\bm{m}$ and reverses sign upon magnetic field reversal, consistent with the behaviors of anomalous transport [Fig.~\ref{fig:ano_berry}(d-f) and Fig.~\textcolor{blue}{S23}~\cite{SuppMater}].  Similar phenomena are also observed in other AMs [Figs.~\textcolor{blue}{S24-S26}~\cite{SuppMater}], where crystal transport remains largely insensitive to spin canting, while anomalous transport shows pronounced sensitivity to the net magnetization. The key distinction in these systems lies in the more intricate band structures, where the anomalous response originates not only from altermagnetic pseudogap surfaces but also from additional topological features such as Weyl nodes and nodal lines [Fig.~\textcolor{blue}{S27}~\cite{SuppMater}].


	We then explore the physical mechanisms underlying the anomalous transport properties induced by spin canting. In this canted configuration, the presence of symmetries $\mathcal{TM}_y$, $\mathcal{TC}_{2y}$, and $\mathcal{P}$ ensures that the Berry curvature components $\Omega_{xy}$ and $\Omega_{yz}$ remain even with respect to both the $\mathcal{M}_y$ mirror plane and the $\Gamma$ point, as shown in Figs.~\ref{fig:ano_berry}(b) and (c). Importantly, both components originate from altermagnetic pseudonodal surfaces. However, a clear difference emerges: $\Omega_{yz}$ contains both positive and negative contributions that tend to cancel upon integration, while $\Omega_{xy}$ is almost entirely positive and substantially larger in magnitude. As a result, the integrated AHC $\sigma_{xy}$ dominates over $\sigma_{yz}$. Furthermore, with increasing canting angle, the pseudonodal surfaces experience enhanced band splitting, which in turn amplifies the Berry curvature magnitude and further increases $\sigma_{xy}$. This further supports the altermagnetic topological origin in \textit{X}V$_2$\textit{Y}$_2$O.

	Finally, we discuss the experimental detection of transverse transport in \textit{X}V$_2$\textit{Y}$_2$O. In AMs, the Hall resistivity can be written as~\cite{Takagi_FeS2_NM2025} $\rho_{ji} = \rho_{ji}^{NHE} + \rho_{ji}^{AHE} + \rho_{ji}^{CHE} =  R_0B + S_H\rho_{ii}^2m + \rho_{ji}^{CHE}$, representing the normal (NHE), AHE, and CHE terms. In zero field ($B=0$) and with $m\approx 0$, the Hall signal directly reflects the CHE. Under finite $B$, $\rho_{yz}$ contains CHE, NHE, and a small AHE, while $\rho_{xy}$ contains only NHE and AHE. Details of the decomposition of these contributions are given in Sec. IV~\cite{SuppMater}. Notably, although measuring $\rho_{yz}$ and $\rho_{zx}$ in layered crystals is challenging, it is feasible~\cite{Singh2024,CY-Guo2022} using two Hall bar geometries [Fig. \textcolor{blue}{S1}~\cite{SuppMater}], especially with the recently synthesized high-quality single crystals~\cite{jiang_NP2025,ZhangFY2024}.

	
	In this study, we demonstrate the pivotal role of altermagnetic topology in governing the transverse transport in AMs, with a particular focus on crystal and anomalous transport contributions modulated by strain engineering and spin canting.   Shear strain modulates crystal transport while preserving zero net magnetization, whereas spin canting, induced by external magnetic field, primarily controls the anomalous response.  By separating crystal and anomalous contributions, these approaches allow independent control over the two transport channels.  This work paves the way for room-temperature altermagnetic spintronics and spin caloritronics, with potential applications in next-generation electronic and thermoelectric devices. The ability to manipulate the altermagnetic topology via strain engineering and spin canting may be key to optimizing AM performance in future technological applications.

		



	\vspace{0.3cm}	
		The authors thanks Chaoxi Cui, Jingyi Duan, Jinjin Liu, Yichen Liu, Jie Chen, Ling Bai and Wanxiang Feng for their helpful discussion.
		This work is supported by the National Natural Science Foundation of China (Grants No. 12404052, No. 12304066,  No. 12474012, No. 12404256, No. 12574064, and No. 12174160), the National Key R$\&$D Program of China (Grant No. 2022YFA1402600), the Basic Research Program of Jiangsu (Grant No. BK20241049 and No. BK20230684), the Natural Science Fund for Colleges and Universities in Jiangsu Province (Grant No. 24KJB140011 and No. 23KJB140008).

	\bibliography{references}

\begin{thebibliography}{117}%
\makeatletter
\providecommand \@ifxundefined [1]{%
 \@ifx{#1\undefined}
}%
\providecommand \@ifnum [1]{%
 \ifnum #1\expandafter \@firstoftwo
 \else \expandafter \@secondoftwo
 \fi
}%
\providecommand \@ifx [1]{%
 \ifx #1\expandafter \@firstoftwo
 \else \expandafter \@secondoftwo
 \fi
}%
\providecommand \natexlab [1]{#1}%
\providecommand \enquote  [1]{``#1''}%
\providecommand \bibnamefont  [1]{#1}%
\providecommand \bibfnamefont [1]{#1}%
\providecommand \citenamefont [1]{#1}%
\providecommand \href@noop [0]{\@secondoftwo}%
\providecommand \href [0]{\begingroup \@sanitize@url \@href}%
\providecommand \@href[1]{\@@startlink{#1}\@@href}%
\providecommand \@@href[1]{\endgroup#1\@@endlink}%
\providecommand \@sanitize@url [0]{\catcode `\\12\catcode `\$12\catcode `\&12\catcode `\#12\catcode `\^12\catcode `\_12\catcode `\%12\relax}%
\providecommand \@@startlink[1]{}%
\providecommand \@@endlink[0]{}%
\providecommand \url  [0]{\begingroup\@sanitize@url \@url }%
\providecommand \@url [1]{\endgroup\@href {#1}{\urlprefix }}%
\providecommand \urlprefix  [0]{URL }%
\providecommand \Eprint [0]{\href }%
\providecommand \doibase [0]{http://dx.doi.org/}%
\providecommand \selectlanguage [0]{\@gobble}%
\providecommand \bibinfo  [0]{\@secondoftwo}%
\providecommand \bibfield  [0]{\@secondoftwo}%
\providecommand \translation [1]{[#1]}%
\providecommand \BibitemOpen [0]{}%
\providecommand \bibitemStop [0]{}%
\providecommand \bibitemNoStop [0]{.\EOS\space}%
\providecommand \EOS [0]{\spacefactor3000\relax}%
\providecommand \BibitemShut  [1]{\csname bibitem#1\endcsname}%
\let\auto@bib@innerbib\@empty
\bibitem [{\citenamefont {Nagaosa}\ \emph {et~al.}(2010)\citenamefont {Nagaosa}, \citenamefont {Sinova}, \citenamefont {Onoda}, \citenamefont {MacDonald},\ and\ \citenamefont {Ong}}]{Nagaosa2010}%
  \BibitemOpen
  \bibfield  {author} {\bibinfo {author} {\bibfnamefont {N.}~\bibnamefont {Nagaosa}}, \bibinfo {author} {\bibfnamefont {J.}~\bibnamefont {Sinova}}, \bibinfo {author} {\bibfnamefont {S.}~\bibnamefont {Onoda}}, \bibinfo {author} {\bibfnamefont {A.~H.}\ \bibnamefont {MacDonald}}, \ and\ \bibinfo {author} {\bibfnamefont {N.~P.}\ \bibnamefont {Ong}},\ }\bibinfo {title} {Anomalous hall effect},\ \href {https://link.aps.org/doi/10.1103/RevModPhys.82.1539} {\bibfield  {journal} {\bibinfo  {journal} {Rev. Mod. Phys.}\ }\textbf {\bibinfo {volume} {82}},\ \bibinfo {pages} {1539} (\bibinfo {year} {2010})}\BibitemShut {NoStop}%
\bibitem [{\citenamefont {Xiao}\ \emph {et~al.}(2006)\citenamefont {Xiao}, \citenamefont {Yao}, \citenamefont {Fang},\ and\ \citenamefont {Niu}}]{XiaoD2006}%
  \BibitemOpen
  \bibfield  {author} {\bibinfo {author} {\bibfnamefont {D.}~\bibnamefont {Xiao}}, \bibinfo {author} {\bibfnamefont {Y.}~\bibnamefont {Yao}}, \bibinfo {author} {\bibfnamefont {Z.}~\bibnamefont {Fang}}, \ and\ \bibinfo {author} {\bibfnamefont {Q.}~\bibnamefont {Niu}},\ }\bibinfo {title} {Berry-Phase Effect in Anomalous Thermoelectric Transport},\ \href {\doibase 10.1103/PhysRevLett.97.026603} {\bibfield  {journal} {\bibinfo  {journal} {Phys. Rev. Lett.}\ }\textbf {\bibinfo {volume} {97}},\ \bibinfo {pages} {026603} (\bibinfo {year} {2006})}\BibitemShut {NoStop}%
\bibitem [{\citenamefont {Qin}\ \emph {et~al.}(2011)\citenamefont {Qin}, \citenamefont {Niu},\ and\ \citenamefont {Shi}}]{Qin2011}%
  \BibitemOpen
  \bibfield  {author} {\bibinfo {author} {\bibfnamefont {T.}~\bibnamefont {Qin}}, \bibinfo {author} {\bibfnamefont {Q.}~\bibnamefont {Niu}}, \ and\ \bibinfo {author} {\bibfnamefont {J.}~\bibnamefont {Shi}},\ }\bibinfo {title} {Energy Magnetization and the Thermal Hall Effect},\ \href {\doibase 10.1103/PhysRevLett.107.236601} {\bibfield  {journal} {\bibinfo  {journal} {Phys. Rev. Lett.}\ }\textbf {\bibinfo {volume} {107}},\ \bibinfo {pages} {236601} (\bibinfo {year} {2011})}\BibitemShut {NoStop}%
\bibitem [{\citenamefont {\ifmmode \check{Z}\else \v{Z}\fi{}uti\ifmmode~\acute{c}\else \'{c}\fi{}}\ \emph {et~al.}(2004)\citenamefont {\ifmmode \check{Z}\else \v{Z}\fi{}uti\ifmmode~\acute{c}\else \'{c}\fi{}}, \citenamefont {Fabian},\ and\ \citenamefont {Das~Sarma}}]{IgorZutic2004}%
  \BibitemOpen
  \bibfield  {author} {\bibinfo {author} {\bibfnamefont {I.}~\bibnamefont {\ifmmode \check{Z}\else \v{Z}\fi{}uti\ifmmode~\acute{c}\else \'{c}\fi{}}}, \bibinfo {author} {\bibfnamefont {J.}~\bibnamefont {Fabian}}, \ and\ \bibinfo {author} {\bibfnamefont {S.}~\bibnamefont {Das~Sarma}},\ }\bibinfo {title} {Spintronics: Fundamentals and applications},\ \href {\doibase 10.1103/RevModPhys.76.323} {\bibfield  {journal} {\bibinfo  {journal} {Rev. Mod. Phys.}\ }\textbf {\bibinfo {volume} {76}},\ \bibinfo {pages} {323} (\bibinfo {year} {2004})}\BibitemShut {NoStop}%
\bibitem [{\citenamefont {Gerrit E. W.~Bauer}\ and\ \citenamefont {van Wees}(2012)}]{Gerrit2012}%
  \BibitemOpen
  \bibfield  {author} {\bibinfo {author} {\bibfnamefont {E.~S.}\ \bibnamefont {Gerrit E. W.~Bauer}}\ and\ \bibinfo {author} {\bibfnamefont {B.~J.}\ \bibnamefont {van Wees}},\ }\bibinfo {title} {Spin caloritronics},\ \href {https://www.nature.com/articles/nmat3301} {\bibfield  {journal} {\bibinfo  {journal} {Nat. mater}\ }\textbf {\bibinfo {volume} {11}},\ \bibinfo {pages} {391} (\bibinfo {year} {2012})}\BibitemShut {NoStop}%
\bibitem [{\citenamefont {Jungwirth}\ \emph {et~al.}(2002)\citenamefont {Jungwirth}, \citenamefont {Niu},\ and\ \citenamefont {MacDonald}}]{Jungwirth2002}%
  \BibitemOpen
  \bibfield  {author} {\bibinfo {author} {\bibfnamefont {T.}~\bibnamefont {Jungwirth}}, \bibinfo {author} {\bibfnamefont {Q.}~\bibnamefont {Niu}}, \ and\ \bibinfo {author} {\bibfnamefont {A.~H.}\ \bibnamefont {MacDonald}},\ }\bibinfo {title} {Anomalous Hall Effect in Ferromagnetic Semiconductors},\ \href {\doibase 10.1103/PhysRevLett.88.207208} {\bibfield  {journal} {\bibinfo  {journal} {Phys. Rev. Lett.}\ }\textbf {\bibinfo {volume} {88}},\ \bibinfo {pages} {207208} (\bibinfo {year} {2002})}\BibitemShut {NoStop}%
\bibitem [{\citenamefont {Fang}\ \emph {et~al.}(2003)\citenamefont {Fang}, \citenamefont {Nagaosa}, \citenamefont {Takahashi}, \citenamefont {Asamitsu}, \citenamefont {Mathieu}, \citenamefont {Ogasawara}, \citenamefont {Yamada}, \citenamefont {Kawasaki}, \citenamefont {Tokura},\ and\ \citenamefont {Terakura}}]{Fang2003}%
  \BibitemOpen
  \bibfield  {author} {\bibinfo {author} {\bibfnamefont {Z.}~\bibnamefont {Fang}}, \bibinfo {author} {\bibfnamefont {N.}~\bibnamefont {Nagaosa}}, \bibinfo {author} {\bibfnamefont {K.~S.}\ \bibnamefont {Takahashi}}, \bibinfo {author} {\bibfnamefont {A.}~\bibnamefont {Asamitsu}}, \bibinfo {author} {\bibfnamefont {R.}~\bibnamefont {Mathieu}}, \bibinfo {author} {\bibfnamefont {T.}~\bibnamefont {Ogasawara}}, \bibinfo {author} {\bibfnamefont {H.}~\bibnamefont {Yamada}}, \bibinfo {author} {\bibfnamefont {M.}~\bibnamefont {Kawasaki}}, \bibinfo {author} {\bibfnamefont {Y.}~\bibnamefont {Tokura}}, \ and\ \bibinfo {author} {\bibfnamefont {K.}~\bibnamefont {Terakura}},\ }\bibinfo {title} {The Anomalous Hall Effect and Magnetic Monopoles in Momentum Space},\ \href {\doibase 10.1126/science.1089408} {\bibfield  {journal} {\bibinfo  {journal} {Science}\ }\textbf {\bibinfo {volume} {302}},\ \bibinfo {pages} {92} (\bibinfo {year} {2003})}\BibitemShut {NoStop}%
\bibitem [{\citenamefont {Yao}\ \emph {et~al.}(2004)\citenamefont {Yao}, \citenamefont {Kleinman}, \citenamefont {MacDonald}, \citenamefont {Sinova}, \citenamefont {Jungwirth}, \citenamefont {Wang}, \citenamefont {Wang},\ and\ \citenamefont {Niu}}]{YaoYG2004}%
  \BibitemOpen
  \bibfield  {author} {\bibinfo {author} {\bibfnamefont {Y.}~\bibnamefont {Yao}}, \bibinfo {author} {\bibfnamefont {L.}~\bibnamefont {Kleinman}}, \bibinfo {author} {\bibfnamefont {A.~H.}\ \bibnamefont {MacDonald}}, \bibinfo {author} {\bibfnamefont {J.}~\bibnamefont {Sinova}}, \bibinfo {author} {\bibfnamefont {T.}~\bibnamefont {Jungwirth}}, \bibinfo {author} {\bibfnamefont {D.-s.}\ \bibnamefont {Wang}}, \bibinfo {author} {\bibfnamefont {E.}~\bibnamefont {Wang}}, \ and\ \bibinfo {author} {\bibfnamefont {Q.}~\bibnamefont {Niu}},\ }\bibinfo {title} {First Principles Calculation of Anomalous Hall Conductivity in Ferromagnetic bcc Fe},\ \href {\doibase 10.1103/PhysRevLett.92.037204} {\bibfield  {journal} {\bibinfo  {journal} {Phys. Rev. Lett.}\ }\textbf {\bibinfo {volume} {92}},\ \bibinfo {pages} {037204} (\bibinfo {year} {2004})}\BibitemShut {NoStop}%
\bibitem [{\citenamefont {Onoda}\ \emph {et~al.}(2006)\citenamefont {Onoda}, \citenamefont {Sugimoto},\ and\ \citenamefont {Nagaosa}}]{Onoda2006}%
  \BibitemOpen
  \bibfield  {author} {\bibinfo {author} {\bibfnamefont {S.}~\bibnamefont {Onoda}}, \bibinfo {author} {\bibfnamefont {N.}~\bibnamefont {Sugimoto}}, \ and\ \bibinfo {author} {\bibfnamefont {N.}~\bibnamefont {Nagaosa}},\ }\bibinfo {title} {Intrinsic Versus Extrinsic Anomalous Hall Effect in Ferromagnets},\ \href {\doibase 10.1103/PhysRevLett.97.126602} {\bibfield  {journal} {\bibinfo  {journal} {Phys. Rev. Lett.}\ }\textbf {\bibinfo {volume} {97}},\ \bibinfo {pages} {126602} (\bibinfo {year} {2006})}\BibitemShut {NoStop}%
\bibitem [{\citenamefont {Onoda}\ \emph {et~al.}(2008)\citenamefont {Onoda}, \citenamefont {Sugimoto},\ and\ \citenamefont {Nagaosa}}]{Onoda2008}%
  \BibitemOpen
  \bibfield  {author} {\bibinfo {author} {\bibfnamefont {S.}~\bibnamefont {Onoda}}, \bibinfo {author} {\bibfnamefont {N.}~\bibnamefont {Sugimoto}}, \ and\ \bibinfo {author} {\bibfnamefont {N.}~\bibnamefont {Nagaosa}},\ }\bibinfo {title} {Quantum transport theory of anomalous electric, thermoelectric, and thermal Hall effects in ferromagnets},\ \href {\doibase 10.1103/PhysRevB.77.165103} {\bibfield  {journal} {\bibinfo  {journal} {Phys. Rev. B}\ }\textbf {\bibinfo {volume} {77}},\ \bibinfo {pages} {165103} (\bibinfo {year} {2008})}\BibitemShut {NoStop}%
\bibitem [{\citenamefont {Onose}\ \emph {et~al.}(2008)\citenamefont {Onose}, \citenamefont {Shiomi},\ and\ \citenamefont {Tokura}}]{Onose_Lor2008}%
  \BibitemOpen
  \bibfield  {author} {\bibinfo {author} {\bibfnamefont {Y.}~\bibnamefont {Onose}}, \bibinfo {author} {\bibfnamefont {Y.}~\bibnamefont {Shiomi}}, \ and\ \bibinfo {author} {\bibfnamefont {Y.}~\bibnamefont {Tokura}},\ }\bibinfo {title} {Lorenz Number Determination of the Dissipationless Nature of the Anomalous Hall Effect in Itinerant Ferromagnets},\ \href {\doibase 10.1103/PhysRevLett.100.016601} {\bibfield  {journal} {\bibinfo  {journal} {Phys. Rev. Lett.}\ }\textbf {\bibinfo {volume} {100}},\ \bibinfo {pages} {016601} (\bibinfo {year} {2008})}\BibitemShut {NoStop}%
\bibitem [{\citenamefont {Shiomi}\ \emph {et~al.}(2009)\citenamefont {Shiomi}, \citenamefont {Onose},\ and\ \citenamefont {Tokura}}]{Shiomi_Lor2009}%
  \BibitemOpen
  \bibfield  {author} {\bibinfo {author} {\bibfnamefont {Y.}~\bibnamefont {Shiomi}}, \bibinfo {author} {\bibfnamefont {Y.}~\bibnamefont {Onose}}, \ and\ \bibinfo {author} {\bibfnamefont {Y.}~\bibnamefont {Tokura}},\ }\bibinfo {title} {Extrinsic anomalous Hall effect in charge and heat transport in pure iron, ${\text{Fe}}_{0.997}{\text{Si}}_{0.003}$, and ${\text{Fe}}_{0.97}{\text{Co}}_{0.03}$},\ \href {\doibase 10.1103/PhysRevB.79.100404} {\bibfield  {journal} {\bibinfo  {journal} {Phys. Rev. B}\ }\textbf {\bibinfo {volume} {79}},\ \bibinfo {pages} {100404} (\bibinfo {year} {2009})}\BibitemShut {NoStop}%
\bibitem [{\citenamefont {Shiomi}\ \emph {et~al.}(2010)\citenamefont {Shiomi}, \citenamefont {Onose},\ and\ \citenamefont {Tokura}}]{Shiomi_Lor2010}%
  \BibitemOpen
  \bibfield  {author} {\bibinfo {author} {\bibfnamefont {Y.}~\bibnamefont {Shiomi}}, \bibinfo {author} {\bibfnamefont {Y.}~\bibnamefont {Onose}}, \ and\ \bibinfo {author} {\bibfnamefont {Y.}~\bibnamefont {Tokura}},\ }\bibinfo {title} {Effect of scattering on intrinsic anomalous Hall effect investigated by Lorenz ratio},\ \href {\doibase 10.1103/PhysRevB.81.054414} {\bibfield  {journal} {\bibinfo  {journal} {Phys. Rev. B}\ }\textbf {\bibinfo {volume} {81}},\ \bibinfo {pages} {054414} (\bibinfo {year} {2010})}\BibitemShut {NoStop}%
\bibitem [{\citenamefont {Ikhlas}\ \emph {et~al.}(2017)\citenamefont {Ikhlas}, \citenamefont {Tomita}, \citenamefont {Koretsune}, \citenamefont {Suzuki}, \citenamefont {Nishio-Hamane}, \citenamefont {Arita}, \citenamefont {Otani},\ and\ \citenamefont {Nakatsuji}}]{Ikhlas2017}%
  \BibitemOpen
  \bibfield  {author} {\bibinfo {author} {\bibfnamefont {M.}~\bibnamefont {Ikhlas}}, \bibinfo {author} {\bibfnamefont {T.}~\bibnamefont {Tomita}}, \bibinfo {author} {\bibfnamefont {T.}~\bibnamefont {Koretsune}}, \bibinfo {author} {\bibfnamefont {M.-T.}\ \bibnamefont {Suzuki}}, \bibinfo {author} {\bibfnamefont {D.}~\bibnamefont {Nishio-Hamane}}, \bibinfo {author} {\bibfnamefont {R.}~\bibnamefont {Arita}}, \bibinfo {author} {\bibfnamefont {Y.}~\bibnamefont {Otani}}, \ and\ \bibinfo {author} {\bibfnamefont {S.}~\bibnamefont {Nakatsuji}},\ }\bibinfo {title} {Large anomalous {Nernst} effect at room temperature in a chiral antiferromagnet},\ \href {\doibase 10.1038/nphys4181} {\bibfield  {journal} {\bibinfo  {journal} {Nat. Phys.}\ }\textbf {\bibinfo {volume} {13}},\ \bibinfo {pages} {1085} (\bibinfo {year} {2017})}\BibitemShut {NoStop}%
\bibitem [{\citenamefont {Li}\ \emph {et~al.}(2017)\citenamefont {Li}, \citenamefont {Xu}, \citenamefont {Ding}, \citenamefont {Wang}, \citenamefont {Shen}, \citenamefont {Lu}, \citenamefont {Zhu},\ and\ \citenamefont {Behnia}}]{LiXiaoKang2017}%
  \BibitemOpen
  \bibfield  {author} {\bibinfo {author} {\bibfnamefont {X.}~\bibnamefont {Li}}, \bibinfo {author} {\bibfnamefont {L.}~\bibnamefont {Xu}}, \bibinfo {author} {\bibfnamefont {L.}~\bibnamefont {Ding}}, \bibinfo {author} {\bibfnamefont {J.}~\bibnamefont {Wang}}, \bibinfo {author} {\bibfnamefont {M.}~\bibnamefont {Shen}}, \bibinfo {author} {\bibfnamefont {X.}~\bibnamefont {Lu}}, \bibinfo {author} {\bibfnamefont {Z.}~\bibnamefont {Zhu}}, \ and\ \bibinfo {author} {\bibfnamefont {K.}~\bibnamefont {Behnia}},\ }\bibinfo {title} {Anomalous Nernst and Righi-Leduc Effects in ${\mathrm{Mn}}_{3}\mathrm{Sn}$: Berry Curvature and Entropy Flow},\ \href {\doibase 10.1103/PhysRevLett.119.056601} {\bibfield  {journal} {\bibinfo  {journal} {Phys. Rev. Lett.}\ }\textbf {\bibinfo {volume} {119}},\ \bibinfo {pages} {056601} (\bibinfo {year} {2017})}\BibitemShut {NoStop}%
\bibitem [{\citenamefont {Guo}\ and\ \citenamefont {Wang}(2017)}]{GuoGY2017}%
  \BibitemOpen
  \bibfield  {author} {\bibinfo {author} {\bibfnamefont {G.-Y.}\ \bibnamefont {Guo}}\ and\ \bibinfo {author} {\bibfnamefont {T.-C.}\ \bibnamefont {Wang}},\ }\bibinfo {title} {Large anomalous Nernst and spin Nernst effects in the noncollinear antiferromagnets ${\mathrm{Mn}}_{3}X$ ($X=\mathrm{Sn},\mathrm{Ge},\mathrm{Ga}$)},\ \href {\doibase 10.1103/PhysRevB.96.224415} {\bibfield  {journal} {\bibinfo  {journal} {Phys. Rev. B}\ }\textbf {\bibinfo {volume} {96}},\ \bibinfo {pages} {224415} (\bibinfo {year} {2017})}\BibitemShut {NoStop}%
\bibitem [{\citenamefont {Xu}\ \emph {et~al.}(2020)\citenamefont {Xu}, \citenamefont {Li}, \citenamefont {Lu}, \citenamefont {Collignon}, \citenamefont {Fu}, \citenamefont {Koo}, \citenamefont {Fauqu{\'e}}, \citenamefont {Yan}, \citenamefont {Zhu},\ and\ \citenamefont {Behnia}}]{LC-Xu2020}%
  \BibitemOpen
  \bibfield  {author} {\bibinfo {author} {\bibfnamefont {L.}~\bibnamefont {Xu}}, \bibinfo {author} {\bibfnamefont {X.}~\bibnamefont {Li}}, \bibinfo {author} {\bibfnamefont {X.}~\bibnamefont {Lu}}, \bibinfo {author} {\bibfnamefont {C.}~\bibnamefont {Collignon}}, \bibinfo {author} {\bibfnamefont {H.}~\bibnamefont {Fu}}, \bibinfo {author} {\bibfnamefont {J.}~\bibnamefont {Koo}}, \bibinfo {author} {\bibfnamefont {B.}~\bibnamefont {Fauqu{\'e}}}, \bibinfo {author} {\bibfnamefont {B.}~\bibnamefont {Yan}}, \bibinfo {author} {\bibfnamefont {Z.}~\bibnamefont {Zhu}}, \ and\ \bibinfo {author} {\bibfnamefont {K.}~\bibnamefont {Behnia}},\ }\bibinfo {title} {Finite-temperature violation of the anomalous transverse Wiedemann-Franz law},\ \href {\doibase 10.1126/sciadv.aaz3522} {\bibfield  {journal} {\bibinfo  {journal} {Sci. Adv.}\ }\textbf {\bibinfo {volume} {6}},\ \bibinfo {pages} {eaaz3522} (\bibinfo {year} {2020})}\BibitemShut {NoStop}%
\bibitem [{\citenamefont {Sugii}\ \emph {et~al.}(2019)\citenamefont {Sugii}, \citenamefont {Imai}, \citenamefont {Shimozawa}, \citenamefont {Ikhlas}, \citenamefont {Kiyohara}, \citenamefont {Tomita}, \citenamefont {Suzuki}, \citenamefont {Koretsune}, \citenamefont {Arita}, \citenamefont {Nakatsuji},\ and\ \citenamefont {Yamashita}}]{sugii2019}%
  \BibitemOpen
  \bibfield  {author} {\bibinfo {author} {\bibfnamefont {K.}~\bibnamefont {Sugii}}, \bibinfo {author} {\bibfnamefont {Y.}~\bibnamefont {Imai}}, \bibinfo {author} {\bibfnamefont {M.}~\bibnamefont {Shimozawa}}, \bibinfo {author} {\bibfnamefont {M.}~\bibnamefont {Ikhlas}}, \bibinfo {author} {\bibfnamefont {N.}~\bibnamefont {Kiyohara}}, \bibinfo {author} {\bibfnamefont {T.}~\bibnamefont {Tomita}}, \bibinfo {author} {\bibfnamefont {M.-T.}\ \bibnamefont {Suzuki}}, \bibinfo {author} {\bibfnamefont {T.}~\bibnamefont {Koretsune}}, \bibinfo {author} {\bibfnamefont {R.}~\bibnamefont {Arita}}, \bibinfo {author} {\bibfnamefont {S.}~\bibnamefont {Nakatsuji}}, \ and\ \bibinfo {author} {\bibfnamefont {M.}~\bibnamefont {Yamashita}},\ }\bibinfo {title} {Anomalous thermal Hall effect in the topological antiferromagnetic state},\ \href {https://arxiv.org/abs/1902.06601} {\bibfield  {journal} {\bibinfo  {journal} {arXiv}\ ,\ \bibinfo {pages} {1902.06601}} (\bibinfo {year} {2019})}\BibitemShut {NoStop}%
\bibitem [{\citenamefont {Zhou}\ \emph {et~al.}(2020)\citenamefont {Zhou}, \citenamefont {Hanke}, \citenamefont {Feng}, \citenamefont {Bl{\"u}gel}, \citenamefont {Mokrousov},\ and\ \citenamefont {Yao}}]{XD-Zhou2020}%
  \BibitemOpen
  \bibfield  {author} {\bibinfo {author} {\bibfnamefont {X.}~\bibnamefont {Zhou}}, \bibinfo {author} {\bibfnamefont {J.-P.}\ \bibnamefont {Hanke}}, \bibinfo {author} {\bibfnamefont {W.}~\bibnamefont {Feng}}, \bibinfo {author} {\bibfnamefont {S.}~\bibnamefont {Bl{\"u}gel}}, \bibinfo {author} {\bibfnamefont {Y.}~\bibnamefont {Mokrousov}}, \ and\ \bibinfo {author} {\bibfnamefont {Y.}~\bibnamefont {Yao}},\ }\bibinfo {title} {Giant Anomalous Nernst Effect in Noncollinear Antiferromagnetic Mn-based Antiperovskite Nitrides},\ \href {https://doi.org/10.1103/PhysRevMaterials.4.024408} {\bibfield  {journal} {\bibinfo  {journal} {Phys. Rev. Mater.}\ }\textbf {\bibinfo {volume} {4}},\ \bibinfo {pages} {024408} (\bibinfo {year} {2020})}\BibitemShut {NoStop}%
\bibitem [{\citenamefont {Zhou}\ \emph {et~al.}(2019)\citenamefont {Zhou}, \citenamefont {Hanke}, \citenamefont {Feng}, \citenamefont {Li}, \citenamefont {Guo}, \citenamefont {Yao}, \citenamefont {Bl\"ugel},\ and\ \citenamefont {Mokrousov}}]{Zhou2019}%
  \BibitemOpen
  \bibfield  {author} {\bibinfo {author} {\bibfnamefont {X.}~\bibnamefont {Zhou}}, \bibinfo {author} {\bibfnamefont {J.-P.}\ \bibnamefont {Hanke}}, \bibinfo {author} {\bibfnamefont {W.}~\bibnamefont {Feng}}, \bibinfo {author} {\bibfnamefont {F.}~\bibnamefont {Li}}, \bibinfo {author} {\bibfnamefont {G.-Y.}\ \bibnamefont {Guo}}, \bibinfo {author} {\bibfnamefont {Y.}~\bibnamefont {Yao}}, \bibinfo {author} {\bibfnamefont {S.}~\bibnamefont {Bl\"ugel}}, \ and\ \bibinfo {author} {\bibfnamefont {Y.}~\bibnamefont {Mokrousov}},\ }\bibinfo {title} {Spin-order dependent anomalous Hall effect and magneto-optical effect in the noncollinear antiferromagnets Mn$_3$\emph{X}N with \emph{X} = Ga, Zn, Ag, or Ni},\ \href {\doibase 10.1103/PhysRevB.99.104428} {\bibfield  {journal} {\bibinfo  {journal} {Phys. Rev. B}\ }\textbf {\bibinfo {volume} {99}},\ \bibinfo {pages} {104428} (\bibinfo {year} {2019})}\BibitemShut {NoStop}%
\bibitem [{\citenamefont {Shiomi}\ \emph {et~al.}(2013)\citenamefont {Shiomi}, \citenamefont {Kanazawa}, \citenamefont {Shibata}, \citenamefont {Onose},\ and\ \citenamefont {Tokura}}]{Shiomi2013}%
  \BibitemOpen
  \bibfield  {author} {\bibinfo {author} {\bibfnamefont {Y.}~\bibnamefont {Shiomi}}, \bibinfo {author} {\bibfnamefont {N.}~\bibnamefont {Kanazawa}}, \bibinfo {author} {\bibfnamefont {K.}~\bibnamefont {Shibata}}, \bibinfo {author} {\bibfnamefont {Y.}~\bibnamefont {Onose}}, \ and\ \bibinfo {author} {\bibfnamefont {Y.}~\bibnamefont {Tokura}},\ }\bibinfo {title} {Topological Nernst effect in a three-dimensional skyrmion-lattice phase},\ \href {\doibase 10.1103/PhysRevB.88.064409} {\bibfield  {journal} {\bibinfo  {journal} {Phys. Rev. B}\ }\textbf {\bibinfo {volume} {88}},\ \bibinfo {pages} {064409} (\bibinfo {year} {2013})}\BibitemShut {NoStop}%
\bibitem [{\citenamefont {Hirschberger}\ \emph {et~al.}(2020)\citenamefont {Hirschberger}, \citenamefont {Spitz}, \citenamefont {Nomoto}, \citenamefont {Kurumaji}, \citenamefont {Gao}, \citenamefont {Masell}, \citenamefont {Nakajima}, \citenamefont {Kikkawa}, \citenamefont {Yamasaki}, \citenamefont {Sagayama}, \citenamefont {Nakao}, \citenamefont {Taguchi}, \citenamefont {Arita}, \citenamefont {Arima},\ and\ \citenamefont {Tokura}}]{Hirschberger2020}%
  \BibitemOpen
  \bibfield  {author} {\bibinfo {author} {\bibfnamefont {M.}~\bibnamefont {Hirschberger}}, \bibinfo {author} {\bibfnamefont {L.}~\bibnamefont {Spitz}}, \bibinfo {author} {\bibfnamefont {T.}~\bibnamefont {Nomoto}}, \bibinfo {author} {\bibfnamefont {T.}~\bibnamefont {Kurumaji}}, \bibinfo {author} {\bibfnamefont {S.}~\bibnamefont {Gao}}, \bibinfo {author} {\bibfnamefont {J.}~\bibnamefont {Masell}}, \bibinfo {author} {\bibfnamefont {T.}~\bibnamefont {Nakajima}}, \bibinfo {author} {\bibfnamefont {A.}~\bibnamefont {Kikkawa}}, \bibinfo {author} {\bibfnamefont {Y.}~\bibnamefont {Yamasaki}}, \bibinfo {author} {\bibfnamefont {H.}~\bibnamefont {Sagayama}}, \bibinfo {author} {\bibfnamefont {H.}~\bibnamefont {Nakao}}, \bibinfo {author} {\bibfnamefont {Y.}~\bibnamefont {Taguchi}}, \bibinfo {author} {\bibfnamefont {R.}~\bibnamefont {Arita}}, \bibinfo {author} {\bibfnamefont {T.-h.}\ \bibnamefont {Arima}}, \ and\ \bibinfo {author} {\bibfnamefont {Y.}~\bibnamefont {Tokura}},\ }\bibinfo {title} {Topological Nernst Effect of the Two-Dimensional Skyrmion Lattice},\ \href {\doibase 10.1103/PhysRevLett.125.076602} {\bibfield  {journal} {\bibinfo  {journal} {Phys. Rev. Lett.}\ }\textbf {\bibinfo {volume} {125}},\ \bibinfo {pages} {076602} (\bibinfo {year} {2020})}\BibitemShut {NoStop}%
\bibitem [{\citenamefont {Zhang}\ \emph {et~al.}(2021)\citenamefont {Zhang}, \citenamefont {Xu},\ and\ \citenamefont {Ke}}]{ZhangH2021}%
  \BibitemOpen
  \bibfield  {author} {\bibinfo {author} {\bibfnamefont {H.}~\bibnamefont {Zhang}}, \bibinfo {author} {\bibfnamefont {C.~Q.}\ \bibnamefont {Xu}}, \ and\ \bibinfo {author} {\bibfnamefont {X.}~\bibnamefont {Ke}},\ }\bibinfo {title} {Topological Nernst effect, anomalous Nernst effect, and anomalous thermal Hall effect in the Dirac semimetal ${\mathrm{Fe}}_{3}{\mathrm{Sn}}_{2}$},\ \href {\doibase 10.1103/PhysRevB.103.L201101} {\bibfield  {journal} {\bibinfo  {journal} {Phys. Rev. B}\ }\textbf {\bibinfo {volume} {103}},\ \bibinfo {pages} {L201101} (\bibinfo {year} {2021})}\BibitemShut {NoStop}%
\bibitem [{\citenamefont {Zhou}\ \emph {et~al.}(2016)\citenamefont {Zhou}, \citenamefont {Liang}, \citenamefont {Weng}, \citenamefont {Chen}, \citenamefont {Yao}, \citenamefont {Chen}, \citenamefont {Dong},\ and\ \citenamefont {Guo}}]{ZhouJ2016}%
  \BibitemOpen
  \bibfield  {author} {\bibinfo {author} {\bibfnamefont {J.}~\bibnamefont {Zhou}}, \bibinfo {author} {\bibfnamefont {Q.-F.}\ \bibnamefont {Liang}}, \bibinfo {author} {\bibfnamefont {H.}~\bibnamefont {Weng}}, \bibinfo {author} {\bibfnamefont {Y.~B.}\ \bibnamefont {Chen}}, \bibinfo {author} {\bibfnamefont {S.-H.}\ \bibnamefont {Yao}}, \bibinfo {author} {\bibfnamefont {Y.-F.}\ \bibnamefont {Chen}}, \bibinfo {author} {\bibfnamefont {J.}~\bibnamefont {Dong}}, \ and\ \bibinfo {author} {\bibfnamefont {G.-Y.}\ \bibnamefont {Guo}},\ }\bibinfo {title} {Predicted Quantum Topological Hall Effect and Noncoplanar Antiferromagnetism in ${\mathrm{K}}_{0.5}{\mathrm{RhO}}_{2}$},\ \href {\doibase 10.1103/PhysRevLett.116.256601} {\bibfield  {journal} {\bibinfo  {journal} {Phys. Rev. Lett.}\ }\textbf {\bibinfo {volume} {116}},\ \bibinfo {pages} {256601} (\bibinfo {year} {2016})}\BibitemShut {NoStop}%
\bibitem [{\citenamefont {Hanke}\ \emph {et~al.}(2017)\citenamefont {Hanke}, \citenamefont {Freimuth}, \citenamefont {Bl\"ugel},\ and\ \citenamefont {Mokrousov}}]{Hanke2017}%
  \BibitemOpen
  \bibfield  {author} {\bibinfo {author} {\bibfnamefont {J.-P.}\ \bibnamefont {Hanke}}, \bibinfo {author} {\bibfnamefont {F.}~\bibnamefont {Freimuth}}, \bibinfo {author} {\bibfnamefont {S.}~\bibnamefont {Bl\"ugel}}, \ and\ \bibinfo {author} {\bibfnamefont {Y.}~\bibnamefont {Mokrousov}},\ }\bibinfo {title} {Prototypical topological orbital ferromagnet $\gamma$-FeMn},\ \href {https://doi.org/10.1038/srep41078} {\bibfield  {journal} {\bibinfo  {journal} {Sci. Rep}\ }\textbf {\bibinfo {volume} {7}},\ \bibinfo {pages} {41078} (\bibinfo {year} {2017})}\BibitemShut {NoStop}%
\bibitem [{\citenamefont {Feng}\ \emph {et~al.}(2020)\citenamefont {Feng}, \citenamefont {Hanke}, \citenamefont {Zhou}, \citenamefont {Guang-Yu}, \citenamefont {Bl\"ugel}, \citenamefont {Mokrousov},\ and\ \citenamefont {Yao}}]{FengWX2020}%
  \BibitemOpen
  \bibfield  {author} {\bibinfo {author} {\bibfnamefont {W.}~\bibnamefont {Feng}}, \bibinfo {author} {\bibfnamefont {J.-P.}\ \bibnamefont {Hanke}}, \bibinfo {author} {\bibfnamefont {X.}~\bibnamefont {Zhou}}, \bibinfo {author} {\bibfnamefont {G.}~\bibnamefont {Guang-Yu}}, \bibinfo {author} {\bibfnamefont {S.}~\bibnamefont {Bl\"ugel}}, \bibinfo {author} {\bibfnamefont {Y.}~\bibnamefont {Mokrousov}}, \ and\ \bibinfo {author} {\bibfnamefont {Y.}~\bibnamefont {Yao}},\ }\bibinfo {title} {Topological magneto-optical effects and their quantization in noncoplanar antiferromagnets},\ \href {https://doi.org/10.1038/s41467-019-13968-8} {\bibfield  {journal} {\bibinfo  {journal} {Nat. Commun.}\ }\textbf {\bibinfo {volume} {11}},\ \bibinfo {pages} {118} (\bibinfo {year} {2020})}\BibitemShut {NoStop}%
\bibitem [{\citenamefont {$\rm\check{S}$mejkal}\ \emph {et~al.}(2020)\citenamefont {$\rm\check{S}$mejkal}, \citenamefont {Gonz\'{a}lez-Hern\'{a}ndez}, \citenamefont {Jungwirth},\ and\ \citenamefont {Sinova}}]{Libor2020}%
  \BibitemOpen
  \bibfield  {author} {\bibinfo {author} {\bibfnamefont {L.}~\bibnamefont {$\rm\check{S}$mejkal}}, \bibinfo {author} {\bibfnamefont {R.}~\bibnamefont {Gonz\'{a}lez-Hern\'{a}ndez}}, \bibinfo {author} {\bibfnamefont {T.}~\bibnamefont {Jungwirth}}, \ and\ \bibinfo {author} {\bibfnamefont {J.}~\bibnamefont {Sinova}},\ }\bibinfo {title} {Crystal time-reversal symmetry breaking and spontaneous Hall effect in collinear antiferromagnets},\ \href {\doibase 10.1126/sciadv.aaz8809} {\bibfield  {journal} {\bibinfo  {journal} {Sci. Adv.}\ }\textbf {\bibinfo {volume} {6}},\ \bibinfo {pages} {eaaz8809} (\bibinfo {year} {2020})}\BibitemShut {NoStop}%
\bibitem [{\citenamefont {Zhou}\ \emph {et~al.}(2024)\citenamefont {Zhou}, \citenamefont {Feng}, \citenamefont {Zhang}, \citenamefont {\ifmmode~\check{S}\else \v{S}\fi{}mejkal}, \citenamefont {Sinova}, \citenamefont {Mokrousov},\ and\ \citenamefont {Yao}}]{xdzhou2024}%
  \BibitemOpen
  \bibfield  {author} {\bibinfo {author} {\bibfnamefont {X.}~\bibnamefont {Zhou}}, \bibinfo {author} {\bibfnamefont {W.}~\bibnamefont {Feng}}, \bibinfo {author} {\bibfnamefont {R.-W.}\ \bibnamefont {Zhang}}, \bibinfo {author} {\bibfnamefont {L.}~\bibnamefont {\ifmmode~\check{S}\else \v{S}\fi{}mejkal}}, \bibinfo {author} {\bibfnamefont {J.}~\bibnamefont {Sinova}}, \bibinfo {author} {\bibfnamefont {Y.}~\bibnamefont {Mokrousov}}, \ and\ \bibinfo {author} {\bibfnamefont {Y.}~\bibnamefont {Yao}},\ }\bibinfo {title} {Crystal Thermal Transport in Altermagnetic ${\mathrm{RuO}}_{2}$},\ \href {\doibase 10.1103/PhysRevLett.132.056701} {\bibfield  {journal} {\bibinfo  {journal} {Phys. Rev. Lett.}\ }\textbf {\bibinfo {volume} {132}},\ \bibinfo {pages} {056701} (\bibinfo {year} {2024})}\BibitemShut {NoStop}%
\bibitem [{\citenamefont {\ifmmode~\check{S}\else \v{S}\fi{}mejkal}\ \emph {et~al.}(2022{\natexlab{a}})\citenamefont {\ifmmode~\check{S}\else \v{S}\fi{}mejkal}, \citenamefont {Sinova},\ and\ \citenamefont {Jungwirth}}]{Libor2022}%
  \BibitemOpen
  \bibfield  {author} {\bibinfo {author} {\bibfnamefont {L.}~\bibnamefont {\ifmmode~\check{S}\else \v{S}\fi{}mejkal}}, \bibinfo {author} {\bibfnamefont {J.}~\bibnamefont {Sinova}}, \ and\ \bibinfo {author} {\bibfnamefont {T.}~\bibnamefont {Jungwirth}},\ }\bibinfo {title} {Beyond Conventional Ferromagnetism and Antiferromagnetism: A Phase with Nonrelativistic Spin and Crystal Rotation Symmetry},\ \href {\doibase 10.1103/PhysRevX.12.031042} {\bibfield  {journal} {\bibinfo  {journal} {Phys. Rev. X}\ }\textbf {\bibinfo {volume} {12}},\ \bibinfo {pages} {031042} (\bibinfo {year} {2022}{\natexlab{a}})}\BibitemShut {NoStop}%
\bibitem [{\citenamefont {\ifmmode~\check{S}\else \v{S}\fi{}mejkal}\ \emph {et~al.}(2022{\natexlab{b}})\citenamefont {\ifmmode~\check{S}\else \v{S}\fi{}mejkal}, \citenamefont {Sinova},\ and\ \citenamefont {Jungwirth}}]{Libor2022b}%
  \BibitemOpen
  \bibfield  {author} {\bibinfo {author} {\bibfnamefont {L.}~\bibnamefont {\ifmmode~\check{S}\else \v{S}\fi{}mejkal}}, \bibinfo {author} {\bibfnamefont {J.}~\bibnamefont {Sinova}}, \ and\ \bibinfo {author} {\bibfnamefont {T.}~\bibnamefont {Jungwirth}},\ }\bibinfo {title} {Emerging Research Landscape of Altermagnetism},\ \href {\doibase 10.1103/PhysRevX.12.040501} {\bibfield  {journal} {\bibinfo  {journal} {Phys. Rev. X}\ }\textbf {\bibinfo {volume} {12}},\ \bibinfo {pages} {040501} (\bibinfo {year} {2022}{\natexlab{b}})}\BibitemShut {NoStop}%
\bibitem [{\citenamefont {Mazin}(2022)}]{Mazin2022}%
  \BibitemOpen
  \bibfield  {author} {\bibinfo {author} {\bibfnamefont {I.}~\bibnamefont {Mazin}} (\bibinfo {collaboration} {The PRX Editors}),\ }\bibinfo {title} {Editorial: Altermagnetism---A New Punch Line of Fundamental Magnetism},\ \href {\doibase 10.1103/PhysRevX.12.040002} {\bibfield  {journal} {\bibinfo  {journal} {Phys. Rev. X}\ }\textbf {\bibinfo {volume} {12}},\ \bibinfo {pages} {040002} (\bibinfo {year} {2022})}\BibitemShut {NoStop}%
\bibitem [{\citenamefont {Hayami}\ \emph {et~al.}(2019)\citenamefont {Hayami}, \citenamefont {Yanagi},\ and\ \citenamefont {Kusunose}}]{HayamiS2019}%
  \BibitemOpen
  \bibfield  {author} {\bibinfo {author} {\bibfnamefont {S.}~\bibnamefont {Hayami}}, \bibinfo {author} {\bibfnamefont {Y.}~\bibnamefont {Yanagi}}, \ and\ \bibinfo {author} {\bibfnamefont {H.}~\bibnamefont {Kusunose}},\ }\bibinfo {title} {Momentum-Dependent Spin Splitting by Collinear Antiferromagnetic Ordering},\ \href {\doibase 10.7566/JPSJ.88.123702} {\bibfield  {journal} {\bibinfo  {journal} {Journal of the Physical Society of Japan}\ }\textbf {\bibinfo {volume} {88}},\ \bibinfo {pages} {123702} (\bibinfo {year} {2019})}\BibitemShut {NoStop}%
\bibitem [{\citenamefont {Zhou}\ \emph {et~al.}(2021)\citenamefont {Zhou}, \citenamefont {Feng}, \citenamefont {Yang}, \citenamefont {Guo},\ and\ \citenamefont {Yao}}]{xdzhou2021}%
  \BibitemOpen
  \bibfield  {author} {\bibinfo {author} {\bibfnamefont {X.}~\bibnamefont {Zhou}}, \bibinfo {author} {\bibfnamefont {W.}~\bibnamefont {Feng}}, \bibinfo {author} {\bibfnamefont {X.}~\bibnamefont {Yang}}, \bibinfo {author} {\bibfnamefont {G.-Y.}\ \bibnamefont {Guo}}, \ and\ \bibinfo {author} {\bibfnamefont {Y.}~\bibnamefont {Yao}},\ }\bibinfo {title} {Crystal chirality magneto-optical effects in collinear antiferromagnets},\ \href {\doibase 10.1103/PhysRevB.104.024401} {\bibfield  {journal} {\bibinfo  {journal} {Phys. Rev. B}\ }\textbf {\bibinfo {volume} {104}},\ \bibinfo {pages} {024401} (\bibinfo {year} {2021})}\BibitemShut {NoStop}%
\bibitem [{\citenamefont {Ma}\ \emph {et~al.}(2021)\citenamefont {Ma}, \citenamefont {Hu}, \citenamefont {Li}, \citenamefont {Liu}, \citenamefont {Yao}, \citenamefont {Jia},\ and\ \citenamefont {Liu}}]{MaHY2021}%
  \BibitemOpen
  \bibfield  {author} {\bibinfo {author} {\bibfnamefont {H.-Y.}\ \bibnamefont {Ma}}, \bibinfo {author} {\bibfnamefont {M.}~\bibnamefont {Hu}}, \bibinfo {author} {\bibfnamefont {N.}~\bibnamefont {Li}}, \bibinfo {author} {\bibfnamefont {J.}~\bibnamefont {Liu}}, \bibinfo {author} {\bibfnamefont {W.}~\bibnamefont {Yao}}, \bibinfo {author} {\bibfnamefont {J.-F.}\ \bibnamefont {Jia}}, \ and\ \bibinfo {author} {\bibfnamefont {J.}~\bibnamefont {Liu}},\ }\bibinfo {title} {Multifunctional antiferromagnetic materials with giant piezomagnetism and noncollinear spin current},\ \href {\doibase 10.1038/s41467-021-23127-7} {\bibfield  {journal} {\bibinfo  {journal} {Nat. Commun.}\ }\textbf {\bibinfo {volume} {12}},\ \bibinfo {pages} {2846} (\bibinfo {year} {2021})}\BibitemShut {NoStop}%
\bibitem [{\citenamefont {Feng}\ \emph {et~al.}(2022)\citenamefont {Feng}, \citenamefont {Zhou}, \citenamefont {Šmejkal}, \citenamefont {Wu}, \citenamefont {Zhu}, \citenamefont {Guo}, \citenamefont {González-Hernández}, \citenamefont {Wang}, \citenamefont {Yan}, \citenamefont {Qin}, \citenamefont {Zhang}, \citenamefont {Wu}, \citenamefont {Chen}, \citenamefont {Meng}, \citenamefont {Liu}, \citenamefont {Xia}, \citenamefont {Sinova}, \citenamefont {Jungwirth},\ and\ \citenamefont {Liu}}]{feng2022}%
  \BibitemOpen
  \bibfield  {author} {\bibinfo {author} {\bibfnamefont {Z.}~\bibnamefont {Feng}}, \bibinfo {author} {\bibfnamefont {X.}~\bibnamefont {Zhou}}, \bibinfo {author} {\bibfnamefont {L.}~\bibnamefont {Šmejkal}}, \bibinfo {author} {\bibfnamefont {L.}~\bibnamefont {Wu}}, \bibinfo {author} {\bibfnamefont {Z.}~\bibnamefont {Zhu}}, \bibinfo {author} {\bibfnamefont {H.}~\bibnamefont {Guo}}, \bibinfo {author} {\bibfnamefont {R.}~\bibnamefont {González-Hernández}}, \bibinfo {author} {\bibfnamefont {X.}~\bibnamefont {Wang}}, \bibinfo {author} {\bibfnamefont {H.}~\bibnamefont {Yan}}, \bibinfo {author} {\bibfnamefont {P.}~\bibnamefont {Qin}}, \bibinfo {author} {\bibfnamefont {X.}~\bibnamefont {Zhang}}, \bibinfo {author} {\bibfnamefont {H.}~\bibnamefont {Wu}}, \bibinfo {author} {\bibfnamefont {H.}~\bibnamefont {Chen}}, \bibinfo {author} {\bibfnamefont {Z.}~\bibnamefont {Meng}}, \bibinfo {author} {\bibfnamefont {L.}~\bibnamefont {Liu}}, \bibinfo {author} {\bibfnamefont {Z.}~\bibnamefont {Xia}}, \bibinfo {author} {\bibfnamefont {J.}~\bibnamefont {Sinova}}, \bibinfo {author} {\bibfnamefont {T.}~\bibnamefont {Jungwirth}}, \ and\ \bibinfo {author} {\bibfnamefont {Z.}~\bibnamefont {Liu}},\ }\bibinfo {title} {An anomalous Hall effect in altermagnetic ruthenium dioxide},\ \href {\doibase 10.1038/s41928-022-00866-z} {\bibfield  {journal} {\bibinfo  {journal} {Nat. Electron.}\ }\textbf {\bibinfo {volume} {5}},\ \bibinfo {pages} {735} (\bibinfo {year} {2022})}\BibitemShut {NoStop}%
\bibitem [{\citenamefont {Bai}\ \emph {et~al.}(2023)\citenamefont {Bai}, \citenamefont {Zhang}, \citenamefont {Zhou}, \citenamefont {Chen}, \citenamefont {Wan}, \citenamefont {Han}, \citenamefont {Zhu}, \citenamefont {Liang}, \citenamefont {Su}, \citenamefont {Han}, \citenamefont {Pan},\ and\ \citenamefont {Song}}]{H-Bai2023}%
  \BibitemOpen
  \bibfield  {author} {\bibinfo {author} {\bibfnamefont {H.}~\bibnamefont {Bai}}, \bibinfo {author} {\bibfnamefont {Y.~C.}\ \bibnamefont {Zhang}}, \bibinfo {author} {\bibfnamefont {Y.~J.}\ \bibnamefont {Zhou}}, \bibinfo {author} {\bibfnamefont {P.}~\bibnamefont {Chen}}, \bibinfo {author} {\bibfnamefont {C.~H.}\ \bibnamefont {Wan}}, \bibinfo {author} {\bibfnamefont {L.}~\bibnamefont {Han}}, \bibinfo {author} {\bibfnamefont {W.~X.}\ \bibnamefont {Zhu}}, \bibinfo {author} {\bibfnamefont {S.~X.}\ \bibnamefont {Liang}}, \bibinfo {author} {\bibfnamefont {Y.~C.}\ \bibnamefont {Su}}, \bibinfo {author} {\bibfnamefont {X.~F.}\ \bibnamefont {Han}}, \bibinfo {author} {\bibfnamefont {F.}~\bibnamefont {Pan}}, \ and\ \bibinfo {author} {\bibfnamefont {C.}~\bibnamefont {Song}},\ }\bibinfo {title} {Efficient Spin-to-Charge Conversion via Altermagnetic Spin Splitting Effect in Antiferromagnet ${\mathrm{RuO}}_{2}$},\ \href {\doibase 10.1103/PhysRevLett.130.216701} {\bibfield  {journal} {\bibinfo  {journal} {Phys. Rev. Lett.}\ }\textbf {\bibinfo {volume} {130}},\ \bibinfo {pages} {216701} (\bibinfo {year} {2023})}\BibitemShut {NoStop}%
\bibitem [{\citenamefont {Bai}\ \emph {et~al.}(2024)\citenamefont {Bai}, \citenamefont {Feng}, \citenamefont {Liu}, \citenamefont {{\v{S}}mejkal}, \citenamefont {Mokrousov},\ and\ \citenamefont {Yao}}]{L-Bai2024}%
  \BibitemOpen
  \bibfield  {author} {\bibinfo {author} {\bibfnamefont {L.}~\bibnamefont {Bai}}, \bibinfo {author} {\bibfnamefont {W.}~\bibnamefont {Feng}}, \bibinfo {author} {\bibfnamefont {S.}~\bibnamefont {Liu}}, \bibinfo {author} {\bibfnamefont {L.}~\bibnamefont {{\v{S}}mejkal}}, \bibinfo {author} {\bibfnamefont {Y.}~\bibnamefont {Mokrousov}}, \ and\ \bibinfo {author} {\bibfnamefont {Y.}~\bibnamefont {Yao}},\ }\bibinfo {title} {Altermagnetism: Exploring New Frontiers in Magnetism and Spintronics},\ \href {\doibase 10.1002/adfm.202409327} {\bibfield  {journal} {\bibinfo  {journal} {Adv. Funct. Mater.}\ }\textbf {\bibinfo {volume} {34}},\ \bibinfo {pages} {2409327} (\bibinfo {year} {2024})}\BibitemShut {NoStop}%
\bibitem [{\citenamefont {Ouassou}\ \emph {et~al.}(2023)\citenamefont {Ouassou}, \citenamefont {Brataas},\ and\ \citenamefont {Linder}}]{Ouassou2023}%
  \BibitemOpen
  \bibfield  {author} {\bibinfo {author} {\bibfnamefont {J.~A.}\ \bibnamefont {Ouassou}}, \bibinfo {author} {\bibfnamefont {A.}~\bibnamefont {Brataas}}, \ and\ \bibinfo {author} {\bibfnamefont {J.}~\bibnamefont {Linder}},\ }\bibinfo {title} {dc Josephson Effect in Altermagnets},\ \href {\doibase 10.1103/PhysRevLett.131.076003} {\bibfield  {journal} {\bibinfo  {journal} {Phys. Rev. Lett.}\ }\textbf {\bibinfo {volume} {131}},\ \bibinfo {pages} {076003} (\bibinfo {year} {2023})}\BibitemShut {NoStop}%
\bibitem [{\citenamefont {Li}\ and\ \citenamefont {Liu}(2023)}]{YXLi2023}%
  \BibitemOpen
  \bibfield  {author} {\bibinfo {author} {\bibfnamefont {Y.-X.}\ \bibnamefont {Li}}\ and\ \bibinfo {author} {\bibfnamefont {C.-C.}\ \bibnamefont {Liu}},\ }\bibinfo {title} {Majorana corner modes and tunable patterns in an altermagnet heterostructure},\ \href {\doibase 10.1103/PhysRevB.108.205410} {\bibfield  {journal} {\bibinfo  {journal} {Phys. Rev. B}\ }\textbf {\bibinfo {volume} {108}},\ \bibinfo {pages} {205410} (\bibinfo {year} {2023})}\BibitemShut {NoStop}%
\bibitem [{\citenamefont {Zhu}\ \emph {et~al.}(2023)\citenamefont {Zhu}, \citenamefont {Zhuang}, \citenamefont {Wu},\ and\ \citenamefont {Yan}}]{DiZhu2023}%
  \BibitemOpen
  \bibfield  {author} {\bibinfo {author} {\bibfnamefont {D.}~\bibnamefont {Zhu}}, \bibinfo {author} {\bibfnamefont {Z.-Y.}\ \bibnamefont {Zhuang}}, \bibinfo {author} {\bibfnamefont {Z.}~\bibnamefont {Wu}}, \ and\ \bibinfo {author} {\bibfnamefont {Z.}~\bibnamefont {Yan}},\ }\bibinfo {title} {Topological superconductivity in two-dimensional altermagnetic metals},\ \href {\doibase 10.1103/PhysRevB.108.184505} {\bibfield  {journal} {\bibinfo  {journal} {Phys. Rev. B}\ }\textbf {\bibinfo {volume} {108}},\ \bibinfo {pages} {184505} (\bibinfo {year} {2023})}\BibitemShut {NoStop}%
\bibitem [{\citenamefont {Zhang}\ \emph {et~al.}(2024{\natexlab{a}})\citenamefont {Zhang}, \citenamefont {Hu},\ and\ \citenamefont {Neupert}}]{zhang2024}%
  \BibitemOpen
  \bibfield  {author} {\bibinfo {author} {\bibfnamefont {S.-B.}\ \bibnamefont {Zhang}}, \bibinfo {author} {\bibfnamefont {L.-H.}\ \bibnamefont {Hu}}, \ and\ \bibinfo {author} {\bibfnamefont {T.}~\bibnamefont {Neupert}},\ }\bibinfo {title} {Finite-momentum Cooper pairing in proximitized altermagnets},\ \href {\doibase 10.1038/s41467-024-45951-3} {\bibfield  {journal} {\bibinfo  {journal} {Nat. Commun.}\ }\textbf {\bibinfo {volume} {15}},\ \bibinfo {pages} {1801} (\bibinfo {year} {2024}{\natexlab{a}})}\BibitemShut {NoStop}%
\bibitem [{\citenamefont {Fender}\ \emph {et~al.}(2025)\citenamefont {Fender}, \citenamefont {Gonzalez},\ and\ \citenamefont {Bediako}}]{Fender2025JACS}%
  \BibitemOpen
  \bibfield  {author} {\bibinfo {author} {\bibfnamefont {S.~S.}\ \bibnamefont {Fender}}, \bibinfo {author} {\bibfnamefont {O.}~\bibnamefont {Gonzalez}}, \ and\ \bibinfo {author} {\bibfnamefont {D.~K.}\ \bibnamefont {Bediako}},\ }\bibinfo {title} {Altermagnetism: A Chemical Perspective},\ \href {\doibase 10.1021/jacs.4c14503} {\bibfield  {journal} {\bibinfo  {journal} {J. Am. Chem. Soc.}\ }\textbf {\bibinfo {volume} {147}},\ \bibinfo {pages} {2257} (\bibinfo {year} {2025})}\BibitemShut {NoStop}%
\bibitem [{\citenamefont {Guo}\ \emph {et~al.}(2025)\citenamefont {Guo}, \citenamefont {Wang}, \citenamefont {Wang}, \citenamefont {Zhang}, \citenamefont {Zhou},\ and\ \citenamefont {Cheng}}]{guo_AM2025}%
  \BibitemOpen
  \bibfield  {author} {\bibinfo {author} {\bibfnamefont {Z.}~\bibnamefont {Guo}}, \bibinfo {author} {\bibfnamefont {X.}~\bibnamefont {Wang}}, \bibinfo {author} {\bibfnamefont {W.}~\bibnamefont {Wang}}, \bibinfo {author} {\bibfnamefont {G.}~\bibnamefont {Zhang}}, \bibinfo {author} {\bibfnamefont {X.}~\bibnamefont {Zhou}}, \ and\ \bibinfo {author} {\bibfnamefont {Z.}~\bibnamefont {Cheng}},\ }\bibinfo {title} {Spin‐{Polarized} {Antiferromagnets} for {Spintronics}},\ \href {https://advanced.onlinelibrary.wiley.com/doi/10.1002/adma.202505779} {\bibfield  {journal} {\bibinfo  {journal} {Adv. Mater.}\ ,\ \bibinfo {pages} {2505779}} (\bibinfo {year} {2025})}\BibitemShut {NoStop}%
\bibitem [{\citenamefont {Lee}\ \emph {et~al.}(2024)\citenamefont {Lee}, \citenamefont {Lee}, \citenamefont {Jung}, \citenamefont {Jung}, \citenamefont {Kim}, \citenamefont {Lee}, \citenamefont {Seok}, \citenamefont {Kim}, \citenamefont {Park}, \citenamefont {\ifmmode~\check{S}\else \v{S}\fi{}mejkal}, \citenamefont {Kang},\ and\ \citenamefont {Kim}}]{Li2024}%
  \BibitemOpen
  \bibfield  {author} {\bibinfo {author} {\bibfnamefont {S.}~\bibnamefont {Lee}}, \bibinfo {author} {\bibfnamefont {S.}~\bibnamefont {Lee}}, \bibinfo {author} {\bibfnamefont {S.}~\bibnamefont {Jung}}, \bibinfo {author} {\bibfnamefont {J.}~\bibnamefont {Jung}}, \bibinfo {author} {\bibfnamefont {D.}~\bibnamefont {Kim}}, \bibinfo {author} {\bibfnamefont {Y.}~\bibnamefont {Lee}}, \bibinfo {author} {\bibfnamefont {B.}~\bibnamefont {Seok}}, \bibinfo {author} {\bibfnamefont {J.}~\bibnamefont {Kim}}, \bibinfo {author} {\bibfnamefont {B.~G.}\ \bibnamefont {Park}}, \bibinfo {author} {\bibfnamefont {L.}~\bibnamefont {\ifmmode~\check{S}\else \v{S}\fi{}mejkal}}, \bibinfo {author} {\bibfnamefont {C.-J.}\ \bibnamefont {Kang}}, \ and\ \bibinfo {author} {\bibfnamefont {C.}~\bibnamefont {Kim}},\ }\bibinfo {title} {Broken Kramers Degeneracy in Altermagnetic MnTe},\ \href {\doibase 10.1103/PhysRevLett.132.036702} {\bibfield  {journal} {\bibinfo  {journal} {Phys. Rev. Lett.}\ }\textbf {\bibinfo {volume} {132}},\ \bibinfo {pages} {036702} (\bibinfo {year} {2024})}\BibitemShut {NoStop}%
\bibitem [{\citenamefont {Krempask\'{y}}\ \emph {et~al.}(2024)\citenamefont {Krempask\'{y}}, \citenamefont {$\rm\check{S}$mejkal}, \citenamefont {D$'$Souza}, \citenamefont {Hajlaoui}, \citenamefont {Springholz}, \citenamefont {Uhl\'{i}$\rm\check{r}$ov\'{a}}, \citenamefont {Alarab}, \citenamefont {Constantinou}, \citenamefont {Strocov}, \citenamefont {Usanov}, \citenamefont {Pudelko}, \citenamefont {Gonz\'{a}lez-Hern\'{a}ndez}, \citenamefont {Birk~Hellenes}, \citenamefont {Jansa}, \citenamefont {Reichlov\'{a}}, \citenamefont {$\rm\check{S}$ob\'{a}$\rm\check{n}$}, \citenamefont {Gonzalez~Betancourt}, \citenamefont {Wadley}, \citenamefont {Sinova}, \citenamefont {Kriegner}, \citenamefont {Min\'{a}r}, \citenamefont {Dil},\ and\ \citenamefont {Jungwirth}}]{krempasky2024}%
  \BibitemOpen
  \bibfield  {author} {\bibinfo {author} {\bibfnamefont {J.}~\bibnamefont {Krempask\'{y}}}, \bibinfo {author} {\bibfnamefont {L.}~\bibnamefont {$\rm\check{S}$mejkal}}, \bibinfo {author} {\bibfnamefont {S.~W.}\ \bibnamefont {D$'$Souza}}, \bibinfo {author} {\bibfnamefont {M.}~\bibnamefont {Hajlaoui}}, \bibinfo {author} {\bibfnamefont {G.}~\bibnamefont {Springholz}}, \bibinfo {author} {\bibfnamefont {K.}~\bibnamefont {Uhl\'{i}$\rm\check{r}$ov\'{a}}}, \bibinfo {author} {\bibfnamefont {F.}~\bibnamefont {Alarab}}, \bibinfo {author} {\bibfnamefont {P.~C.}\ \bibnamefont {Constantinou}}, \bibinfo {author} {\bibfnamefont {V.}~\bibnamefont {Strocov}}, \bibinfo {author} {\bibfnamefont {D.}~\bibnamefont {Usanov}}, \bibinfo {author} {\bibfnamefont {W.~R.}\ \bibnamefont {Pudelko}}, \bibinfo {author} {\bibfnamefont {R.}~\bibnamefont {Gonz\'{a}lez-Hern\'{a}ndez}}, \bibinfo {author} {\bibfnamefont {A.}~\bibnamefont {Birk~Hellenes}}, \bibinfo {author} {\bibfnamefont {Z.}~\bibnamefont {Jansa}}, \bibinfo {author} {\bibfnamefont {H.}~\bibnamefont {Reichlov\'{a}}}, \bibinfo {author} {\bibfnamefont {Z.}~\bibnamefont {$\rm\check{S}$ob\'{a}$\rm\check{n}$}}, \bibinfo {author} {\bibfnamefont {R.~D.}\ \bibnamefont {Gonzalez~Betancourt}}, \bibinfo {author} {\bibfnamefont {P.}~\bibnamefont {Wadley}}, \bibinfo {author} {\bibfnamefont {J.}~\bibnamefont {Sinova}}, \bibinfo {author} {\bibfnamefont {D.}~\bibnamefont {Kriegner}}, \bibinfo {author} {\bibfnamefont {J.}~\bibnamefont {Min\'{a}r}}, \bibinfo {author} {\bibfnamefont {J.~H.}\ \bibnamefont {Dil}}, \ and\ \bibinfo {author} {\bibfnamefont {T.}~\bibnamefont {Jungwirth}},\ }\bibinfo {title} {Altermagnetic lifting of {Kramers} spin degeneracy},\ \href {\doibase 10.1038/s41586-023-06907-7} {\bibfield  {journal} {\bibinfo  {journal} {Nature}\ }\textbf {\bibinfo {volume} {626}},\ \bibinfo {pages} {517} (\bibinfo {year} {2024})}\BibitemShut {NoStop}%
\bibitem [{\citenamefont {Reimers}\ \emph {et~al.}(2024)\citenamefont {Reimers}, \citenamefont {Odenbreit}, \citenamefont {Šmejkal}, \citenamefont {Strocov}, \citenamefont {Constantinou}, \citenamefont {Hellenes}, \citenamefont {Jaeschke~Ubiergo}, \citenamefont {Campos}, \citenamefont {Bharadwaj}, \citenamefont {Chakraborty}, \citenamefont {Denneulin}, \citenamefont {Shi}, \citenamefont {Dunin-Borkowski}, \citenamefont {Das}, \citenamefont {Kläui}, \citenamefont {Sinova},\ and\ \citenamefont {Jourdan}}]{Reimers2024}%
  \BibitemOpen
  \bibfield  {author} {\bibinfo {author} {\bibfnamefont {S.}~\bibnamefont {Reimers}}, \bibinfo {author} {\bibfnamefont {L.}~\bibnamefont {Odenbreit}}, \bibinfo {author} {\bibfnamefont {L.}~\bibnamefont {Šmejkal}}, \bibinfo {author} {\bibfnamefont {V.~N.}\ \bibnamefont {Strocov}}, \bibinfo {author} {\bibfnamefont {P.}~\bibnamefont {Constantinou}}, \bibinfo {author} {\bibfnamefont {A.~B.}\ \bibnamefont {Hellenes}}, \bibinfo {author} {\bibfnamefont {R.}~\bibnamefont {Jaeschke~Ubiergo}}, \bibinfo {author} {\bibfnamefont {W.~H.}\ \bibnamefont {Campos}}, \bibinfo {author} {\bibfnamefont {V.~K.}\ \bibnamefont {Bharadwaj}}, \bibinfo {author} {\bibfnamefont {A.}~\bibnamefont {Chakraborty}}, \bibinfo {author} {\bibfnamefont {T.}~\bibnamefont {Denneulin}}, \bibinfo {author} {\bibfnamefont {W.}~\bibnamefont {Shi}}, \bibinfo {author} {\bibfnamefont {R.~E.}\ \bibnamefont {Dunin-Borkowski}}, \bibinfo {author} {\bibfnamefont {S.}~\bibnamefont {Das}}, \bibinfo {author} {\bibfnamefont {M.}~\bibnamefont {Kläui}}, \bibinfo {author} {\bibfnamefont {J.}~\bibnamefont {Sinova}}, \ and\ \bibinfo {author} {\bibfnamefont {M.}~\bibnamefont {Jourdan}},\ }\bibinfo {title} {Direct observation of altermagnetic band splitting in {CrSb} thin films},\ \href {\doibase 10.1038/s41467-024-46476-5} {\bibfield  {journal} {\bibinfo  {journal} {Nat. Commun.}\ }\textbf {\bibinfo {volume} {15}},\ \bibinfo {pages} {2116} (\bibinfo {year} {2024})}\BibitemShut {NoStop}%
\bibitem [{\citenamefont {Ding}\ \emph {et~al.}(2024)\citenamefont {Ding}, \citenamefont {Jiang}, \citenamefont {Chen}, \citenamefont {Tao}, \citenamefont {Liu}, \citenamefont {Li}, \citenamefont {Liu}, \citenamefont {Sun}, \citenamefont {Cheng}, \citenamefont {Liu}, \citenamefont {Yang}, \citenamefont {Zhang}, \citenamefont {Deng}, \citenamefont {Jing}, \citenamefont {Huang}, \citenamefont {Shi}, \citenamefont {Ye}, \citenamefont {Qiao}, \citenamefont {Wang}, \citenamefont {Guo}, \citenamefont {Feng},\ and\ \citenamefont {Shen}}]{DingJY2024}%
  \BibitemOpen
  \bibfield  {author} {\bibinfo {author} {\bibfnamefont {J.}~\bibnamefont {Ding}}, \bibinfo {author} {\bibfnamefont {Z.}~\bibnamefont {Jiang}}, \bibinfo {author} {\bibfnamefont {X.}~\bibnamefont {Chen}}, \bibinfo {author} {\bibfnamefont {Z.}~\bibnamefont {Tao}}, \bibinfo {author} {\bibfnamefont {Z.}~\bibnamefont {Liu}}, \bibinfo {author} {\bibfnamefont {T.}~\bibnamefont {Li}}, \bibinfo {author} {\bibfnamefont {J.}~\bibnamefont {Liu}}, \bibinfo {author} {\bibfnamefont {J.}~\bibnamefont {Sun}}, \bibinfo {author} {\bibfnamefont {J.}~\bibnamefont {Cheng}}, \bibinfo {author} {\bibfnamefont {J.}~\bibnamefont {Liu}}, \bibinfo {author} {\bibfnamefont {Y.}~\bibnamefont {Yang}}, \bibinfo {author} {\bibfnamefont {R.}~\bibnamefont {Zhang}}, \bibinfo {author} {\bibfnamefont {L.}~\bibnamefont {Deng}}, \bibinfo {author} {\bibfnamefont {W.}~\bibnamefont {Jing}}, \bibinfo {author} {\bibfnamefont {Y.}~\bibnamefont {Huang}}, \bibinfo {author} {\bibfnamefont {Y.}~\bibnamefont {Shi}}, \bibinfo {author} {\bibfnamefont {M.}~\bibnamefont {Ye}}, \bibinfo {author} {\bibfnamefont {S.}~\bibnamefont {Qiao}}, \bibinfo {author} {\bibfnamefont {Y.}~\bibnamefont {Wang}}, \bibinfo {author} {\bibfnamefont {Y.}~\bibnamefont {Guo}}, \bibinfo {author} {\bibfnamefont {D.}~\bibnamefont {Feng}}, \ and\ \bibinfo {author} {\bibfnamefont {D.}~\bibnamefont {Shen}},\ }\bibinfo {title} {Large Band Splitting in $g$-Wave Altermagnet CrSb},\ \href {\doibase 10.1103/PhysRevLett.133.206401} {\bibfield  {journal} {\bibinfo  {journal} {Phys. Rev. Lett.}\ }\textbf {\bibinfo {volume} {133}},\ \bibinfo {pages} {206401} (\bibinfo {year} {2024})}\BibitemShut {NoStop}%
\bibitem [{\citenamefont {Yang}\ \emph {et~al.}(2025)\citenamefont {Yang}, \citenamefont {Li}, \citenamefont {Yang}, \citenamefont {Li}, \citenamefont {Zheng}, \citenamefont {Zhu}, \citenamefont {Pan}, \citenamefont {Xu}, \citenamefont {Cao}, \citenamefont {Zhao}, \citenamefont {Jana}, \citenamefont {Zhang}, \citenamefont {Ye}, \citenamefont {Song}, \citenamefont {Hu}, \citenamefont {Yang}, \citenamefont {Fujii}, \citenamefont {Vobornik}, \citenamefont {Shi}, \citenamefont {Yuan}, \citenamefont {Zhang}, \citenamefont {Xu},\ and\ \citenamefont {Liu}}]{YangGW2024}%
  \BibitemOpen
  \bibfield  {author} {\bibinfo {author} {\bibfnamefont {G.}~\bibnamefont {Yang}}, \bibinfo {author} {\bibfnamefont {Z.}~\bibnamefont {Li}}, \bibinfo {author} {\bibfnamefont {S.}~\bibnamefont {Yang}}, \bibinfo {author} {\bibfnamefont {J.}~\bibnamefont {Li}}, \bibinfo {author} {\bibfnamefont {H.}~\bibnamefont {Zheng}}, \bibinfo {author} {\bibfnamefont {W.}~\bibnamefont {Zhu}}, \bibinfo {author} {\bibfnamefont {Z.}~\bibnamefont {Pan}}, \bibinfo {author} {\bibfnamefont {Y.}~\bibnamefont {Xu}}, \bibinfo {author} {\bibfnamefont {S.}~\bibnamefont {Cao}}, \bibinfo {author} {\bibfnamefont {W.}~\bibnamefont {Zhao}}, \bibinfo {author} {\bibfnamefont {A.}~\bibnamefont {Jana}}, \bibinfo {author} {\bibfnamefont {J.}~\bibnamefont {Zhang}}, \bibinfo {author} {\bibfnamefont {M.}~\bibnamefont {Ye}}, \bibinfo {author} {\bibfnamefont {Y.}~\bibnamefont {Song}}, \bibinfo {author} {\bibfnamefont {L.-H.}\ \bibnamefont {Hu}}, \bibinfo {author} {\bibfnamefont {L.}~\bibnamefont {Yang}}, \bibinfo {author} {\bibfnamefont {J.}~\bibnamefont {Fujii}}, \bibinfo {author} {\bibfnamefont {I.}~\bibnamefont {Vobornik}}, \bibinfo {author} {\bibfnamefont {M.}~\bibnamefont {Shi}}, \bibinfo {author} {\bibfnamefont {H.}~\bibnamefont {Yuan}}, \bibinfo {author} {\bibfnamefont {Y.}~\bibnamefont {Zhang}}, \bibinfo {author} {\bibfnamefont {Y.}~\bibnamefont {Xu}}, \ and\ \bibinfo {author} {\bibfnamefont {Y.}~\bibnamefont {Liu}},\ }\bibinfo {title} {Three-dimensional mapping of the altermagnetic spin splitting in {CrSb}},\ \href {\doibase 10.1038/s41467-025-56647-7} {\bibfield  {journal} {\bibinfo  {journal} {Nat. Commun.}\ }\textbf {\bibinfo {volume} {16}},\ \bibinfo {pages} {1442} (\bibinfo {year} {2025})}\BibitemShut {NoStop}%
\bibitem [{\citenamefont {Zeng}\ \emph {et~al.}(2024)\citenamefont {Zeng}, \citenamefont {Zhu}, \citenamefont {Zhu}, \citenamefont {Liu}, \citenamefont {Ma}, \citenamefont {Hao}, \citenamefont {Liu}, \citenamefont {Qu}, \citenamefont {Yang}, \citenamefont {Jiang}, \citenamefont {Yamagami}, \citenamefont {Arita}, \citenamefont {Zhang}, \citenamefont {Shao}, \citenamefont {Dai}, \citenamefont {Shimada}, \citenamefont {Liu}, \citenamefont {Ye}, \citenamefont {Huang}, \citenamefont {Liu},\ and\ \citenamefont {Liu}}]{ZengM2024}%
  \BibitemOpen
  \bibfield  {author} {\bibinfo {author} {\bibfnamefont {M.}~\bibnamefont {Zeng}}, \bibinfo {author} {\bibfnamefont {M.-Y.}\ \bibnamefont {Zhu}}, \bibinfo {author} {\bibfnamefont {Y.-P.}\ \bibnamefont {Zhu}}, \bibinfo {author} {\bibfnamefont {X.-R.}\ \bibnamefont {Liu}}, \bibinfo {author} {\bibfnamefont {X.-M.}\ \bibnamefont {Ma}}, \bibinfo {author} {\bibfnamefont {Y.-J.}\ \bibnamefont {Hao}}, \bibinfo {author} {\bibfnamefont {P.}~\bibnamefont {Liu}}, \bibinfo {author} {\bibfnamefont {G.}~\bibnamefont {Qu}}, \bibinfo {author} {\bibfnamefont {Y.}~\bibnamefont {Yang}}, \bibinfo {author} {\bibfnamefont {Z.}~\bibnamefont {Jiang}}, \bibinfo {author} {\bibfnamefont {K.}~\bibnamefont {Yamagami}}, \bibinfo {author} {\bibfnamefont {M.}~\bibnamefont {Arita}}, \bibinfo {author} {\bibfnamefont {X.}~\bibnamefont {Zhang}}, \bibinfo {author} {\bibfnamefont {T.-H.}\ \bibnamefont {Shao}}, \bibinfo {author} {\bibfnamefont {Y.}~\bibnamefont {Dai}}, \bibinfo {author} {\bibfnamefont {K.}~\bibnamefont {Shimada}}, \bibinfo {author} {\bibfnamefont {Z.}~\bibnamefont {Liu}}, \bibinfo {author} {\bibfnamefont {M.}~\bibnamefont {Ye}}, \bibinfo {author} {\bibfnamefont {Y.}~\bibnamefont {Huang}}, \bibinfo {author} {\bibfnamefont {Q.}~\bibnamefont {Liu}}, \ and\ \bibinfo {author} {\bibfnamefont {C.}~\bibnamefont {Liu}},\ }\bibinfo {title} {Observation of Spin Splitting in Room-Temperature Metallic Antiferromagnet CrSb},\ \href {\doibase https://doi.org/10.1002/advs.202406529} {\bibfield  {journal} {\bibinfo  {journal} {Adv. Sci.}\ }\textbf {\bibinfo {volume} {11}},\ \bibinfo {pages} {2406529} (\bibinfo {year} {2024})}\BibitemShut {NoStop}%
\bibitem [{\citenamefont {Zhou}\ \emph {et~al.}(2025)\citenamefont {Zhou}, \citenamefont {Cheng}, \citenamefont {Hu}, \citenamefont {Chu}, \citenamefont {Bai}, \citenamefont {Han}, \citenamefont {Liu}, \citenamefont {Pan},\ and\ \citenamefont {Song}}]{zhouzhiyuan2025}%
  \BibitemOpen
  \bibfield  {author} {\bibinfo {author} {\bibfnamefont {Z.}~\bibnamefont {Zhou}}, \bibinfo {author} {\bibfnamefont {X.}~\bibnamefont {Cheng}}, \bibinfo {author} {\bibfnamefont {M.}~\bibnamefont {Hu}}, \bibinfo {author} {\bibfnamefont {R.}~\bibnamefont {Chu}}, \bibinfo {author} {\bibfnamefont {H.}~\bibnamefont {Bai}}, \bibinfo {author} {\bibfnamefont {L.}~\bibnamefont {Han}}, \bibinfo {author} {\bibfnamefont {J.}~\bibnamefont {Liu}}, \bibinfo {author} {\bibfnamefont {F.}~\bibnamefont {Pan}}, \ and\ \bibinfo {author} {\bibfnamefont {C.}~\bibnamefont {Song}},\ }\bibinfo {title} {Manipulation of the altermagnetic order in CrSb via crystal symmetry},\ \href {\doibase 10.1038/s41586-024-08436-3} {\bibfield  {journal} {\bibinfo  {journal} {Nature}\ }\textbf {\bibinfo {volume} {638}},\ \bibinfo {pages} {645} (\bibinfo {year} {2025})}\BibitemShut {NoStop}%
\bibitem [{\citenamefont {Jiang}\ \emph {et~al.}(2025)\citenamefont {Jiang}, \citenamefont {Hu}, \citenamefont {Bai}, \citenamefont {Song}, \citenamefont {Mu}, \citenamefont {Qu}, \citenamefont {Li}, \citenamefont {Zhu}, \citenamefont {Pi}, \citenamefont {Wei}, \citenamefont {Sun}, \citenamefont {Huang}, \citenamefont {Zheng}, \citenamefont {Peng}, \citenamefont {He}, \citenamefont {Li}, \citenamefont {Luo}, \citenamefont {Li}, \citenamefont {Chen}, \citenamefont {Li}, \citenamefont {Weng},\ and\ \citenamefont {Qian}}]{jiang_NP2025}%
  \BibitemOpen
  \bibfield  {author} {\bibinfo {author} {\bibfnamefont {B.}~\bibnamefont {Jiang}}, \bibinfo {author} {\bibfnamefont {M.}~\bibnamefont {Hu}}, \bibinfo {author} {\bibfnamefont {J.}~\bibnamefont {Bai}}, \bibinfo {author} {\bibfnamefont {Z.}~\bibnamefont {Song}}, \bibinfo {author} {\bibfnamefont {C.}~\bibnamefont {Mu}}, \bibinfo {author} {\bibfnamefont {G.}~\bibnamefont {Qu}}, \bibinfo {author} {\bibfnamefont {W.}~\bibnamefont {Li}}, \bibinfo {author} {\bibfnamefont {W.}~\bibnamefont {Zhu}}, \bibinfo {author} {\bibfnamefont {H.}~\bibnamefont {Pi}}, \bibinfo {author} {\bibfnamefont {Z.}~\bibnamefont {Wei}}, \bibinfo {author} {\bibfnamefont {Y.-J.}\ \bibnamefont {Sun}}, \bibinfo {author} {\bibfnamefont {Y.}~\bibnamefont {Huang}}, \bibinfo {author} {\bibfnamefont {X.}~\bibnamefont {Zheng}}, \bibinfo {author} {\bibfnamefont {Y.}~\bibnamefont {Peng}}, \bibinfo {author} {\bibfnamefont {L.}~\bibnamefont {He}}, \bibinfo {author} {\bibfnamefont {S.}~\bibnamefont {Li}}, \bibinfo {author} {\bibfnamefont {J.}~\bibnamefont {Luo}}, \bibinfo {author} {\bibfnamefont {Z.}~\bibnamefont {Li}}, \bibinfo {author} {\bibfnamefont {G.}~\bibnamefont {Chen}}, \bibinfo {author} {\bibfnamefont {H.}~\bibnamefont {Li}}, \bibinfo {author} {\bibfnamefont {H.}~\bibnamefont {Weng}}, \ and\ \bibinfo {author} {\bibfnamefont {T.}~\bibnamefont {Qian}},\ }\bibinfo {title} {A metallic room-temperature d-wave altermagnet},\ \href {https://www.nature.com/articles/s41567-025-02822-y} {\bibfield  {journal} {\bibinfo  {journal} {Nat. Phys.}\ }\textbf {\bibinfo {volume} {21}},\ \bibinfo {pages} {754–} (\bibinfo {year} {2025})}\BibitemShut {NoStop}%
\bibitem [{\citenamefont {Zhang}\ \emph {et~al.}(2025)\citenamefont {Zhang}, \citenamefont {Cheng}, \citenamefont {Yin}, \citenamefont {Liu}, \citenamefont {Deng}, \citenamefont {Qiao}, \citenamefont {Shi}, \citenamefont {Zhang}, \citenamefont {Lin}, \citenamefont {Liu}, \citenamefont {Ye}, \citenamefont {Huang}, \citenamefont {Meng}, \citenamefont {Zhang}, \citenamefont {Okuda}, \citenamefont {Shimada}, \citenamefont {Cui}, \citenamefont {Zhao}, \citenamefont {Cao}, \citenamefont {Qiao}, \citenamefont {Liu},\ and\ \citenamefont {Chen}}]{ZhangFY2024}%
  \BibitemOpen
  \bibfield  {author} {\bibinfo {author} {\bibfnamefont {F.}~\bibnamefont {Zhang}}, \bibinfo {author} {\bibfnamefont {X.}~\bibnamefont {Cheng}}, \bibinfo {author} {\bibfnamefont {Z.}~\bibnamefont {Yin}}, \bibinfo {author} {\bibfnamefont {C.}~\bibnamefont {Liu}}, \bibinfo {author} {\bibfnamefont {L.}~\bibnamefont {Deng}}, \bibinfo {author} {\bibfnamefont {Y.}~\bibnamefont {Qiao}}, \bibinfo {author} {\bibfnamefont {Z.}~\bibnamefont {Shi}}, \bibinfo {author} {\bibfnamefont {S.}~\bibnamefont {Zhang}}, \bibinfo {author} {\bibfnamefont {J.}~\bibnamefont {Lin}}, \bibinfo {author} {\bibfnamefont {Z.}~\bibnamefont {Liu}}, \bibinfo {author} {\bibfnamefont {M.}~\bibnamefont {Ye}}, \bibinfo {author} {\bibfnamefont {Y.}~\bibnamefont {Huang}}, \bibinfo {author} {\bibfnamefont {X.}~\bibnamefont {Meng}}, \bibinfo {author} {\bibfnamefont {C.}~\bibnamefont {Zhang}}, \bibinfo {author} {\bibfnamefont {T.}~\bibnamefont {Okuda}}, \bibinfo {author} {\bibfnamefont {K.}~\bibnamefont {Shimada}}, \bibinfo {author} {\bibfnamefont {S.}~\bibnamefont {Cui}}, \bibinfo {author} {\bibfnamefont {Y.}~\bibnamefont {Zhao}}, \bibinfo {author} {\bibfnamefont {G.-H.}\ \bibnamefont {Cao}}, \bibinfo {author} {\bibfnamefont {S.}~\bibnamefont {Qiao}}, \bibinfo {author} {\bibfnamefont {J.}~\bibnamefont {Liu}}, \ and\ \bibinfo {author} {\bibfnamefont {C.}~\bibnamefont {Chen}},\ }\bibinfo {title} {Crystal-symmetry-paired spin–valley locking in a layered room-temperature metallic altermagnet candidate},\ \href {https://doi.org/10.1038/s41567-025-02864-2} {\bibfield  {journal} {\bibinfo  {journal} {Nat. Phys.}\ }\textbf {\bibinfo {volume} {21}},\ \bibinfo {pages} {760} (\bibinfo {year} {2025})}\BibitemShut {NoStop}%
\bibitem [{\citenamefont {Bai}\ \emph {et~al.}(2022)\citenamefont {Bai}, \citenamefont {Han}, \citenamefont {Feng}, \citenamefont {Zhou}, \citenamefont {Su}, \citenamefont {Wang}, \citenamefont {Liao}, \citenamefont {Zhu}, \citenamefont {Chen}, \citenamefont {Pan}, \citenamefont {Fan},\ and\ \citenamefont {Song}}]{H-Bai2022}%
  \BibitemOpen
  \bibfield  {author} {\bibinfo {author} {\bibfnamefont {H.}~\bibnamefont {Bai}}, \bibinfo {author} {\bibfnamefont {L.}~\bibnamefont {Han}}, \bibinfo {author} {\bibfnamefont {X.~Y.}\ \bibnamefont {Feng}}, \bibinfo {author} {\bibfnamefont {Y.~J.}\ \bibnamefont {Zhou}}, \bibinfo {author} {\bibfnamefont {R.~X.}\ \bibnamefont {Su}}, \bibinfo {author} {\bibfnamefont {Q.}~\bibnamefont {Wang}}, \bibinfo {author} {\bibfnamefont {L.~Y.}\ \bibnamefont {Liao}}, \bibinfo {author} {\bibfnamefont {W.~X.}\ \bibnamefont {Zhu}}, \bibinfo {author} {\bibfnamefont {X.~Z.}\ \bibnamefont {Chen}}, \bibinfo {author} {\bibfnamefont {F.}~\bibnamefont {Pan}}, \bibinfo {author} {\bibfnamefont {X.~L.}\ \bibnamefont {Fan}}, \ and\ \bibinfo {author} {\bibfnamefont {C.}~\bibnamefont {Song}},\ }\bibinfo {title} {Observation of Spin Splitting Torque in a Collinear Antiferromagnet ${\mathrm{RuO}}_{2}$},\ \href {\doibase 10.1103/PhysRevLett.128.197202} {\bibfield  {journal} {\bibinfo  {journal} {Phys. Rev. Lett.}\ }\textbf {\bibinfo {volume} {128}},\ \bibinfo {pages} {197202} (\bibinfo {year} {2022})}\BibitemShut {NoStop}%
\bibitem [{\citenamefont {Karube}\ \emph {et~al.}(2022)\citenamefont {Karube}, \citenamefont {Tanaka}, \citenamefont {Sugawara}, \citenamefont {Kadoguchi}, \citenamefont {Kohda},\ and\ \citenamefont {Nitta}}]{Karube2022}%
  \BibitemOpen
  \bibfield  {author} {\bibinfo {author} {\bibfnamefont {S.}~\bibnamefont {Karube}}, \bibinfo {author} {\bibfnamefont {T.}~\bibnamefont {Tanaka}}, \bibinfo {author} {\bibfnamefont {D.}~\bibnamefont {Sugawara}}, \bibinfo {author} {\bibfnamefont {N.}~\bibnamefont {Kadoguchi}}, \bibinfo {author} {\bibfnamefont {M.}~\bibnamefont {Kohda}}, \ and\ \bibinfo {author} {\bibfnamefont {J.}~\bibnamefont {Nitta}},\ }\bibinfo {title} {Observation of Spin-Splitter Torque in Collinear Antiferromagnetic ${\mathrm{RuO}}_{2}$},\ \href {\doibase 10.1103/PhysRevLett.129.137201} {\bibfield  {journal} {\bibinfo  {journal} {Phys. Rev. Lett.}\ }\textbf {\bibinfo {volume} {129}},\ \bibinfo {pages} {137201} (\bibinfo {year} {2022})}\BibitemShut {NoStop}%
\bibitem [{\citenamefont {Lin}\ \emph {et~al.}(2024)\citenamefont {Lin}, \citenamefont {Chen}, \citenamefont {Lu}, \citenamefont {Liang}, \citenamefont {Feng}, \citenamefont {Yamagami}, \citenamefont {Osiecki}, \citenamefont {Leandersson}, \citenamefont {Thiagarajan}, \citenamefont {Liu}, \citenamefont {Felser},\ and\ \citenamefont {Ma}}]{LinZH2024}%
  \BibitemOpen
  \bibfield  {author} {\bibinfo {author} {\bibfnamefont {Z.}~\bibnamefont {Lin}}, \bibinfo {author} {\bibfnamefont {D.}~\bibnamefont {Chen}}, \bibinfo {author} {\bibfnamefont {W.}~\bibnamefont {Lu}}, \bibinfo {author} {\bibfnamefont {X.}~\bibnamefont {Liang}}, \bibinfo {author} {\bibfnamefont {S.}~\bibnamefont {Feng}}, \bibinfo {author} {\bibfnamefont {K.}~\bibnamefont {Yamagami}}, \bibinfo {author} {\bibfnamefont {J.}~\bibnamefont {Osiecki}}, \bibinfo {author} {\bibfnamefont {M.}~\bibnamefont {Leandersson}}, \bibinfo {author} {\bibfnamefont {B.}~\bibnamefont {Thiagarajan}}, \bibinfo {author} {\bibfnamefont {J.}~\bibnamefont {Liu}}, \bibinfo {author} {\bibfnamefont {C.}~\bibnamefont {Felser}}, \ and\ \bibinfo {author} {\bibfnamefont {J.}~\bibnamefont {Ma}},\ }\bibinfo {title} {Observation of Giant Spin Splitting and d-wave Spin Texture in Room Temperature Altermagnet RuO2},\ \href {https://arxiv.org/abs/2402.04995} {\bibfield  {journal} {\bibinfo  {journal} {arXiv}\ ,\ \bibinfo {pages} {2402.04995}} (\bibinfo {year} {2024})}\BibitemShut {NoStop}%
\bibitem [{\citenamefont {Fedchenko}\ \emph {et~al.}(2024)\citenamefont {Fedchenko}, \citenamefont {Min\'{a}r}, \citenamefont {Akashdeep}, \citenamefont {D’Souza}, \citenamefont {Vasilyev}, \citenamefont {Tkach}, \citenamefont {Odenbreit}, \citenamefont {Nguyen}, \citenamefont {Kutnyakhov}, \citenamefont {Wind}, \citenamefont {Wenthaus}, \citenamefont {Scholz}, \citenamefont {Rossnagel}, \citenamefont {Hoesch}, \citenamefont {Aeschlimann}, \citenamefont {Stadtmüller}, \citenamefont {Kläui}, \citenamefont {Schönhense}, \citenamefont {Jungwirth}, \citenamefont {Hellenes}, \citenamefont {Jakob}, \citenamefont {$\rm\check{S}$mejkal}, \citenamefont {Sinova},\ and\ \citenamefont {Elmers}}]{Fedchenko2024}%
  \BibitemOpen
  \bibfield  {author} {\bibinfo {author} {\bibfnamefont {O.}~\bibnamefont {Fedchenko}}, \bibinfo {author} {\bibfnamefont {J.}~\bibnamefont {Min\'{a}r}}, \bibinfo {author} {\bibfnamefont {A.}~\bibnamefont {Akashdeep}}, \bibinfo {author} {\bibfnamefont {S.~W.}\ \bibnamefont {D’Souza}}, \bibinfo {author} {\bibfnamefont {D.}~\bibnamefont {Vasilyev}}, \bibinfo {author} {\bibfnamefont {O.}~\bibnamefont {Tkach}}, \bibinfo {author} {\bibfnamefont {L.}~\bibnamefont {Odenbreit}}, \bibinfo {author} {\bibfnamefont {Q.}~\bibnamefont {Nguyen}}, \bibinfo {author} {\bibfnamefont {D.}~\bibnamefont {Kutnyakhov}}, \bibinfo {author} {\bibfnamefont {N.}~\bibnamefont {Wind}}, \bibinfo {author} {\bibfnamefont {L.}~\bibnamefont {Wenthaus}}, \bibinfo {author} {\bibfnamefont {M.}~\bibnamefont {Scholz}}, \bibinfo {author} {\bibfnamefont {K.}~\bibnamefont {Rossnagel}}, \bibinfo {author} {\bibfnamefont {M.}~\bibnamefont {Hoesch}}, \bibinfo {author} {\bibfnamefont {M.}~\bibnamefont {Aeschlimann}}, \bibinfo {author} {\bibfnamefont {B.}~\bibnamefont {Stadtmüller}}, \bibinfo {author} {\bibfnamefont {M.}~\bibnamefont {Kläui}}, \bibinfo {author} {\bibfnamefont {G.}~\bibnamefont {Schönhense}}, \bibinfo {author} {\bibfnamefont {T.}~\bibnamefont {Jungwirth}}, \bibinfo {author} {\bibfnamefont {A.~B.}\ \bibnamefont {Hellenes}}, \bibinfo {author} {\bibfnamefont {G.}~\bibnamefont {Jakob}}, \bibinfo {author} {\bibfnamefont {L.}~\bibnamefont {$\rm\check{S}$mejkal}}, \bibinfo {author} {\bibfnamefont {J.}~\bibnamefont {Sinova}}, \ and\ \bibinfo {author} {\bibfnamefont {H.-J.}\ \bibnamefont {Elmers}},\ }\bibinfo {title} {Observation of time-reversal symmetry breaking in the band structure of altermagnetic RuO$_2$},\ \href {\doibase 10.1126/sciadv.adj4883} {\bibfield  {journal} {\bibinfo  {journal} {Sci. Adv.}\ }\textbf {\bibinfo {volume} {10}},\ \bibinfo {pages} {eadj4883} (\bibinfo {year} {2024})}\BibitemShut {NoStop}%
\bibitem [{\citenamefont {Chen}\ \emph {et~al.}(2025)\citenamefont {Chen}, \citenamefont {Wang}, \citenamefont {Qin}, \citenamefont {Meng}, \citenamefont {Zhou}, \citenamefont {Wang}, \citenamefont {Liu}, \citenamefont {Zhao}, \citenamefont {Duan}, \citenamefont {Zhang}, \citenamefont {Liu}, \citenamefont {Shao}, \citenamefont {Jiang},\ and\ \citenamefont {Liu}}]{chen_AM2025}%
  \BibitemOpen
  \bibfield  {author} {\bibinfo {author} {\bibfnamefont {H.}~\bibnamefont {Chen}}, \bibinfo {author} {\bibfnamefont {Z.}~\bibnamefont {Wang}}, \bibinfo {author} {\bibfnamefont {P.}~\bibnamefont {Qin}}, \bibinfo {author} {\bibfnamefont {Z.}~\bibnamefont {Meng}}, \bibinfo {author} {\bibfnamefont {X.}~\bibnamefont {Zhou}}, \bibinfo {author} {\bibfnamefont {X.}~\bibnamefont {Wang}}, \bibinfo {author} {\bibfnamefont {L.}~\bibnamefont {Liu}}, \bibinfo {author} {\bibfnamefont {G.}~\bibnamefont {Zhao}}, \bibinfo {author} {\bibfnamefont {Z.}~\bibnamefont {Duan}}, \bibinfo {author} {\bibfnamefont {T.}~\bibnamefont {Zhang}}, \bibinfo {author} {\bibfnamefont {J.}~\bibnamefont {Liu}}, \bibinfo {author} {\bibfnamefont {D.}~\bibnamefont {Shao}}, \bibinfo {author} {\bibfnamefont {C.}~\bibnamefont {Jiang}}, \ and\ \bibinfo {author} {\bibfnamefont {Z.}~\bibnamefont {Liu}},\ }\bibinfo {title} {Spin‐{Splitting} {Magnetoresistance} in {Altermagnetic} {RuO}$_{\textrm{2}}$ {Thin} {Films}},\ \href {\doibase 10.1002/adma.202507764} {\bibfield  {journal} {\bibinfo  {journal} {Adv. Mater.}\ ,\ \bibinfo {pages} {2507764}} (\bibinfo {year} {2025})}\BibitemShut {NoStop}%
\bibitem [{\citenamefont {Jeong}\ \emph {et~al.}(2025{\natexlab{a}})\citenamefont {Jeong}, \citenamefont {Lee}, \citenamefont {Lin}, \citenamefont {Yang}, \citenamefont {Choi}, \citenamefont {Oh}, \citenamefont {Song}, \citenamefont {Lee}, \citenamefont {Nair}, \citenamefont {Choudhary} \emph {et~al.}}]{Jeong2025}%
  \BibitemOpen
  \bibfield  {author} {\bibinfo {author} {\bibfnamefont {S.~G.}\ \bibnamefont {Jeong}}, \bibinfo {author} {\bibfnamefont {S.}~\bibnamefont {Lee}}, \bibinfo {author} {\bibfnamefont {B.}~\bibnamefont {Lin}}, \bibinfo {author} {\bibfnamefont {Z.}~\bibnamefont {Yang}}, \bibinfo {author} {\bibfnamefont {I.~H.}\ \bibnamefont {Choi}}, \bibinfo {author} {\bibfnamefont {J.~Y.}\ \bibnamefont {Oh}}, \bibinfo {author} {\bibfnamefont {S.}~\bibnamefont {Song}}, \bibinfo {author} {\bibfnamefont {S.~w.}\ \bibnamefont {Lee}}, \bibinfo {author} {\bibfnamefont {S.}~\bibnamefont {Nair}}, \bibinfo {author} {\bibfnamefont {R.}~\bibnamefont {Choudhary}},  \emph {et~al.},\ }\bibinfo {title} {Metallicity and anomalous Hall effect in epitaxially strained, atomically thin RuO2 films},\ \href {\doibase 10.1073/pnas.2500831122} {\bibfield  {journal} {\bibinfo  {journal} {Proc. Natl. Acad. Sci. U.S.A.}\ }\textbf {\bibinfo {volume} {122}},\ \bibinfo {pages} {e2500831122} (\bibinfo {year} {2025}{\natexlab{a}})}\BibitemShut {NoStop}%
\bibitem [{\citenamefont {Berlijn}\ \emph {et~al.}(2017)\citenamefont {Berlijn}, \citenamefont {Snijders}, \citenamefont {Delaire}, \citenamefont {Zhou}, \citenamefont {Maier}, \citenamefont {Cao}, \citenamefont {Chi}, \citenamefont {Matsuda}, \citenamefont {Wang}, \citenamefont {Koehler}, \citenamefont {Kent},\ and\ \citenamefont {Weitering}}]{Berlijn2017}%
  \BibitemOpen
  \bibfield  {author} {\bibinfo {author} {\bibfnamefont {T.}~\bibnamefont {Berlijn}}, \bibinfo {author} {\bibfnamefont {P.~C.}\ \bibnamefont {Snijders}}, \bibinfo {author} {\bibfnamefont {O.}~\bibnamefont {Delaire}}, \bibinfo {author} {\bibfnamefont {H.-D.}\ \bibnamefont {Zhou}}, \bibinfo {author} {\bibfnamefont {T.~A.}\ \bibnamefont {Maier}}, \bibinfo {author} {\bibfnamefont {H.-B.}\ \bibnamefont {Cao}}, \bibinfo {author} {\bibfnamefont {S.-X.}\ \bibnamefont {Chi}}, \bibinfo {author} {\bibfnamefont {M.}~\bibnamefont {Matsuda}}, \bibinfo {author} {\bibfnamefont {Y.}~\bibnamefont {Wang}}, \bibinfo {author} {\bibfnamefont {M.~R.}\ \bibnamefont {Koehler}}, \bibinfo {author} {\bibfnamefont {P.~R.~C.}\ \bibnamefont {Kent}}, \ and\ \bibinfo {author} {\bibfnamefont {H.~H.}\ \bibnamefont {Weitering}},\ }\bibinfo {title} {Itinerant Antiferromagnetism in ${\mathrm{RuO}}_{2}$},\ \href {\doibase 10.1103/PhysRevLett.118.077201} {\bibfield  {journal} {\bibinfo  {journal} {Phys. Rev. Lett.}\ }\textbf {\bibinfo {volume} {118}},\ \bibinfo {pages} {077201} (\bibinfo {year} {2017})}\BibitemShut {NoStop}%
\bibitem [{\citenamefont {Zhu}\ \emph {et~al.}(2019)\citenamefont {Zhu}, \citenamefont {Strempfer}, \citenamefont {Rao}, \citenamefont {Occhialini}, \citenamefont {Pelliciari}, \citenamefont {Choi}, \citenamefont {Kawaguchi}, \citenamefont {You}, \citenamefont {Mitchell}, \citenamefont {Shao-Horn},\ and\ \citenamefont {Comin}}]{ZhuZH2019}%
  \BibitemOpen
  \bibfield  {author} {\bibinfo {author} {\bibfnamefont {Z.~H.}\ \bibnamefont {Zhu}}, \bibinfo {author} {\bibfnamefont {J.}~\bibnamefont {Strempfer}}, \bibinfo {author} {\bibfnamefont {R.~R.}\ \bibnamefont {Rao}}, \bibinfo {author} {\bibfnamefont {C.~A.}\ \bibnamefont {Occhialini}}, \bibinfo {author} {\bibfnamefont {J.}~\bibnamefont {Pelliciari}}, \bibinfo {author} {\bibfnamefont {Y.}~\bibnamefont {Choi}}, \bibinfo {author} {\bibfnamefont {T.}~\bibnamefont {Kawaguchi}}, \bibinfo {author} {\bibfnamefont {H.}~\bibnamefont {You}}, \bibinfo {author} {\bibfnamefont {J.~F.}\ \bibnamefont {Mitchell}}, \bibinfo {author} {\bibfnamefont {Y.}~\bibnamefont {Shao-Horn}}, \ and\ \bibinfo {author} {\bibfnamefont {R.}~\bibnamefont {Comin}},\ }\bibinfo {title} {Anomalous Antiferromagnetism in Metallic ${\mathrm{RuO}}_{2}$ Determined by Resonant X-ray Scattering},\ \href {\doibase 10.1103/PhysRevLett.122.017202} {\bibfield  {journal} {\bibinfo  {journal} {Phys. Rev. Lett.}\ }\textbf {\bibinfo {volume} {122}},\ \bibinfo {pages} {017202} (\bibinfo {year} {2019})}\BibitemShut {NoStop}%
\bibitem [{\citenamefont {Liu}\ \emph {et~al.}(2024)\citenamefont {Liu}, \citenamefont {Zhan}, \citenamefont {Li}, \citenamefont {Liu}, \citenamefont {Cheng}, \citenamefont {Shi}, \citenamefont {Deng}, \citenamefont {Zhang}, \citenamefont {Li}, \citenamefont {Ding}, \citenamefont {Jiang}, \citenamefont {Ye}, \citenamefont {Liu}, \citenamefont {Jiang}, \citenamefont {Wang}, \citenamefont {Li}, \citenamefont {Xie}, \citenamefont {Wang}, \citenamefont {Qiao}, \citenamefont {Wen}, \citenamefont {Sun},\ and\ \citenamefont {Shen}}]{LiuJY2024}%
  \BibitemOpen
  \bibfield  {author} {\bibinfo {author} {\bibfnamefont {J.}~\bibnamefont {Liu}}, \bibinfo {author} {\bibfnamefont {J.}~\bibnamefont {Zhan}}, \bibinfo {author} {\bibfnamefont {T.}~\bibnamefont {Li}}, \bibinfo {author} {\bibfnamefont {J.}~\bibnamefont {Liu}}, \bibinfo {author} {\bibfnamefont {S.}~\bibnamefont {Cheng}}, \bibinfo {author} {\bibfnamefont {Y.}~\bibnamefont {Shi}}, \bibinfo {author} {\bibfnamefont {L.}~\bibnamefont {Deng}}, \bibinfo {author} {\bibfnamefont {M.}~\bibnamefont {Zhang}}, \bibinfo {author} {\bibfnamefont {C.}~\bibnamefont {Li}}, \bibinfo {author} {\bibfnamefont {J.}~\bibnamefont {Ding}}, \bibinfo {author} {\bibfnamefont {Q.}~\bibnamefont {Jiang}}, \bibinfo {author} {\bibfnamefont {M.}~\bibnamefont {Ye}}, \bibinfo {author} {\bibfnamefont {Z.}~\bibnamefont {Liu}}, \bibinfo {author} {\bibfnamefont {Z.}~\bibnamefont {Jiang}}, \bibinfo {author} {\bibfnamefont {S.}~\bibnamefont {Wang}}, \bibinfo {author} {\bibfnamefont {Q.}~\bibnamefont {Li}}, \bibinfo {author} {\bibfnamefont {Y.}~\bibnamefont {Xie}}, \bibinfo {author} {\bibfnamefont {Y.}~\bibnamefont {Wang}}, \bibinfo {author} {\bibfnamefont {S.}~\bibnamefont {Qiao}}, \bibinfo {author} {\bibfnamefont {J.}~\bibnamefont {Wen}}, \bibinfo {author} {\bibfnamefont {Y.}~\bibnamefont {Sun}}, \ and\ \bibinfo {author} {\bibfnamefont {D.}~\bibnamefont {Shen}},\ }\bibinfo {title} {Absence of Altermagnetic Spin Splitting Character in Rutile Oxide ${\mathrm{RuO}}_{2}$},\ \href {\doibase 10.1103/PhysRevLett.133.176401} {\bibfield  {journal} {\bibinfo  {journal} {Phys. Rev. Lett.}\ }\textbf {\bibinfo {volume} {133}},\ \bibinfo {pages} {176401} (\bibinfo {year} {2024})}\BibitemShut {NoStop}%
\bibitem [{\citenamefont {Hiraishi}\ \emph {et~al.}(2024)\citenamefont {Hiraishi}, \citenamefont {Okabe}, \citenamefont {Koda}, \citenamefont {Kadono}, \citenamefont {Muroi}, \citenamefont {Hirai},\ and\ \citenamefont {Hiroi}}]{Hiraishi2024}%
  \BibitemOpen
  \bibfield  {author} {\bibinfo {author} {\bibfnamefont {M.}~\bibnamefont {Hiraishi}}, \bibinfo {author} {\bibfnamefont {H.}~\bibnamefont {Okabe}}, \bibinfo {author} {\bibfnamefont {A.}~\bibnamefont {Koda}}, \bibinfo {author} {\bibfnamefont {R.}~\bibnamefont {Kadono}}, \bibinfo {author} {\bibfnamefont {T.}~\bibnamefont {Muroi}}, \bibinfo {author} {\bibfnamefont {D.}~\bibnamefont {Hirai}}, \ and\ \bibinfo {author} {\bibfnamefont {Z.}~\bibnamefont {Hiroi}},\ }\bibinfo {title} {Nonmagnetic Ground State in ${\mathrm{RuO}}_{2}$ Revealed by Muon Spin Rotation},\ \href {\doibase 10.1103/PhysRevLett.132.166702} {\bibfield  {journal} {\bibinfo  {journal} {Phys. Rev. Lett.}\ }\textbf {\bibinfo {volume} {132}},\ \bibinfo {pages} {166702} (\bibinfo {year} {2024})}\BibitemShut {NoStop}%
\bibitem [{\citenamefont {Zhang}\ \emph {et~al.}(2024{\natexlab{b}})\citenamefont {Zhang}, \citenamefont {Bai}, \citenamefont {Chen}, \citenamefont {Han}, \citenamefont {Liang}, \citenamefont {Chu}, \citenamefont {Dai}, \citenamefont {Pan},\ and\ \citenamefont {Song}}]{ZhangYC2024}%
  \BibitemOpen
  \bibfield  {author} {\bibinfo {author} {\bibfnamefont {Y.-C.}\ \bibnamefont {Zhang}}, \bibinfo {author} {\bibfnamefont {H.}~\bibnamefont {Bai}}, \bibinfo {author} {\bibfnamefont {C.}~\bibnamefont {Chen}}, \bibinfo {author} {\bibfnamefont {L.}~\bibnamefont {Han}}, \bibinfo {author} {\bibfnamefont {S.-X.}\ \bibnamefont {Liang}}, \bibinfo {author} {\bibfnamefont {R.-Y.}\ \bibnamefont {Chu}}, \bibinfo {author} {\bibfnamefont {J.-K.}\ \bibnamefont {Dai}}, \bibinfo {author} {\bibfnamefont {F.}~\bibnamefont {Pan}}, \ and\ \bibinfo {author} {\bibfnamefont {C.}~\bibnamefont {Song}},\ }\bibinfo {title} {Probing the N\'eel order in altermagnetic RuO$_2$ films by X-ray magnetic linear dichroism},\ \href {https://arxiv.org/abs/2412.17016} {\bibfield  {journal} {\bibinfo  {journal} {arXiv}\ ,\ \bibinfo {pages} {2412.17016}} (\bibinfo {year} {2024}{\natexlab{b}})}\BibitemShut {NoStop}%
\bibitem [{\citenamefont {Noh}\ \emph {et~al.}(2025)\citenamefont {Noh}, \citenamefont {Kim}, \citenamefont {Lee}, \citenamefont {Jung}, \citenamefont {Seo}, \citenamefont {So}, \citenamefont {Lee}, \citenamefont {Lee}, \citenamefont {Park}, \citenamefont {Yang}, \citenamefont {Oh}, \citenamefont {Jin}, \citenamefont {Sohn},\ and\ \citenamefont {Yoo}}]{Noh2025}%
  \BibitemOpen
  \bibfield  {author} {\bibinfo {author} {\bibfnamefont {S.}~\bibnamefont {Noh}}, \bibinfo {author} {\bibfnamefont {G.-H.}\ \bibnamefont {Kim}}, \bibinfo {author} {\bibfnamefont {J.}~\bibnamefont {Lee}}, \bibinfo {author} {\bibfnamefont {H.}~\bibnamefont {Jung}}, \bibinfo {author} {\bibfnamefont {U.}~\bibnamefont {Seo}}, \bibinfo {author} {\bibfnamefont {G.}~\bibnamefont {So}}, \bibinfo {author} {\bibfnamefont {J.}~\bibnamefont {Lee}}, \bibinfo {author} {\bibfnamefont {S.}~\bibnamefont {Lee}}, \bibinfo {author} {\bibfnamefont {M.}~\bibnamefont {Park}}, \bibinfo {author} {\bibfnamefont {S.}~\bibnamefont {Yang}}, \bibinfo {author} {\bibfnamefont {Y.~S.}\ \bibnamefont {Oh}}, \bibinfo {author} {\bibfnamefont {H.}~\bibnamefont {Jin}}, \bibinfo {author} {\bibfnamefont {C.}~\bibnamefont {Sohn}}, \ and\ \bibinfo {author} {\bibfnamefont {J.-W.}\ \bibnamefont {Yoo}},\ }\bibinfo {title} {Tunneling Magnetoresistance in Altermagnetic ${\mathrm{RuO}}_{2}$-Based Magnetic Tunnel Junctions},\ \href {\doibase 10.1103/nrk5-5zrj} {\bibfield  {journal} {\bibinfo  {journal} {Phys. Rev. Lett.}\ }\textbf {\bibinfo {volume} {134}},\ \bibinfo {pages} {246703} (\bibinfo {year} {2025})}\BibitemShut {NoStop}%
\bibitem [{\citenamefont {Cui}\ \emph {et~al.}(2023)\citenamefont {Cui}, \citenamefont {Zhu}, \citenamefont {Yao}, \citenamefont {Cui},\ and\ \citenamefont {Yang}}]{QR-Cui2023}%
  \BibitemOpen
  \bibfield  {author} {\bibinfo {author} {\bibfnamefont {Q.}~\bibnamefont {Cui}}, \bibinfo {author} {\bibfnamefont {Y.}~\bibnamefont {Zhu}}, \bibinfo {author} {\bibfnamefont {X.}~\bibnamefont {Yao}}, \bibinfo {author} {\bibfnamefont {P.}~\bibnamefont {Cui}}, \ and\ \bibinfo {author} {\bibfnamefont {H.}~\bibnamefont {Yang}},\ }\bibinfo {title} {Giant spin-Hall and tunneling magnetoresistance effects based on a two-dimensional nonrelativistic antiferromagnetic metal},\ \href {\doibase 10.1103/PhysRevB.108.024410} {\bibfield  {journal} {\bibinfo  {journal} {Phys. Rev. B}\ }\textbf {\bibinfo {volume} {108}},\ \bibinfo {pages} {024410} (\bibinfo {year} {2023})}\BibitemShut {NoStop}%
\bibitem [{\citenamefont {Zhu}\ \emph {et~al.}(2024)\citenamefont {Zhu}, \citenamefont {Chen}, \citenamefont {Li}, \citenamefont {Qiao}, \citenamefont {Ma}, \citenamefont {Liu}, \citenamefont {Hu}, \citenamefont {Gao},\ and\ \citenamefont {Ren}}]{Zhuyu2024}%
  \BibitemOpen
  \bibfield  {author} {\bibinfo {author} {\bibfnamefont {Y.}~\bibnamefont {Zhu}}, \bibinfo {author} {\bibfnamefont {T.}~\bibnamefont {Chen}}, \bibinfo {author} {\bibfnamefont {Y.}~\bibnamefont {Li}}, \bibinfo {author} {\bibfnamefont {L.}~\bibnamefont {Qiao}}, \bibinfo {author} {\bibfnamefont {X.}~\bibnamefont {Ma}}, \bibinfo {author} {\bibfnamefont {C.}~\bibnamefont {Liu}}, \bibinfo {author} {\bibfnamefont {T.}~\bibnamefont {Hu}}, \bibinfo {author} {\bibfnamefont {H.}~\bibnamefont {Gao}}, \ and\ \bibinfo {author} {\bibfnamefont {W.}~\bibnamefont {Ren}},\ }\bibinfo {title} {Multipiezo Effect in Altermagnetic V$_2$SeTeO Monolayer},\ \href {\doibase 10.1021/acs.nanolett.3c04330} {\bibfield  {journal} {\bibinfo  {journal} {Nano Lett.}\ }\textbf {\bibinfo {volume} {24}},\ \bibinfo {pages} {472} (\bibinfo {year} {2024})}\BibitemShut {NoStop}%
\bibitem [{\citenamefont {Samanta}\ \emph {et~al.}(2020)\citenamefont {Samanta}, \citenamefont {Le{\v{z}}ai{\'c}}, \citenamefont {Merte}, \citenamefont {Freimuth}, \citenamefont {Bl{\"u}gel},\ and\ \citenamefont {Mokrousov}}]{Samanta2020}%
  \BibitemOpen
  \bibfield  {author} {\bibinfo {author} {\bibfnamefont {K.}~\bibnamefont {Samanta}}, \bibinfo {author} {\bibfnamefont {M.}~\bibnamefont {Le{\v{z}}ai{\'c}}}, \bibinfo {author} {\bibfnamefont {M.}~\bibnamefont {Merte}}, \bibinfo {author} {\bibfnamefont {F.}~\bibnamefont {Freimuth}}, \bibinfo {author} {\bibfnamefont {S.}~\bibnamefont {Bl{\"u}gel}}, \ and\ \bibinfo {author} {\bibfnamefont {Y.}~\bibnamefont {Mokrousov}},\ }\bibinfo {title} {Crystal Hall and crystal magneto-optical effect in thin films of SrRuO$_3$},\ \href {https://doi.org/10.1063/5.0005017} {\bibfield  {journal} {\bibinfo  {journal} {J. Appl. Phys.}\ }\textbf {\bibinfo {volume} {127}} (\bibinfo {year} {2020})}\BibitemShut {NoStop}%
\bibitem [{\citenamefont {Mazin}\ \emph {et~al.}(2021)\citenamefont {Mazin}, \citenamefont {Koepernik}, \citenamefont {Johannes}, \citenamefont {Gonz\'{a}lez-Hern\'{a}ndez},\ and\ \citenamefont {$\rm\check{S}$mejkal}}]{Mazin2021}%
  \BibitemOpen
  \bibfield  {author} {\bibinfo {author} {\bibfnamefont {I.~I.}\ \bibnamefont {Mazin}}, \bibinfo {author} {\bibfnamefont {K.}~\bibnamefont {Koepernik}}, \bibinfo {author} {\bibfnamefont {M.~D.}\ \bibnamefont {Johannes}}, \bibinfo {author} {\bibfnamefont {R.}~\bibnamefont {Gonz\'{a}lez-Hern\'{a}ndez}}, \ and\ \bibinfo {author} {\bibfnamefont {L.}~\bibnamefont {$\rm\check{S}$mejkal}},\ }\bibinfo {title} {Prediction of unconventional magnetism in doped FeSb$_2$},\ \href {\doibase 10.1073/pnas.2108924118} {\bibfield  {journal} {\bibinfo  {journal} {Proc. Natl. Acad. Sci. U.S.A.}\ }\textbf {\bibinfo {volume} {118}},\ \bibinfo {pages} {e2108924118} (\bibinfo {year} {2021})}\BibitemShut {NoStop}%
\bibitem [{\citenamefont {Shao}\ \emph {et~al.}(2021)\citenamefont {Shao}, \citenamefont {Ding}, \citenamefont {Gurung}, \citenamefont {Zhang},\ and\ \citenamefont {Tsymbal}}]{ShaoDF2021}%
  \BibitemOpen
  \bibfield  {author} {\bibinfo {author} {\bibfnamefont {D.-F.}\ \bibnamefont {Shao}}, \bibinfo {author} {\bibfnamefont {J.}~\bibnamefont {Ding}}, \bibinfo {author} {\bibfnamefont {G.}~\bibnamefont {Gurung}}, \bibinfo {author} {\bibfnamefont {S.-H.}\ \bibnamefont {Zhang}}, \ and\ \bibinfo {author} {\bibfnamefont {E.~Y.}\ \bibnamefont {Tsymbal}},\ }\bibinfo {title} {Interfacial Crystal Hall Effect Reversible by Ferroelectric Polarization},\ \href {https://link.aps.org/doi/10.1103/PhysRevApplied.15.024057} {\bibfield  {journal} {\bibinfo  {journal} {Phys. Rev. Appl.}\ }\textbf {\bibinfo {volume} {15}},\ \bibinfo {pages} {024057} (\bibinfo {year} {2021})}\BibitemShut {NoStop}%
\bibitem [{\citenamefont {\ifmmode~\check{S}\else \v{S}\fi{}mejkal}\ \emph {et~al.}(2022{\natexlab{c}})\citenamefont {\ifmmode~\check{S}\else \v{S}\fi{}mejkal}, \citenamefont {MacDonald}, \citenamefont {Sinova}, \citenamefont {Nakatsuji},\ and\ \citenamefont {Jungwirth}}]{Libor_NRM2022}%
  \BibitemOpen
  \bibfield  {author} {\bibinfo {author} {\bibfnamefont {L.}~\bibnamefont {\ifmmode~\check{S}\else \v{S}\fi{}mejkal}}, \bibinfo {author} {\bibfnamefont {A.~H.}\ \bibnamefont {MacDonald}}, \bibinfo {author} {\bibfnamefont {J.}~\bibnamefont {Sinova}}, \bibinfo {author} {\bibfnamefont {S.}~\bibnamefont {Nakatsuji}}, \ and\ \bibinfo {author} {\bibfnamefont {T.}~\bibnamefont {Jungwirth}},\ }\bibinfo {title} {Anomalous {Hall} antiferromagnets},\ \href {\doibase 10.1038/s41578-022-00430-3} {\bibfield  {journal} {\bibinfo  {journal} {Nat. Rev. Mater.}\ }\textbf {\bibinfo {volume} {7}},\ \bibinfo {pages} {482} (\bibinfo {year} {2022}{\natexlab{c}})}\BibitemShut {NoStop}%
\bibitem [{\citenamefont {Gonzalez~Betancourt}\ \emph {et~al.}(2023)\citenamefont {Gonzalez~Betancourt}, \citenamefont {Zub\'a\ifmmode~\check{c}\else \v{c}\fi{}}, \citenamefont {Gonzalez-Hernandez}, \citenamefont {Geishendorf}, \citenamefont {\ifmmode \check{S}\else \v{S}\fi{}ob\'a\ifmmode~\check{n}\else \v{n}\fi{}}, \citenamefont {Springholz}, \citenamefont {Olejn\'{\i}k}, \citenamefont {\ifmmode~\check{S}\else \v{S}\fi{}mejkal}, \citenamefont {Sinova}, \citenamefont {Jungwirth}, \citenamefont {Goennenwein}, \citenamefont {Thomas}, \citenamefont {Reichlov\'a}, \citenamefont {\ifmmode~\check{Z}\else \v{Z}\fi{}elezn\'y},\ and\ \citenamefont {Kriegner}}]{Gonzalez_CrSb_PRL2023}%
  \BibitemOpen
  \bibfield  {author} {\bibinfo {author} {\bibfnamefont {R.~D.}\ \bibnamefont {Gonzalez~Betancourt}}, \bibinfo {author} {\bibfnamefont {J.}~\bibnamefont {Zub\'a\ifmmode~\check{c}\else \v{c}\fi{}}}, \bibinfo {author} {\bibfnamefont {R.}~\bibnamefont {Gonzalez-Hernandez}}, \bibinfo {author} {\bibfnamefont {K.}~\bibnamefont {Geishendorf}}, \bibinfo {author} {\bibfnamefont {Z.}~\bibnamefont {\ifmmode \check{S}\else \v{S}\fi{}ob\'a\ifmmode~\check{n}\else \v{n}\fi{}}}, \bibinfo {author} {\bibfnamefont {G.}~\bibnamefont {Springholz}}, \bibinfo {author} {\bibfnamefont {K.}~\bibnamefont {Olejn\'{\i}k}}, \bibinfo {author} {\bibfnamefont {L.}~\bibnamefont {\ifmmode~\check{S}\else \v{S}\fi{}mejkal}}, \bibinfo {author} {\bibfnamefont {J.}~\bibnamefont {Sinova}}, \bibinfo {author} {\bibfnamefont {T.}~\bibnamefont {Jungwirth}}, \bibinfo {author} {\bibfnamefont {S.~T.~B.}\ \bibnamefont {Goennenwein}}, \bibinfo {author} {\bibfnamefont {A.}~\bibnamefont {Thomas}}, \bibinfo {author} {\bibfnamefont {H.}~\bibnamefont {Reichlov\'a}}, \bibinfo {author} {\bibfnamefont {J.}~\bibnamefont {\ifmmode~\check{Z}\else \v{Z}\fi{}elezn\'y}}, \ and\ \bibinfo {author} {\bibfnamefont {D.}~\bibnamefont {Kriegner}},\ }\bibinfo {title} {Spontaneous {Anomalous} {Hall} {Effect} {Arising} from an {Unconventional} {Compensated} {Magnetic} {Phase} in a {Semiconductor}},\ \href {https://link.aps.org/doi/10.1103/PhysRevLett.130.036702} {\bibfield  {journal} {\bibinfo  {journal} {Phys. Rev. Lett.}\ }\textbf {\bibinfo {volume} {130}},\ \bibinfo {pages} {036702} (\bibinfo {year} {2023})}\BibitemShut {NoStop}%
\bibitem [{\citenamefont {Takagi}\ \emph {et~al.}(2025)\citenamefont {Takagi}, \citenamefont {Hirakida}, \citenamefont {Settai}, \citenamefont {Oiwa}, \citenamefont {Takagi}, \citenamefont {Kitaori}, \citenamefont {Yamauchi}, \citenamefont {Inoue}, \citenamefont {Yamaura}, \citenamefont {Nishio-Hamane}, \citenamefont {Itoh}, \citenamefont {Aji}, \citenamefont {Saito}, \citenamefont {Nakajima}, \citenamefont {Nomoto}, \citenamefont {Arita},\ and\ \citenamefont {Seki}}]{Takagi_FeS2_NM2025}%
  \BibitemOpen
  \bibfield  {author} {\bibinfo {author} {\bibfnamefont {R.}~\bibnamefont {Takagi}}, \bibinfo {author} {\bibfnamefont {R.}~\bibnamefont {Hirakida}}, \bibinfo {author} {\bibfnamefont {Y.}~\bibnamefont {Settai}}, \bibinfo {author} {\bibfnamefont {R.}~\bibnamefont {Oiwa}}, \bibinfo {author} {\bibfnamefont {H.}~\bibnamefont {Takagi}}, \bibinfo {author} {\bibfnamefont {A.}~\bibnamefont {Kitaori}}, \bibinfo {author} {\bibfnamefont {K.}~\bibnamefont {Yamauchi}}, \bibinfo {author} {\bibfnamefont {H.}~\bibnamefont {Inoue}}, \bibinfo {author} {\bibfnamefont {J.-i.}\ \bibnamefont {Yamaura}}, \bibinfo {author} {\bibfnamefont {D.}~\bibnamefont {Nishio-Hamane}}, \bibinfo {author} {\bibfnamefont {S.}~\bibnamefont {Itoh}}, \bibinfo {author} {\bibfnamefont {S.}~\bibnamefont {Aji}}, \bibinfo {author} {\bibfnamefont {H.}~\bibnamefont {Saito}}, \bibinfo {author} {\bibfnamefont {T.}~\bibnamefont {Nakajima}}, \bibinfo {author} {\bibfnamefont {T.}~\bibnamefont {Nomoto}}, \bibinfo {author} {\bibfnamefont {R.}~\bibnamefont {Arita}}, \ and\ \bibinfo {author} {\bibfnamefont {S.}~\bibnamefont {Seki}},\ }\bibinfo {title} {Spontaneous {Hall} effect induced by collinear antiferromagnetic order at room temperature},\ \href {https://www.nature.com/articles/s41563-024-02058-w} {\bibfield  {journal} {\bibinfo  {journal} {Nat. Mater.}\ }\textbf {\bibinfo {volume} {24}},\ \bibinfo {pages} {63} (\bibinfo {year} {2025})}\BibitemShut {NoStop}%
\bibitem [{\citenamefont {Kawamura}\ \emph {et~al.}(2024)\citenamefont {Kawamura}, \citenamefont {Yoshimi}, \citenamefont {Hashimoto}, \citenamefont {Kobayashi},\ and\ \citenamefont {Misawa}}]{Kawamura2024}%
  \BibitemOpen
  \bibfield  {author} {\bibinfo {author} {\bibfnamefont {T.}~\bibnamefont {Kawamura}}, \bibinfo {author} {\bibfnamefont {K.}~\bibnamefont {Yoshimi}}, \bibinfo {author} {\bibfnamefont {K.}~\bibnamefont {Hashimoto}}, \bibinfo {author} {\bibfnamefont {A.}~\bibnamefont {Kobayashi}}, \ and\ \bibinfo {author} {\bibfnamefont {T.}~\bibnamefont {Misawa}},\ }\bibinfo {title} {Compensated Ferrimagnets with Colossal Spin Splitting in Organic Compounds},\ \href {\doibase 10.1103/PhysRevLett.132.156502} {\bibfield  {journal} {\bibinfo  {journal} {Phys. Rev. Lett.}\ }\textbf {\bibinfo {volume} {132}},\ \bibinfo {pages} {156502} (\bibinfo {year} {2024})}\BibitemShut {NoStop}%
\bibitem [{\citenamefont {Yuan}\ \emph {et~al.}(2024)\citenamefont {Yuan}, \citenamefont {Georgescu},\ and\ \citenamefont {Rondinelli}}]{LD-Yuan2024}%
  \BibitemOpen
  \bibfield  {author} {\bibinfo {author} {\bibfnamefont {L.-D.}\ \bibnamefont {Yuan}}, \bibinfo {author} {\bibfnamefont {A.~B.}\ \bibnamefont {Georgescu}}, \ and\ \bibinfo {author} {\bibfnamefont {J.~M.}\ \bibnamefont {Rondinelli}},\ }\bibinfo {title} {Nonrelativistic Spin Splitting at the Brillouin Zone Center in Compensated Magnets},\ \href {\doibase 10.1103/PhysRevLett.133.216701} {\bibfield  {journal} {\bibinfo  {journal} {Phys. Rev. Lett.}\ }\textbf {\bibinfo {volume} {133}},\ \bibinfo {pages} {216701} (\bibinfo {year} {2024})}\BibitemShut {NoStop}%
\bibitem [{\citenamefont {Liu}\ \emph {et~al.}(2025)\citenamefont {Liu}, \citenamefont {Guo}, \citenamefont {Li},\ and\ \citenamefont {Liu}}]{YC-Liu2025}%
  \BibitemOpen
  \bibfield  {author} {\bibinfo {author} {\bibfnamefont {Y.}~\bibnamefont {Liu}}, \bibinfo {author} {\bibfnamefont {S.-D.}\ \bibnamefont {Guo}}, \bibinfo {author} {\bibfnamefont {Y.}~\bibnamefont {Li}}, \ and\ \bibinfo {author} {\bibfnamefont {C.-C.}\ \bibnamefont {Liu}},\ }\bibinfo {title} {Two-Dimensional Fully Compensated Ferrimagnetism},\ \href {\doibase 10.1103/PhysRevLett.134.116703} {\bibfield  {journal} {\bibinfo  {journal} {Phys. Rev. Lett.}\ }\textbf {\bibinfo {volume} {134}},\ \bibinfo {pages} {116703} (\bibinfo {year} {2025})}\BibitemShut {NoStop}%
\bibitem [{\citenamefont {Feng}\ \emph {et~al.}(2025{\natexlab{a}})\citenamefont {Feng}, \citenamefont {Zhou}, \citenamefont {Chen}, \citenamefont {Xu}, \citenamefont {Yang},\ and\ \citenamefont {Li}}]{JQ-Feng2025}%
  \BibitemOpen
  \bibfield  {author} {\bibinfo {author} {\bibfnamefont {J.}~\bibnamefont {Feng}}, \bibinfo {author} {\bibfnamefont {X.}~\bibnamefont {Zhou}}, \bibinfo {author} {\bibfnamefont {J.}~\bibnamefont {Chen}}, \bibinfo {author} {\bibfnamefont {M.}~\bibnamefont {Xu}}, \bibinfo {author} {\bibfnamefont {X.}~\bibnamefont {Yang}}, \ and\ \bibinfo {author} {\bibfnamefont {Y.}~\bibnamefont {Li}},\ }\bibinfo {title} {Ferroelectric antiferromagnetic lifting of spin-valley degeneracy},\ \href {\doibase 10.1103/jyq4-d4gs} {\bibfield  {journal} {\bibinfo  {journal} {Phys. Rev. B}\ }\textbf {\bibinfo {volume} {111}},\ \bibinfo {pages} {214446} (\bibinfo {year} {2025}{\natexlab{a}})}\BibitemShut {NoStop}%
\bibitem [{Sup()}]{SuppMater}%
  \BibitemOpen
  \href@noop {} {\bibinfo  {journal} {See Supplemental Material at http://link.aps.org/xxx, which contains: (a-c) the details of spin group of XV$_2$Y$_2$O, RuO$_2$ and MnF$_2$, CrSb and MnTe, without perturbation, (d) the detail of theory and computational details, (e-g) band structures and transverse transport phenomena without perturbation, with shear strain, and with spin canting, including spin group and magnetic group analysis, as well as exchange parameter. And it includes Refs.~{\normalsize\cite{phonopy-phono3py-JPCM,phonopy-phono3py-JPSJ,TB2J,FengJQ2025,ZhangRW2024,Mazurenko2005,Logemann2017,Morrish1995,Machala2011,Abrajevitch2021,Masina2024,Zysler2001,cornell2003iron,Kubaniova2019,Takagi_FeS2_NM2025,Libor2020,Kresse1993,Kresse1996,Perdew1996,Arash2008,wannierberri,Anisimov_LDAU1991,Dudarev_LDAU1998}}}\ }\BibitemShut {NoStop}%
\bibitem [{\citenamefont {Ashcroft}\ and\ \citenamefont {Mermin}(1976)}]{ashcroft1976solid}%
  \BibitemOpen
\bibfield  {journal} {  }\bibfield  {author} {\bibinfo {author} {\bibfnamefont {N.~W.}\ \bibnamefont {Ashcroft}}\ and\ \bibinfo {author} {\bibfnamefont {N.~D.}\ \bibnamefont {Mermin}},\ }\href@noop {} {\bibinfo {title} {Solid state physics}}\ (\bibinfo  {publisher} {SaundersCollege Publishing, Philadelphia},\ \bibinfo {year} {1976})\BibitemShut {NoStop}%
\bibitem [{\citenamefont {van Houten}\ \emph {et~al.}(1992)\citenamefont {van Houten}, \citenamefont {Molenkamp}, \citenamefont {Beenakker},\ and\ \citenamefont {Foxon}}]{Houten1992}%
  \BibitemOpen
  \bibfield  {author} {\bibinfo {author} {\bibfnamefont {H.}~\bibnamefont {van Houten}}, \bibinfo {author} {\bibfnamefont {L.~W.}\ \bibnamefont {Molenkamp}}, \bibinfo {author} {\bibfnamefont {C.~W.~J.}\ \bibnamefont {Beenakker}}, \ and\ \bibinfo {author} {\bibfnamefont {C.~T.}\ \bibnamefont {Foxon}},\ }\bibinfo {title} {Thermo-electric properties of quantum point contacts},\ \href {\doibase 10.1088/0268-1242/7/3b/052} {\bibfield  {journal} {\bibinfo  {journal} {Semicond. Sci. Technol.}\ }\textbf {\bibinfo {volume} {7}},\ \bibinfo {pages} {B215} (\bibinfo {year} {1992})}\BibitemShut {NoStop}%
\bibitem [{\citenamefont {Behnia}(2015)}]{behnia2015fundamentals}%
  \BibitemOpen
  \bibfield  {author} {\bibinfo {author} {\bibfnamefont {K.}~\bibnamefont {Behnia}},\ }\href@noop {} {\bibinfo {title} {Fundamentals of thermoelectricity}}\ (\bibinfo  {publisher} {Oxford University Press},\ \bibinfo {year} {2015})\BibitemShut {NoStop}%
\bibitem [{\citenamefont {Lei}\ \emph {et~al.}(2025)\citenamefont {Lei}, \citenamefont {Li}, \citenamefont {Duan}, \citenamefont {Long}, \citenamefont {Wang},\ and\ \citenamefont {Ouyang}}]{strainLeiBC2025}%
  \BibitemOpen
  \bibfield  {author} {\bibinfo {author} {\bibfnamefont {B.}~\bibnamefont {Lei}}, \bibinfo {author} {\bibfnamefont {A.}~\bibnamefont {Li}}, \bibinfo {author} {\bibfnamefont {H.}~\bibnamefont {Duan}}, \bibinfo {author} {\bibfnamefont {M.}~\bibnamefont {Long}}, \bibinfo {author} {\bibfnamefont {Y.}~\bibnamefont {Wang}}, \ and\ \bibinfo {author} {\bibfnamefont {F.}~\bibnamefont {Ouyang}},\ }\bibinfo {title} {Shear-strain-induced switchable spin splitting and piezomagnetic properties in altermagnetic materials},\ \href {\doibase 10.15302/frontphys.2025.064204} {\bibfield  {journal} {\bibinfo  {journal} {Front. Phys.}\ }\textbf {\bibinfo {volume} {20}},\ \bibinfo {pages} {64204} (\bibinfo {year} {2025})}\BibitemShut {NoStop}%
\bibitem [{\citenamefont {Jeong}\ \emph {et~al.}(2025{\natexlab{b}})\citenamefont {Jeong}, \citenamefont {Lee}, \citenamefont {Lin}, \citenamefont {Yang}, \citenamefont {Choi}, \citenamefont {Oh}, \citenamefont {Song}, \citenamefont {Lee}, \citenamefont {Nair}, \citenamefont {Choudhary}, \citenamefont {Parikh}, \citenamefont {Park}, \citenamefont {Choi}, \citenamefont {Lee}, \citenamefont {LeBeau}, \citenamefont {Low},\ and\ \citenamefont {Jalan}}]{strainGyo2025}%
  \BibitemOpen
  \bibfield  {author} {\bibinfo {author} {\bibfnamefont {S.~G.}\ \bibnamefont {Jeong}}, \bibinfo {author} {\bibfnamefont {S.}~\bibnamefont {Lee}}, \bibinfo {author} {\bibfnamefont {B.}~\bibnamefont {Lin}}, \bibinfo {author} {\bibfnamefont {Z.}~\bibnamefont {Yang}}, \bibinfo {author} {\bibfnamefont {I.~H.}\ \bibnamefont {Choi}}, \bibinfo {author} {\bibfnamefont {J.~Y.}\ \bibnamefont {Oh}}, \bibinfo {author} {\bibfnamefont {S.}~\bibnamefont {Song}}, \bibinfo {author} {\bibfnamefont {S.~W.}\ \bibnamefont {Lee}}, \bibinfo {author} {\bibfnamefont {S.}~\bibnamefont {Nair}}, \bibinfo {author} {\bibfnamefont {R.}~\bibnamefont {Choudhary}}, \bibinfo {author} {\bibfnamefont {J.}~\bibnamefont {Parikh}}, \bibinfo {author} {\bibfnamefont {S.}~\bibnamefont {Park}}, \bibinfo {author} {\bibfnamefont {W.~S.}\ \bibnamefont {Choi}}, \bibinfo {author} {\bibfnamefont {J.~S.}\ \bibnamefont {Lee}}, \bibinfo {author} {\bibfnamefont {J.~M.}\ \bibnamefont {LeBeau}}, \bibinfo {author} {\bibfnamefont {T.}~\bibnamefont {Low}}, \ and\ \bibinfo {author} {\bibfnamefont {B.}~\bibnamefont {Jalan}},\ }\bibinfo {title} {Metallicity and anomalous {Hall} effect in epitaxially strained, atomically thin {RuO}$_{\textrm{2}}$ films},\ \href {https://pnas.org/doi/10.1073/pnas.2500831122} {\bibfield  {journal} {\bibinfo  {journal} {Proc. Natl. Acad. Sci. U.S.A.}\ }\textbf {\bibinfo {volume} {122}} (\bibinfo {year} {2025}{\natexlab{b}})}\BibitemShut {NoStop}%
\bibitem [{\citenamefont {Suzuki}\ \emph {et~al.}(2016)\citenamefont {Suzuki}, \citenamefont {Chisnell}, \citenamefont {Devarakonda}, \citenamefont {Liu}, \citenamefont {Feng}, \citenamefont {Xiao}, \citenamefont {Lynn},\ and\ \citenamefont {Checkelsky}}]{cant_suzuki_NP2016}%
  \BibitemOpen
  \bibfield  {author} {\bibinfo {author} {\bibfnamefont {T.}~\bibnamefont {Suzuki}}, \bibinfo {author} {\bibfnamefont {R.}~\bibnamefont {Chisnell}}, \bibinfo {author} {\bibfnamefont {A.}~\bibnamefont {Devarakonda}}, \bibinfo {author} {\bibfnamefont {Y.-T.}\ \bibnamefont {Liu}}, \bibinfo {author} {\bibfnamefont {W.}~\bibnamefont {Feng}}, \bibinfo {author} {\bibfnamefont {D.}~\bibnamefont {Xiao}}, \bibinfo {author} {\bibfnamefont {J.}~\bibnamefont {Lynn}}, \ and\ \bibinfo {author} {\bibfnamefont {J.}~\bibnamefont {Checkelsky}},\ }\bibinfo {title} {Large anomalous {Hall} effect in a half-{Heusler} antiferromagnet},\ \href {\doibase 10.1038/nphys3831} {\bibfield  {journal} {\bibinfo  {journal} {Nature Physics}\ }\textbf {\bibinfo {volume} {12}},\ \bibinfo {pages} {1119} (\bibinfo {year} {2016})}\BibitemShut {NoStop}%
\bibitem [{\citenamefont {Takahashi}\ \emph {et~al.}(2018)\citenamefont {Takahashi}, \citenamefont {Ishizuka}, \citenamefont {Murata}, \citenamefont {Wang}, \citenamefont {Tokura}, \citenamefont {Nagaosa},\ and\ \citenamefont {Kawasaki}}]{cant_takahashi_Science2018}%
  \BibitemOpen
  \bibfield  {author} {\bibinfo {author} {\bibfnamefont {K.~S.}\ \bibnamefont {Takahashi}}, \bibinfo {author} {\bibfnamefont {H.}~\bibnamefont {Ishizuka}}, \bibinfo {author} {\bibfnamefont {T.}~\bibnamefont {Murata}}, \bibinfo {author} {\bibfnamefont {Q.~Y.}\ \bibnamefont {Wang}}, \bibinfo {author} {\bibfnamefont {Y.}~\bibnamefont {Tokura}}, \bibinfo {author} {\bibfnamefont {N.}~\bibnamefont {Nagaosa}}, \ and\ \bibinfo {author} {\bibfnamefont {M.}~\bibnamefont {Kawasaki}},\ }\bibinfo {title} {Anomalous {Hall} effect derived from multiple {Weyl} nodes in high-mobility {EuTiO}$_{\textrm{3}}$ films},\ \href {\doibase 10.1126/sciadv.aar7880} {\bibfield  {journal} {\bibinfo  {journal} {Sci. Adv.}\ }\textbf {\bibinfo {volume} {4}},\ \bibinfo {pages} {eaar7880} (\bibinfo {year} {2018})}\BibitemShut {NoStop}%
\bibitem [{\citenamefont {Borisenko}\ \emph {et~al.}(2019)\citenamefont {Borisenko}, \citenamefont {Evtushinsky}, \citenamefont {Gibson}, \citenamefont {Yaresko}, \citenamefont {Koepernik}, \citenamefont {Kim}, \citenamefont {Ali}, \citenamefont {Van Den~Brink}, \citenamefont {Hoesch}, \citenamefont {Fedorov}, \citenamefont {Haubold}, \citenamefont {Kushnirenko}, \citenamefont {Soldatov}, \citenamefont {Sch$\rm\ddot{a}$fer},\ and\ \citenamefont {Cava}}]{Borisenko2019}%
  \BibitemOpen
  \bibfield  {author} {\bibinfo {author} {\bibfnamefont {S.}~\bibnamefont {Borisenko}}, \bibinfo {author} {\bibfnamefont {D.}~\bibnamefont {Evtushinsky}}, \bibinfo {author} {\bibfnamefont {Q.}~\bibnamefont {Gibson}}, \bibinfo {author} {\bibfnamefont {A.}~\bibnamefont {Yaresko}}, \bibinfo {author} {\bibfnamefont {K.}~\bibnamefont {Koepernik}}, \bibinfo {author} {\bibfnamefont {T.}~\bibnamefont {Kim}}, \bibinfo {author} {\bibfnamefont {M.}~\bibnamefont {Ali}}, \bibinfo {author} {\bibfnamefont {J.}~\bibnamefont {Van Den~Brink}}, \bibinfo {author} {\bibfnamefont {M.}~\bibnamefont {Hoesch}}, \bibinfo {author} {\bibfnamefont {A.}~\bibnamefont {Fedorov}}, \bibinfo {author} {\bibfnamefont {E.}~\bibnamefont {Haubold}}, \bibinfo {author} {\bibfnamefont {Y.}~\bibnamefont {Kushnirenko}}, \bibinfo {author} {\bibfnamefont {I.}~\bibnamefont {Soldatov}}, \bibinfo {author} {\bibfnamefont {R.}~\bibnamefont {Sch$\rm\ddot{a}$fer}}, \ and\ \bibinfo {author} {\bibfnamefont {R.~J.}\ \bibnamefont {Cava}},\ }\bibinfo {title} {Time-reversal symmetry breaking type-{II} {Weyl} state in {YbMnBi$_2$}},\ \href {\doibase 10.1038/s41467-019-11393-5} {\bibfield  {journal} {\bibinfo  {journal} {Nat. Commun.}\ }\textbf {\bibinfo {volume} {10}},\ \bibinfo {pages} {3424} (\bibinfo {year} {2019})}\BibitemShut {NoStop}%
\bibitem [{\citenamefont {Le}\ \emph {et~al.}(2021)\citenamefont {Le}, \citenamefont {Felser},\ and\ \citenamefont {Sun}}]{cant_LC_PRB2021}%
  \BibitemOpen
  \bibfield  {author} {\bibinfo {author} {\bibfnamefont {C.}~\bibnamefont {Le}}, \bibinfo {author} {\bibfnamefont {C.}~\bibnamefont {Felser}}, \ and\ \bibinfo {author} {\bibfnamefont {Y.}~\bibnamefont {Sun}},\ }\bibinfo {title} {Design strong anomalous Hall effect via spin canting in antiferromagnetic nodal line materials},\ \href {\doibase 10.1103/PhysRevB.104.125145} {\bibfield  {journal} {\bibinfo  {journal} {Phys. Rev. B}\ }\textbf {\bibinfo {volume} {104}},\ \bibinfo {pages} {125145} (\bibinfo {year} {2021})}\BibitemShut {NoStop}%
\bibitem [{\citenamefont {Pan}\ \emph {et~al.}(2022)\citenamefont {Pan}, \citenamefont {Le}, \citenamefont {He}, \citenamefont {Watzman}, \citenamefont {Yao}, \citenamefont {Gooth}, \citenamefont {Heremans}, \citenamefont {Sun},\ and\ \citenamefont {Felser}}]{cant_pan_NM2022}%
  \BibitemOpen
  \bibfield  {author} {\bibinfo {author} {\bibfnamefont {Y.}~\bibnamefont {Pan}}, \bibinfo {author} {\bibfnamefont {C.}~\bibnamefont {Le}}, \bibinfo {author} {\bibfnamefont {B.}~\bibnamefont {He}}, \bibinfo {author} {\bibfnamefont {S.~J.}\ \bibnamefont {Watzman}}, \bibinfo {author} {\bibfnamefont {M.}~\bibnamefont {Yao}}, \bibinfo {author} {\bibfnamefont {J.}~\bibnamefont {Gooth}}, \bibinfo {author} {\bibfnamefont {J.~P.}\ \bibnamefont {Heremans}}, \bibinfo {author} {\bibfnamefont {Y.}~\bibnamefont {Sun}}, \ and\ \bibinfo {author} {\bibfnamefont {C.}~\bibnamefont {Felser}},\ }\bibinfo {title} {Giant anomalous {Nernst} signal in the antiferromagnet {YbMnBi2}},\ \href {\doibase 10.1038/s41563-021-01149-2} {\bibfield  {journal} {\bibinfo  {journal} {Nat. Mater.}\ }\textbf {\bibinfo {volume} {21}},\ \bibinfo {pages} {203} (\bibinfo {year} {2022})}\BibitemShut {NoStop}%
\bibitem [{\citenamefont {Das}\ \emph {et~al.}(2022)\citenamefont {Das}, \citenamefont {Ross}, \citenamefont {Ma}, \citenamefont {Becker}, \citenamefont {Schmitt}, \citenamefont {Van~Duijn}, \citenamefont {Galindez-Ruales}, \citenamefont {Fuhrmann}, \citenamefont {Syskaki}, \citenamefont {Ebels}, \citenamefont {Baltz}, \citenamefont {Barra}, \citenamefont {Chen}, \citenamefont {Jakob}, \citenamefont {Cao}, \citenamefont {Sinova}, \citenamefont {Gomonay}, \citenamefont {Lebrun},\ and\ \citenamefont {Kläui}}]{cant_das_NC2022}%
  \BibitemOpen
  \bibfield  {author} {\bibinfo {author} {\bibfnamefont {S.}~\bibnamefont {Das}}, \bibinfo {author} {\bibfnamefont {A.}~\bibnamefont {Ross}}, \bibinfo {author} {\bibfnamefont {X.~X.}\ \bibnamefont {Ma}}, \bibinfo {author} {\bibfnamefont {S.}~\bibnamefont {Becker}}, \bibinfo {author} {\bibfnamefont {C.}~\bibnamefont {Schmitt}}, \bibinfo {author} {\bibfnamefont {F.}~\bibnamefont {Van~Duijn}}, \bibinfo {author} {\bibfnamefont {E.~F.}\ \bibnamefont {Galindez-Ruales}}, \bibinfo {author} {\bibfnamefont {F.}~\bibnamefont {Fuhrmann}}, \bibinfo {author} {\bibfnamefont {M.-A.}\ \bibnamefont {Syskaki}}, \bibinfo {author} {\bibfnamefont {U.}~\bibnamefont {Ebels}}, \bibinfo {author} {\bibfnamefont {V.}~\bibnamefont {Baltz}}, \bibinfo {author} {\bibfnamefont {A.-L.}\ \bibnamefont {Barra}}, \bibinfo {author} {\bibfnamefont {H.~Y.}\ \bibnamefont {Chen}}, \bibinfo {author} {\bibfnamefont {G.}~\bibnamefont {Jakob}}, \bibinfo {author} {\bibfnamefont {S.~X.}\ \bibnamefont {Cao}}, \bibinfo {author} {\bibfnamefont {J.}~\bibnamefont {Sinova}}, \bibinfo {author} {\bibfnamefont {O.}~\bibnamefont {Gomonay}}, \bibinfo {author} {\bibfnamefont {R.}~\bibnamefont {Lebrun}}, \ and\ \bibinfo {author} {\bibfnamefont {M.}~\bibnamefont {Kläui}},\ }\bibinfo {title} {Anisotropic long-range spin transport in canted antiferromagnetic orthoferrite {YFeO3}},\ \href {\doibase 10.1038/s41467-022-33520-5} {\bibfield  {journal} {\bibinfo  {journal} {Nat. Commun.}\ }\textbf {\bibinfo {volume} {13}},\ \bibinfo {pages} {6140} (\bibinfo {year} {2022})}\BibitemShut {NoStop}%
\bibitem [{\citenamefont {Li}\ \emph {et~al.}(2023)\citenamefont {Li}, \citenamefont {Koo}, \citenamefont {Zhu}, \citenamefont {Behnia},\ and\ \citenamefont {Yan}}]{cant_li_NC2023}%
  \BibitemOpen
  \bibfield  {author} {\bibinfo {author} {\bibfnamefont {X.}~\bibnamefont {Li}}, \bibinfo {author} {\bibfnamefont {J.}~\bibnamefont {Koo}}, \bibinfo {author} {\bibfnamefont {Z.}~\bibnamefont {Zhu}}, \bibinfo {author} {\bibfnamefont {K.}~\bibnamefont {Behnia}}, \ and\ \bibinfo {author} {\bibfnamefont {B.}~\bibnamefont {Yan}},\ }\bibinfo {title} {Field-linear anomalous {Hall} effect and {Berry} curvature induced by spin chirality in the kagome antiferromagnet {Mn3Sn}},\ \href {\doibase 10.1038/s41467-023-37076-w} {\bibfield  {journal} {\bibinfo  {journal} {Nat. Commun.}\ }\textbf {\bibinfo {volume} {14}},\ \bibinfo {pages} {1642} (\bibinfo {year} {2023})}\BibitemShut {NoStop}%
\bibitem [{\citenamefont {El~Kanj}\ \emph {et~al.}(2023)\citenamefont {El~Kanj}, \citenamefont {Gomonay}, \citenamefont {Boventer}, \citenamefont {Bortolotti}, \citenamefont {Cros}, \citenamefont {Anane},\ and\ \citenamefont {Lebrun}}]{cant_EK_SA2023}%
  \BibitemOpen
  \bibfield  {author} {\bibinfo {author} {\bibfnamefont {A.}~\bibnamefont {El~Kanj}}, \bibinfo {author} {\bibfnamefont {O.}~\bibnamefont {Gomonay}}, \bibinfo {author} {\bibfnamefont {I.}~\bibnamefont {Boventer}}, \bibinfo {author} {\bibfnamefont {P.}~\bibnamefont {Bortolotti}}, \bibinfo {author} {\bibfnamefont {V.}~\bibnamefont {Cros}}, \bibinfo {author} {\bibfnamefont {A.}~\bibnamefont {Anane}}, \ and\ \bibinfo {author} {\bibfnamefont {R.}~\bibnamefont {Lebrun}},\ }\bibinfo {title} {Antiferromagnetic magnon spintronic based on nonreciprocal and nondegenerated ultra-fast spin-waves in the canted antiferromagnet $\alpha$-{Fe}$_{\textrm{2}}$ {O}$_{\textrm{3}}$},\ \href {\doibase 10.1126/sciadv.adh1601} {\bibfield  {journal} {\bibinfo  {journal} {Sci. Adv.}\ }\textbf {\bibinfo {volume} {9}},\ \bibinfo {pages} {eadh1601} (\bibinfo {year} {2023})}\BibitemShut {NoStop}%
\bibitem [{\citenamefont {Zhang}\ \emph {et~al.}(2024{\natexlab{c}})\citenamefont {Zhang}, \citenamefont {Gao}, \citenamefont {Chien}, \citenamefont {Liu}, \citenamefont {Curtis}, \citenamefont {Sung}, \citenamefont {Ma}, \citenamefont {Ren}, \citenamefont {Cao}, \citenamefont {Narang}, \citenamefont {Von~Hoegen}, \citenamefont {Baldini},\ and\ \citenamefont {Nelson}}]{cant_Zhang_NP2024}%
  \BibitemOpen
  \bibfield  {author} {\bibinfo {author} {\bibfnamefont {Z.}~\bibnamefont {Zhang}}, \bibinfo {author} {\bibfnamefont {F.~Y.}\ \bibnamefont {Gao}}, \bibinfo {author} {\bibfnamefont {Y.-C.}\ \bibnamefont {Chien}}, \bibinfo {author} {\bibfnamefont {Z.-J.}\ \bibnamefont {Liu}}, \bibinfo {author} {\bibfnamefont {J.~B.}\ \bibnamefont {Curtis}}, \bibinfo {author} {\bibfnamefont {E.~R.}\ \bibnamefont {Sung}}, \bibinfo {author} {\bibfnamefont {X.}~\bibnamefont {Ma}}, \bibinfo {author} {\bibfnamefont {W.}~\bibnamefont {Ren}}, \bibinfo {author} {\bibfnamefont {S.}~\bibnamefont {Cao}}, \bibinfo {author} {\bibfnamefont {P.}~\bibnamefont {Narang}}, \bibinfo {author} {\bibfnamefont {A.}~\bibnamefont {Von~Hoegen}}, \bibinfo {author} {\bibfnamefont {E.}~\bibnamefont {Baldini}}, \ and\ \bibinfo {author} {\bibfnamefont {K.~A.}\ \bibnamefont {Nelson}},\ }\bibinfo {title} {Terahertz-field-driven magnon upconversion in an antiferromagnet},\ \href {\doibase 10.1038/s41567-023-02350-7} {\bibfield  {journal} {\bibinfo  {journal} {Nat. Phys.}\ }\textbf {\bibinfo {volume} {20}},\ \bibinfo {pages} {788} (\bibinfo {year} {2024}{\natexlab{c}})}\BibitemShut {NoStop}%
\bibitem [{\citenamefont {Zhang}\ \emph {et~al.}(2024{\natexlab{d}})\citenamefont {Zhang}, \citenamefont {Gao}, \citenamefont {Curtis}, \citenamefont {Liu}, \citenamefont {Chien}, \citenamefont {Von~Hoegen}, \citenamefont {Wong}, \citenamefont {Kurihara}, \citenamefont {Suemoto}, \citenamefont {Narang}, \citenamefont {Baldini},\ and\ \citenamefont {Nelson}}]{cant_Zhang_NP2024_2}%
  \BibitemOpen
  \bibfield  {author} {\bibinfo {author} {\bibfnamefont {Z.}~\bibnamefont {Zhang}}, \bibinfo {author} {\bibfnamefont {F.~Y.}\ \bibnamefont {Gao}}, \bibinfo {author} {\bibfnamefont {J.~B.}\ \bibnamefont {Curtis}}, \bibinfo {author} {\bibfnamefont {Z.-J.}\ \bibnamefont {Liu}}, \bibinfo {author} {\bibfnamefont {Y.-C.}\ \bibnamefont {Chien}}, \bibinfo {author} {\bibfnamefont {A.}~\bibnamefont {Von~Hoegen}}, \bibinfo {author} {\bibfnamefont {M.~T.}\ \bibnamefont {Wong}}, \bibinfo {author} {\bibfnamefont {T.}~\bibnamefont {Kurihara}}, \bibinfo {author} {\bibfnamefont {T.}~\bibnamefont {Suemoto}}, \bibinfo {author} {\bibfnamefont {P.}~\bibnamefont {Narang}}, \bibinfo {author} {\bibfnamefont {E.}~\bibnamefont {Baldini}}, \ and\ \bibinfo {author} {\bibfnamefont {K.~A.}\ \bibnamefont {Nelson}},\ }\bibinfo {title} {Terahertz field-induced nonlinear coupling of two magnon modes in an antiferromagnet},\ \href {\doibase 10.1038/s41567-024-02386-3} {\bibfield  {journal} {\bibinfo  {journal} {Nat. Phys.}\ }\textbf {\bibinfo {volume} {20}},\ \bibinfo {pages} {801} (\bibinfo {year} {2024}{\natexlab{d}})}\BibitemShut {NoStop}%
\bibitem [{\citenamefont {Leenders}\ \emph {et~al.}(2024)\citenamefont {Leenders}, \citenamefont {Afanasiev}, \citenamefont {Kimel},\ and\ \citenamefont {Mikhaylovskiy}}]{cant_leenders_N2024}%
  \BibitemOpen
  \bibfield  {author} {\bibinfo {author} {\bibfnamefont {R.~A.}\ \bibnamefont {Leenders}}, \bibinfo {author} {\bibfnamefont {D.}~\bibnamefont {Afanasiev}}, \bibinfo {author} {\bibfnamefont {A.~V.}\ \bibnamefont {Kimel}}, \ and\ \bibinfo {author} {\bibfnamefont {R.~V.}\ \bibnamefont {Mikhaylovskiy}},\ }\bibinfo {title} {Canted spin order as a platform for ultrafast conversion of magnons},\ \href {\doibase 10.1038/s41586-024-07448-3} {\bibfield  {journal} {\bibinfo  {journal} {Nature}\ }\textbf {\bibinfo {volume} {630}},\ \bibinfo {pages} {335} (\bibinfo {year} {2024})}\BibitemShut {NoStop}%
\bibitem [{\citenamefont {Yang}\ \emph {et~al.}(2020)\citenamefont {Yang}, \citenamefont {Corasaniti}, \citenamefont {Le}, \citenamefont {Liao}, \citenamefont {Wang}, \citenamefont {Du}, \citenamefont {Petrovic}, \citenamefont {Qiu}, \citenamefont {Hu},\ and\ \citenamefont {Degiorgi}}]{YangR2020}%
  \BibitemOpen
  \bibfield  {author} {\bibinfo {author} {\bibfnamefont {R.}~\bibnamefont {Yang}}, \bibinfo {author} {\bibfnamefont {M.}~\bibnamefont {Corasaniti}}, \bibinfo {author} {\bibfnamefont {C.~C.}\ \bibnamefont {Le}}, \bibinfo {author} {\bibfnamefont {Z.~Y.}\ \bibnamefont {Liao}}, \bibinfo {author} {\bibfnamefont {A.~F.}\ \bibnamefont {Wang}}, \bibinfo {author} {\bibfnamefont {Q.}~\bibnamefont {Du}}, \bibinfo {author} {\bibfnamefont {C.}~\bibnamefont {Petrovic}}, \bibinfo {author} {\bibfnamefont {X.~G.}\ \bibnamefont {Qiu}}, \bibinfo {author} {\bibfnamefont {J.~P.}\ \bibnamefont {Hu}}, \ and\ \bibinfo {author} {\bibfnamefont {L.}~\bibnamefont {Degiorgi}},\ }\bibinfo {title} {Spin-Canting-Induced Band Reconstruction in the Dirac Material ${\mathrm{Ca}}_{1\ensuremath{-}x}{\mathrm{Na}}_{x}{\mathrm{MnBi}}_{2}$},\ \href {\doibase 10.1103/PhysRevLett.124.137201} {\bibfield  {journal} {\bibinfo  {journal} {Phys. Rev. Lett.}\ }\textbf {\bibinfo {volume} {124}},\ \bibinfo {pages} {137201} (\bibinfo {year} {2020})}\BibitemShut {NoStop}%
\bibitem [{\citenamefont {Singh}\ \emph {et~al.}(2024)\citenamefont {Singh}, \citenamefont {Sau}, \citenamefont {Rai}, \citenamefont {Panda}, \citenamefont {Kumar},\ and\ \citenamefont {Kumar}}]{Singh2024}%
  \BibitemOpen
  \bibfield  {author} {\bibinfo {author} {\bibfnamefont {M.}~\bibnamefont {Singh}}, \bibinfo {author} {\bibfnamefont {J.}~\bibnamefont {Sau}}, \bibinfo {author} {\bibfnamefont {B.}~\bibnamefont {Rai}}, \bibinfo {author} {\bibfnamefont {A.}~\bibnamefont {Panda}}, \bibinfo {author} {\bibfnamefont {M.}~\bibnamefont {Kumar}}, \ and\ \bibinfo {author} {\bibfnamefont {N.}~\bibnamefont {Kumar}},\ }\bibinfo {title} {Tuning intrinsic anomalous Hall effect from large to zero in two ferromagnetic states of $\mathrm{Sm}{\mathrm{Mn}}_{2}{\mathrm{Ge}}_{2}$},\ \href {\doibase 10.1103/PhysRevMaterials.8.084201} {\bibfield  {journal} {\bibinfo  {journal} {Phys. Rev. Mater.}\ }\textbf {\bibinfo {volume} {8}},\ \bibinfo {pages} {084201} (\bibinfo {year} {2024})}\BibitemShut {NoStop}%
\bibitem [{\citenamefont {Guo}\ \emph {et~al.}(2022)\citenamefont {Guo}, \citenamefont {Putzke}, \citenamefont {Konyzheva}, \citenamefont {Huang}, \citenamefont {Gutierrez-Amigo}, \citenamefont {Errea}, \citenamefont {Chen}, \citenamefont {Vergniory}, \citenamefont {Felser}, \citenamefont {Fischer} \emph {et~al.}}]{CY-Guo2022}%
  \BibitemOpen
  \bibfield  {author} {\bibinfo {author} {\bibfnamefont {C.}~\bibnamefont {Guo}}, \bibinfo {author} {\bibfnamefont {C.}~\bibnamefont {Putzke}}, \bibinfo {author} {\bibfnamefont {S.}~\bibnamefont {Konyzheva}}, \bibinfo {author} {\bibfnamefont {X.}~\bibnamefont {Huang}}, \bibinfo {author} {\bibfnamefont {M.}~\bibnamefont {Gutierrez-Amigo}}, \bibinfo {author} {\bibfnamefont {I.}~\bibnamefont {Errea}}, \bibinfo {author} {\bibfnamefont {D.}~\bibnamefont {Chen}}, \bibinfo {author} {\bibfnamefont {M.~G.}\ \bibnamefont {Vergniory}}, \bibinfo {author} {\bibfnamefont {C.}~\bibnamefont {Felser}}, \bibinfo {author} {\bibfnamefont {M.~H.}\ \bibnamefont {Fischer}},  \emph {et~al.},\ }\bibinfo {title} {Switchable chiral transport in charge-ordered kagome metal CsV3Sb5},\ \href {\doibase https://doi.org/10.1038/s41586-022-05127-9} {\bibfield  {journal} {\bibinfo  {journal} {Nature}\ }\textbf {\bibinfo {volume} {611}},\ \bibinfo {pages} {461} (\bibinfo {year} {2022})}\BibitemShut {NoStop}%
\bibitem [{\citenamefont {Togo}\ \emph {et~al.}(2023)\citenamefont {Togo}, \citenamefont {Chaput}, \citenamefont {Tadano},\ and\ \citenamefont {Tanaka}}]{phonopy-phono3py-JPCM}%
  \BibitemOpen
  \bibfield  {author} {\bibinfo {author} {\bibfnamefont {A.}~\bibnamefont {Togo}}, \bibinfo {author} {\bibfnamefont {L.}~\bibnamefont {Chaput}}, \bibinfo {author} {\bibfnamefont {T.}~\bibnamefont {Tadano}}, \ and\ \bibinfo {author} {\bibfnamefont {I.}~\bibnamefont {Tanaka}},\ }\bibinfo {title} {Implementation strategies in phonopy and phono3py},\ \href {\doibase 10.1088/1361-648X/acd831} {\bibfield  {journal} {\bibinfo  {journal} {J. Phys. Condens. Matter}\ }\textbf {\bibinfo {volume} {35}},\ \bibinfo {pages} {353001} (\bibinfo {year} {2023})}\BibitemShut {NoStop}%
\bibitem [{\citenamefont {Togo}(2023)}]{phonopy-phono3py-JPSJ}%
  \BibitemOpen
  \bibfield  {author} {\bibinfo {author} {\bibfnamefont {A.}~\bibnamefont {Togo}},\ }\bibinfo {title} {First-principles Phonon Calculations with Phonopy and Phono3py},\ \href {\doibase 10.7566/JPSJ.92.012001} {\bibfield  {journal} {\bibinfo  {journal} {J. Phys. Soc. Jpn.}\ }\textbf {\bibinfo {volume} {92}},\ \bibinfo {pages} {012001} (\bibinfo {year} {2023})}\BibitemShut {NoStop}%
\bibitem [{\citenamefont {He}\ \emph {et~al.}(2021)\citenamefont {He}, \citenamefont {Helbig}, \citenamefont {Verstraete},\ and\ \citenamefont {Bousquet}}]{TB2J}%
  \BibitemOpen
  \bibfield  {author} {\bibinfo {author} {\bibfnamefont {X.}~\bibnamefont {He}}, \bibinfo {author} {\bibfnamefont {N.}~\bibnamefont {Helbig}}, \bibinfo {author} {\bibfnamefont {M.~J.}\ \bibnamefont {Verstraete}}, \ and\ \bibinfo {author} {\bibfnamefont {E.}~\bibnamefont {Bousquet}},\ }\bibinfo {title} {TB2J: A python package for computing magnetic interaction parameters},\ \href {\doibase https://doi.org/10.1016/j.cpc.2021.107938} {\bibfield  {journal} {\bibinfo  {journal} {Comput. Phys. Commun.}\ }\textbf {\bibinfo {volume} {264}},\ \bibinfo {pages} {107938} (\bibinfo {year} {2021})}\BibitemShut {NoStop}%
\bibitem [{\citenamefont {Feng}\ \emph {et~al.}(2025{\natexlab{b}})\citenamefont {Feng}, \citenamefont {Zhou}, \citenamefont {Chen}, \citenamefont {Xu}, \citenamefont {Yang},\ and\ \citenamefont {Li}}]{FengJQ2025}%
  \BibitemOpen
  \bibfield  {author} {\bibinfo {author} {\bibfnamefont {J.}~\bibnamefont {Feng}}, \bibinfo {author} {\bibfnamefont {X.}~\bibnamefont {Zhou}}, \bibinfo {author} {\bibfnamefont {J.}~\bibnamefont {Chen}}, \bibinfo {author} {\bibfnamefont {M.}~\bibnamefont {Xu}}, \bibinfo {author} {\bibfnamefont {X.}~\bibnamefont {Yang}}, \ and\ \bibinfo {author} {\bibfnamefont {Y.}~\bibnamefont {Li}},\ }\bibinfo {title} {Ferroelectric antiferromagnetic lifting of spin-valley degeneracy},\ \href {\doibase 10.1103/jyq4-d4gs} {\bibfield  {journal} {\bibinfo  {journal} {Phys. Rev. B}\ }\textbf {\bibinfo {volume} {111}},\ \bibinfo {pages} {214446} (\bibinfo {year} {2025}{\natexlab{b}})}\BibitemShut {NoStop}%
\bibitem [{\citenamefont {Zhang}\ \emph {et~al.}(2024{\natexlab{e}})\citenamefont {Zhang}, \citenamefont {Cui}, \citenamefont {Li}, \citenamefont {Duan}, \citenamefont {Li}, \citenamefont {Yu},\ and\ \citenamefont {Yao}}]{ZhangRW2024}%
  \BibitemOpen
  \bibfield  {author} {\bibinfo {author} {\bibfnamefont {R.-W.}\ \bibnamefont {Zhang}}, \bibinfo {author} {\bibfnamefont {C.}~\bibnamefont {Cui}}, \bibinfo {author} {\bibfnamefont {R.}~\bibnamefont {Li}}, \bibinfo {author} {\bibfnamefont {J.}~\bibnamefont {Duan}}, \bibinfo {author} {\bibfnamefont {L.}~\bibnamefont {Li}}, \bibinfo {author} {\bibfnamefont {Z.-M.}\ \bibnamefont {Yu}}, \ and\ \bibinfo {author} {\bibfnamefont {Y.}~\bibnamefont {Yao}},\ }\bibinfo {title} {Predictable Gate-Field Control of Spin in Altermagnets with Spin-Layer Coupling},\ \href {\doibase 10.1103/PhysRevLett.133.056401} {\bibfield  {journal} {\bibinfo  {journal} {Phys. Rev. Lett.}\ }\textbf {\bibinfo {volume} {133}},\ \bibinfo {pages} {056401} (\bibinfo {year} {2024}{\natexlab{e}})}\BibitemShut {NoStop}%
\bibitem [{\citenamefont {Mazurenko}\ and\ \citenamefont {Anisimov}(2005)}]{Mazurenko2005}%
  \BibitemOpen
  \bibfield  {author} {\bibinfo {author} {\bibfnamefont {V.~V.}\ \bibnamefont {Mazurenko}}\ and\ \bibinfo {author} {\bibfnamefont {V.~I.}\ \bibnamefont {Anisimov}},\ }\bibinfo {title} {Weak ferromagnetism in antiferromagnets: $\ensuremath{\alpha}\text{\ensuremath{-}}{\mathrm{Fe}}_{2}{\mathrm{O}}_{3}$ and ${\mathrm{La}}_{2}\mathrm{Cu}{\mathrm{O}}_{4}$},\ \href {https://link.aps.org/doi/10.1103/PhysRevB.71.184434} {\bibfield  {journal} {\bibinfo  {journal} {Phys. Rev. B}\ }\textbf {\bibinfo {volume} {71}},\ \bibinfo {pages} {184434} (\bibinfo {year} {2005})}\BibitemShut {NoStop}%
\bibitem [{\citenamefont {Logemann}\ \emph {et~al.}(2017)\citenamefont {Logemann}, \citenamefont {Rudenko}, \citenamefont {Katsnelson},\ and\ \citenamefont {Kirilyuk}}]{Logemann2017}%
  \BibitemOpen
  \bibfield  {author} {\bibinfo {author} {\bibfnamefont {R.}~\bibnamefont {Logemann}}, \bibinfo {author} {\bibfnamefont {A.~N.}\ \bibnamefont {Rudenko}}, \bibinfo {author} {\bibfnamefont {M.~I.}\ \bibnamefont {Katsnelson}}, \ and\ \bibinfo {author} {\bibfnamefont {A.}~\bibnamefont {Kirilyuk}},\ }\bibinfo {title} {Exchange interactions in transition metal oxides: the role of oxygen spin polarization},\ \href {https://dx.doi.org/10.1088/1361-648X/aa7b00} {\bibfield  {journal} {\bibinfo  {journal} {J. Phys.: Condens. Matter}\ }\textbf {\bibinfo {volume} {29}},\ \bibinfo {pages} {335801} (\bibinfo {year} {2017})}\BibitemShut {NoStop}%
\bibitem [{\citenamefont {Morrish}(1995)}]{Morrish1995}%
  \BibitemOpen
  \bibfield  {author} {\bibinfo {author} {\bibfnamefont {A.~H.}\ \bibnamefont {Morrish}},\ }\href {\doibase 10.1142/2518} {\bibinfo {title} {Canted Antiferromagnetism: Hematite}}\ (\bibinfo  {publisher} {World Scientific},\ \bibinfo {year} {1995})\BibitemShut {NoStop}%
\bibitem [{\citenamefont {Machala}\ \emph {et~al.}(2011)\citenamefont {Machala}, \citenamefont {Tu$\check{c}$ek},\ and\ \citenamefont {Zbo$\check{r}$il}}]{Machala2011}%
  \BibitemOpen
  \bibfield  {author} {\bibinfo {author} {\bibfnamefont {L.}~\bibnamefont {Machala}}, \bibinfo {author} {\bibfnamefont {J.}~\bibnamefont {Tu$\check{c}$ek}}, \ and\ \bibinfo {author} {\bibfnamefont {R.}~\bibnamefont {Zbo$\check{r}$il}},\ }\bibinfo {title} {Polymorphous Transformations of Nanometric Iron(III) Oxide: A Review},\ \href {https://doi.org/10.1021/cm200397g} {\bibfield  {journal} {\bibinfo  {journal} {Chem. Mater.}\ }\textbf {\bibinfo {volume} {23}},\ \bibinfo {pages} {3255} (\bibinfo {year} {2011})}\BibitemShut {NoStop}%
\bibitem [{\citenamefont {Abrajevitch}\ \emph {et~al.}(2021)\citenamefont {Abrajevitch}, \citenamefont {Roberts}, \citenamefont {Pillans},\ and\ \citenamefont {Hori}}]{Abrajevitch2021}%
  \BibitemOpen
  \bibfield  {author} {\bibinfo {author} {\bibfnamefont {A.}~\bibnamefont {Abrajevitch}}, \bibinfo {author} {\bibfnamefont {A.~P.}\ \bibnamefont {Roberts}}, \bibinfo {author} {\bibfnamefont {B.~J.}\ \bibnamefont {Pillans}}, \ and\ \bibinfo {author} {\bibfnamefont {R.~S.}\ \bibnamefont {Hori}},\ }\bibinfo {title} {Unexpected Magnetic Behavior of Natural Hematite-Bearing Rocks at Low Temperatures},\ \href {https://agupubs.onlinelibrary.wiley.com/doi/abs/10.1029/2021GC010094} {\bibfield  {journal} {\bibinfo  {journal} {Geochemistry, Geophysics, Geosystems}\ }\textbf {\bibinfo {volume} {22}},\ \bibinfo {pages} {e2021GC010094} (\bibinfo {year} {2021})}\BibitemShut {NoStop}%
\bibitem [{\citenamefont {Masina}\ \emph {et~al.}(2024)\citenamefont {Masina}, \citenamefont {Chithwayo}, \citenamefont {Moyo}, \citenamefont {Dlamini},\ and\ \citenamefont {Wamwangi}}]{Masina2024}%
  \BibitemOpen
  \bibfield  {author} {\bibinfo {author} {\bibfnamefont {C.}~\bibnamefont {Masina}}, \bibinfo {author} {\bibfnamefont {A.}~\bibnamefont {Chithwayo}}, \bibinfo {author} {\bibfnamefont {T.}~\bibnamefont {Moyo}}, \bibinfo {author} {\bibfnamefont {S.}~\bibnamefont {Dlamini}}, \ and\ \bibinfo {author} {\bibfnamefont {D.}~\bibnamefont {Wamwangi}},\ }\bibinfo {title} {High room temperature coercivity from $\alpha$-Fe2O3 nanoparticles embedded in silica},\ \href {https://www.sciencedirect.com/science/article/pii/S0304885324008126} {\bibfield  {journal} {\bibinfo  {journal} {J. Magn. Magn. Mater.}\ }\textbf {\bibinfo {volume} {610}},\ \bibinfo {pages} {172521} (\bibinfo {year} {2024})}\BibitemShut {NoStop}%
\bibitem [{\citenamefont {Zysler}\ \emph {et~al.}(2001)\citenamefont {Zysler}, \citenamefont {Vasquez-Mansilla}, \citenamefont {Arciprete}, \citenamefont {Dimitrijewits}, \citenamefont {Rodriguez-Sierra},\ and\ \citenamefont {Saragovi}}]{Zysler2001}%
  \BibitemOpen
  \bibfield  {author} {\bibinfo {author} {\bibfnamefont {R.}~\bibnamefont {Zysler}}, \bibinfo {author} {\bibfnamefont {M.}~\bibnamefont {Vasquez-Mansilla}}, \bibinfo {author} {\bibfnamefont {C.}~\bibnamefont {Arciprete}}, \bibinfo {author} {\bibfnamefont {M.}~\bibnamefont {Dimitrijewits}}, \bibinfo {author} {\bibfnamefont {D.}~\bibnamefont {Rodriguez-Sierra}}, \ and\ \bibinfo {author} {\bibfnamefont {C.}~\bibnamefont {Saragovi}},\ }\bibinfo {title} {Structure and magnetic properties of thermally treated nanohematite},\ \href {https://www.sciencedirect.com/science/article/pii/S0304885300013652} {\bibfield  {journal} {\bibinfo  {journal} {J. Magn. Magn. Mater.}\ }\textbf {\bibinfo {volume} {224}},\ \bibinfo {pages} {39} (\bibinfo {year} {2001})}\BibitemShut {NoStop}%
\bibitem [{\citenamefont {Cornell}\ \emph {et~al.}(2003)\citenamefont {Cornell}, \citenamefont {Schwertmann} \emph {et~al.}}]{cornell2003iron}%
  \BibitemOpen
  \bibfield  {author} {\bibinfo {author} {\bibfnamefont {R.~M.}\ \bibnamefont {Cornell}}, \bibinfo {author} {\bibfnamefont {U.}~\bibnamefont {Schwertmann}},  \emph {et~al.},\ }\href@noop {} {\bibinfo {title} {The iron oxides: structure, properties, reactions, occurrences, and uses}},\ Vol.\ \bibinfo {volume} {664}\ (\bibinfo  {publisher} {Wiley-vch Weinheim},\ \bibinfo {year} {2003})\BibitemShut {NoStop}%
\bibitem [{\citenamefont {Kub$\acute{a}$niov$\acute{a}$}\ \emph {et~al.}(2019)\citenamefont {Kub$\acute{a}$niov$\acute{a}$}, \citenamefont {Kub$\acute{i}\check{c}$kov$\acute{a}$}, \citenamefont {Kmje$\check{c}$}, \citenamefont {Z$\acute{a}$v$\check{e}$ta}, \citenamefont {Ni$\check{z}\check{n}$ansk$\acute{y}$}, \citenamefont {Br$\acute{a}$zda}, \citenamefont {Klementov$\acute{a}$},\ and\ \citenamefont {Kohout}}]{Kubaniova2019}%
  \BibitemOpen
  \bibfield  {author} {\bibinfo {author} {\bibfnamefont {D.}~\bibnamefont {Kub$\acute{a}$niov$\acute{a}$}}, \bibinfo {author} {\bibfnamefont {L.}~\bibnamefont {Kub$\acute{i}\check{c}$kov$\acute{a}$}}, \bibinfo {author} {\bibfnamefont {T.}~\bibnamefont {Kmje$\check{c}$}}, \bibinfo {author} {\bibfnamefont {K.}~\bibnamefont {Z$\acute{a}$v$\check{e}$ta}}, \bibinfo {author} {\bibfnamefont {D.}~\bibnamefont {Ni$\check{z}\check{n}$ansk$\acute{y}$}}, \bibinfo {author} {\bibfnamefont {P.}~\bibnamefont {Br$\acute{a}$zda}}, \bibinfo {author} {\bibfnamefont {M.}~\bibnamefont {Klementov$\acute{a}$}}, \ and\ \bibinfo {author} {\bibfnamefont {J.}~\bibnamefont {Kohout}},\ }\bibinfo {title} {Hematite: Morin temperature of nanoparticles with different size},\ \href {https://www.sciencedirect.com/science/article/pii/S0304885318327896} {\bibfield  {journal} {\bibinfo  {journal} {J. Magn. Magn. Mater.}\ }\textbf {\bibinfo {volume} {475}},\ \bibinfo {pages} {611} (\bibinfo {year} {2019})}\BibitemShut {NoStop}%
\bibitem [{\citenamefont {Kresse}\ and\ \citenamefont {Hafner}(1993)}]{Kresse1993}%
  \BibitemOpen
  \bibfield  {author} {\bibinfo {author} {\bibfnamefont {G.}~\bibnamefont {Kresse}}\ and\ \bibinfo {author} {\bibfnamefont {J.}~\bibnamefont {Hafner}},\ }\bibinfo {title} {Ab initio molecular dynamics for liquid metals},\ \href {https://link.aps.org/doi/10.1103/PhysRevB.47.558} {\bibfield  {journal} {\bibinfo  {journal} {Phys. Rev. B}\ }\textbf {\bibinfo {volume} {47}},\ \bibinfo {pages} {558} (\bibinfo {year} {1993})}\BibitemShut {NoStop}%
\bibitem [{\citenamefont {Kresse}\ and\ \citenamefont {Furthm{\"u}ller}(1996)}]{Kresse1996}%
  \BibitemOpen
  \bibfield  {author} {\bibinfo {author} {\bibfnamefont {G.}~\bibnamefont {Kresse}}\ and\ \bibinfo {author} {\bibfnamefont {J.}~\bibnamefont {Furthm{\"u}ller}},\ }\bibinfo {title} {Efficient iterative schemes for ab initio total-energy calculations using a plane-wave basis set},\ \href {https://link.aps.org/doi/10.1103/PhysRevB.54.11169} {\bibfield  {journal} {\bibinfo  {journal} {Phys. Rev. B}\ }\textbf {\bibinfo {volume} {54}},\ \bibinfo {pages} {11169} (\bibinfo {year} {1996})}\BibitemShut {NoStop}%
\bibitem [{\citenamefont {Perdew}\ \emph {et~al.}(1996)\citenamefont {Perdew}, \citenamefont {Burke},\ and\ \citenamefont {Ernzerhof}}]{Perdew1996}%
  \BibitemOpen
  \bibfield  {author} {\bibinfo {author} {\bibfnamefont {J.~P.}\ \bibnamefont {Perdew}}, \bibinfo {author} {\bibfnamefont {K.}~\bibnamefont {Burke}}, \ and\ \bibinfo {author} {\bibfnamefont {M.}~\bibnamefont {Ernzerhof}},\ }\bibinfo {title} {Generalized gradient approximation made simple},\ \href {https://link.aps.org/doi/10.1103/PhysRevLett.77.3865} {\bibfield  {journal} {\bibinfo  {journal} {Phys. Rev. Lett.}\ }\textbf {\bibinfo {volume} {77}},\ \bibinfo {pages} {3865} (\bibinfo {year} {1996})}\BibitemShut {NoStop}%
\bibitem [{\citenamefont {Mostofi}\ \emph {et~al.}(2008)\citenamefont {Mostofi}, \citenamefont {Yates}, \citenamefont {Lee}, \citenamefont {Souza}, \citenamefont {Vanderbilt},\ and\ \citenamefont {Marzari}}]{Arash2008}%
  \BibitemOpen
  \bibfield  {author} {\bibinfo {author} {\bibfnamefont {A.~A.}\ \bibnamefont {Mostofi}}, \bibinfo {author} {\bibfnamefont {J.~R.}\ \bibnamefont {Yates}}, \bibinfo {author} {\bibfnamefont {Y.-S.}\ \bibnamefont {Lee}}, \bibinfo {author} {\bibfnamefont {I.}~\bibnamefont {Souza}}, \bibinfo {author} {\bibfnamefont {D.}~\bibnamefont {Vanderbilt}}, \ and\ \bibinfo {author} {\bibfnamefont {N.}~\bibnamefont {Marzari}},\ }\bibinfo {title} {wannier90: A tool for obtaining maximally-localised Wannier functions},\ \href {\doibase https://doi.org/10.1016/j.cpc.2007.11.016} {\bibfield  {journal} {\bibinfo  {journal} {Comput. Phys. Commun.}\ }\textbf {\bibinfo {volume} {178}},\ \bibinfo {pages} {685} (\bibinfo {year} {2008})}\BibitemShut {NoStop}%
\bibitem [{\citenamefont {Tsirkin}(2021)}]{wannierberri}%
  \BibitemOpen
  \bibfield  {author} {\bibinfo {author} {\bibfnamefont {S.~S.}\ \bibnamefont {Tsirkin}},\ }\bibinfo {title} {High performance {Wannier} interpolation of {Berry} curvature and related quantities with {WannierBerri} code},\ \href {\doibase 10.1038/s41524-021-00498-5} {\bibfield  {journal} {\bibinfo  {journal} {npj Computational Materials}\ }\textbf {\bibinfo {volume} {7}},\ \bibinfo {pages} {33} (\bibinfo {year} {2021})}\BibitemShut {NoStop}%
\bibitem [{\citenamefont {Anisimov}\ \emph {et~al.}(1991)\citenamefont {Anisimov}, \citenamefont {Zaanen},\ and\ \citenamefont {Andersen}}]{Anisimov_LDAU1991}%
  \BibitemOpen
  \bibfield  {author} {\bibinfo {author} {\bibfnamefont {V.~I.}\ \bibnamefont {Anisimov}}, \bibinfo {author} {\bibfnamefont {J.}~\bibnamefont {Zaanen}}, \ and\ \bibinfo {author} {\bibfnamefont {O.~K.}\ \bibnamefont {Andersen}},\ }\bibinfo {title} {Band theory and Mott insulators: Hubbard U instead of Stoner I},\ \href {\doibase 10.1103/PhysRevB.44.943} {\bibfield  {journal} {\bibinfo  {journal} {Phys. Rev. B}\ }\textbf {\bibinfo {volume} {44}},\ \bibinfo {pages} {943} (\bibinfo {year} {1991})}\BibitemShut {NoStop}%
\bibitem [{\citenamefont {Dudarev}\ \emph {et~al.}(1998)\citenamefont {Dudarev}, \citenamefont {Botton}, \citenamefont {Savrasov}, \citenamefont {Humphreys},\ and\ \citenamefont {Sutton}}]{Dudarev_LDAU1998}%
  \BibitemOpen
  \bibfield  {author} {\bibinfo {author} {\bibfnamefont {S.~L.}\ \bibnamefont {Dudarev}}, \bibinfo {author} {\bibfnamefont {G.~A.}\ \bibnamefont {Botton}}, \bibinfo {author} {\bibfnamefont {S.~Y.}\ \bibnamefont {Savrasov}}, \bibinfo {author} {\bibfnamefont {C.~J.}\ \bibnamefont {Humphreys}}, \ and\ \bibinfo {author} {\bibfnamefont {A.~P.}\ \bibnamefont {Sutton}},\ }\bibinfo {title} {Electron-energy-loss spectra and the structural stability of nickel oxide: An LSDA+U study},\ \href {\doibase 10.1103/PhysRevB.57.1505} {\bibfield  {journal} {\bibinfo  {journal} {Phys. Rev. B}\ }\textbf {\bibinfo {volume} {57}},\ \bibinfo {pages} {1505} (\bibinfo {year} {1998})}\BibitemShut {NoStop}%
\end{thebibliography}%

\end{document}